\documentclass[aps,pr,reprint,groupedaddress,nofootinbib,byrevtex]{revtex4-1}
\usepackage{amsmath,amssymb}
\usepackage{graphicx}
\usepackage{mathrsfs}
\begin{document}
\title{ Wave packet sizes in quantum mechanical scatterings : \\new perspective  } 
\author{K. Ishikawa${}^{1,2}$ and  {O. Jinnouchi${}^{3}$}}
\affiliation{${}^1$Department of Physics, Faculty of Science, Hokkaido
University, Sapporo 060-0810, Japan, \\{${}^2$Natural Science Center,Keio University, Yokohama 223-8521, Japan},\\{${}^3$
Department of Physics, Faculty of Science, Institute of Science Tokyo, Tokyo 152-8550,Japan},\\ }  
\date{\today}

%\author{ K. Ishikawa and O.Jinnouchi }
%\maketitle

%\affiliation{  Department of Physics, Hokkaido University, 
%             Sapporo 060-0810, Japan }
%\centering{
%\begin{center}
%  Department of Physics, Faculty of Science, 
%Hokkaido University Sapporo 060-0810, Japan
%\end{center}
%\date{\today}0
%%%%%%%%%%%%%%%%%%%%%%%%%%%%%%%%%%%%%%%%%%%% abst %%%%%%%%%%%%%%%%%%%%%%%%%%%%%%%%%%%%%%%%%%%%%%%%%%%
\begin{abstract}

 %Particles have  a probability of existence of 1 and are expressed by wave packets in the  quantum  mechanics. The wave packets satisfy the Schr\"{o}dinger equation  and
%are either bound states or continuum states of finite coherence lengths. These  make transitions according to absolute  probabilities  in unambiguous manner. The magnitude %depends on the wave-packet sizes. Because the  sizes  are not universal constant but depend on   the environments and experimental situations,  correct  estimations are %required.   The sizes of propagating wave packets are determined by their coherence lengths in environments, and those of bound states are determined by the wave %functions. The systematic analysis of the wave packet sizes is presented.        

 Particles exist with a unit probability and are represented by wave packets in quantum mechanics. These wave packets satisfy the Schrödinger equation and are classified as either bound states or continuum states with finite coherence lengths. They undergo transitions according to unambiguous absolute probabilities, the magnitudes of which depend on the sizes of the wave packets. Because these sizes are not universal constants but depend on the environment and experimental conditions, their accurate estimation is essential. The sizes of propagating wave packets are determined by their environmental coherence lengths, whereas those of bound states are determined by their wave functions. Here, we present a systematic analysis of wave-packet sizes.

 %In quantum mechanics, particles are represented by wave packets with a total existence probability of 1. Satisfying the Schrödinger equation, these wave packets form either  %bound or continuum states with finite coherence lengths. They undergo transitions according to unambiguous absolute probabilities that depend directly on the wave-packet  %sizes. Since these sizes are not universal constants but are determined by environmental and experimental factors, precise estimation is required. While the sizes of  %propagating wave packets are governed by environmental coherence lengths, those of bound states are determined by their wave functions. This study presents a systematic  %analysis of wave-packet sizes.
 
%Absolute values  of scattering probabilities  are not provided  by plane waves but by normalized states, wave packets. Their sizes govern the magnitudes and properties of %scattering processes.  The microscopic processes give origin to 
%wide natural phenomena, where  sizes of initial and final states vary depending on situations.  Their magnitudes   are inevitable  in precise   comparisons of the theory with %experiments, and determined by  their interaction with matter  in enviorements. The wave packet properties and magnitudes of parameters are  elucidated.   

\end{abstract}
%\pacs{ 73.43.Lp}
\maketitle
\tableofcontents

\section{Introduction}
Matter  forms  stable objects and undergoes  transitions, both of which are driven  by interactions between particles and described by  quantum mechanics.   
These phenomena have conventionally  been formulated with  plane-wave formalisms. However, a  plane wave has a diverging norm and cannot  represnet a particle existing with a unit probability.    
A normalized state represents a particle, and  transitions to other normalized states with  a probability given  by the squared   absolute value of a scalar  product between  these states.    
The   transition probability $P(T)$ between  states over  a  time interval $T$ depends    linearly on  $T$ with two constant terms,
\begin{eqnarray}
P(T)=\Gamma T+P^{(d)} \label{absolute_probability}.
\end{eqnarray}
While  the transition rate $\Gamma$ has traditionaly been computed using the plane-wave formalism   as a standard physical quantity, the constant term  
$P^{(d)}$ has largely been overlooked.   The  term $P^{(d)}$ derives the transitions that reveal  rich  properties unobtainable from the transition rate  alone.   Because  these  may depend explicitly on  the wave-packet sizes,  a systematic study  
of the wace-packet sizes is  highly desirable.

 The absolute transition probability for  each transion is  positive semi-definite and  bounded by unity. This fundamental requirement  is naturally satisfied by physical wave packets.    
  Wave packets  yield physical  states  characterized by the mathematically  rigorous probability. Depending on the magnitude of $P^{(d)}$, physical rocesses exhibit  distinct  behaviors:
  \begin{itemize}
  \item $P_I $ of   $P^{(d)} \ll1$. The transitions  are governed by $\Gamma$ and expressed by the plane-wave formalism.
  \item $P_{II}$  of $P^{(d)} \sim 1$.   $\Gamma T  \sim 0$ and the transitions are governed by  $P^{(d)}$.  
 \end{itemize}
  Consequently, a wave-packet formalism is essential for a proper description for particle transitions.

Studies of wave-packets   have  a long history. Early works utilized wave packets  to bridge  the  uncertainty relation of physical quantities with  their  commutation relations   \cite{schroedinger} \cite{heisenberg} \cite{dirac},  well before  they were routinely integrated   into  scattering theory.  

Initially, research focused primarily on $P_{I}$-type processes.  By analyzing  the  asymptotic behaviors of normalized solutions at $ t \rightarrow \pm \infty $, the existence of   S-matrix was proved  in \cite{Reed-Simon,Kato}, although  concrete expressions for  the transition probabilities  were  not provided. Wave packets were subsequently applied to field theory in \cite{Feynman,LSZ}, as well as in  \cite{ Goldberger,newton,taylor,Sasakawa} and others. Difficulties inherent in the plane-wave formalism, such divergence of $|\delta(\Delta E)|^2$, were partially resolved. However, these studies were restricted to  states with  fixed  positions and a modified interaction Hamiltonian, $e^{-\epsilon |t|}H_{int}$, which represented the amplitudes of the short-range correlations.    Around 1960, the prevailing consensus was that: 
\begin{itemize}
\item  Wave packets   had  semi-microscopic sizes. 
\item Amplitudes  were  well    approximated by those of plane waves.
\item  Transition probabilities of such  wave packets were  almost equivalent to those obtained using  plane waves.  
\end{itemize}
 Since that era,  significant experimental advancements have been achieved   in spatial sizes, beam energies, and  quantitative precision, leading to the establishment of  unified theory of fundamental interactions.  Although the  majority of processes have been  successfully described by unified gauge theory within  the plane-wave framework, a increasing number  of  phenomena and experimental observations  cannot be adequately  explained  by this tradional formalism.   Notable examples include  low-intensity beam interference of  electrons \cite{Tonomura}  and single-photon interference   \cite{Tsuchiya}, where the observed  interference patterns  can hardly be described  by the tradional   $\Gamma$ alone.

 % One such example is the long-range correlation  theoretically predicted   by Einstein-Podolsky-Rosen (EPR)  \cite{EPR} and later experimentally confirmed   by Aspect et al.%\cite{Aspet}. This    long-range correlation plays a key role in modern  quantum mechanics and is expressed by  the absolute value of the transition probability  under the  interaction %Hamiltonian $H_{int}$.  

%   \cite{Ishikawa-Shimomura, Ishikawa-Tobita, Ishikawa-Tobita-ptp, Ishikawa-Tobita-ptep, Ishikawa-Tobita-anp,Ishikawa-Tajima-Tobita-ptep, Ishikawa-Oda,  Ishikawa-Oda-%Nishiwaki,Ishikawa-Nishiwaki-Jinnouchi-Oda}. 

%{\bf Add brief summary of these works and comparisons with the plane wave formalism: one or two paragraphs}

A novel wave-packet formalism  was recently proposed to analyze  both $P_{I}$ and  $P_{II}$ processes   \cite{Ishikawa-Tobita-ptep,Ishikawa-Tobita-anp,Ishikawa-Tajima-Tobita-ptep}. With this framework, the absolute transition probability  Eq.$(\ref{absolute_probability})$ under the un-modified  interaction Hamiltonian $H_{int}$ became  explicitly calculable.  It has been shown  that    a non-negligible  $P^{(d)}$  leads to significant discrepancies  between experimental observations  and standard evaluation  of $\Gamma$. Such deviations  have been reported across various contexts.  In particular, anomalies observed in neutrino experiments   \cite{Ishikawa-Tobita-ptep,Ishikawa-Tobita-anp,Ishikawa-Tajima-Tobita-ptep}, pion life time \cite{Ishikawa-Jinnouchi}, positron annihiltions \cite{Ishikawa-Jinnouchi,Ushioda}, vector meson decays near the thresholds \cite{Ishikawa-Nishiwaki-Jinnouchi-Oda},photosynthesis \cite{Maeda}, the solar-corona \cite{corona}, magnetization \cite{magnetization}, and the blue color of the sky \cite{sky}, among others are hard to understand from the standard theory.  
These discrepancies serve as key signatures of either new dynamical degrees of freedom or  the non-trivial contribution      of $P^{(d)}$.  \footnote{It is possible to  resolve inconsistencies with new dynamical freedoms, such as sterlile neutrino. Researchs in this direction is currently underway. }  Applying $P^{(d)}$ with appropriate wave-packet sizes successfully accounte for these anomalous observations and  uncover  the underlying physical mechanisms across  diverse   phenomena. 
%\footnote{It is possible to  resolve inconsistencies with new dynamical freedoms, such as sterlile neutrino. }
 Furthermore,  the transition probability  under the  unmodified interaction Hamiltonian $H_{int}$  properly account for the long-range correlations in addition to  short range ones
   \cite{Ishikawa-Shimomura, Ishikawa-Tobita, Ishikawa-Tobita-ptp, Ishikawa-Oda,  Ishikawa-Oda-Nishiwaki}. 

The essential feature  of this  new formalism   become explicit when evaluating  the transition   probability over  a complete set of arbitrary positions and momenta   \cite{Ishikawa-Shimomura}. 
Indeed, the probabilities of  the final states shifting in  position are non-negligible.  This  result indicates that  quantum processes span  large spatial regions over  wide time intervals, rainging from $10^{-9}$ second to several seconds or longe, \cite{Ishikawa-Shimomura, Ishikawa-Tobita, Ishikawa-Tobita-ptp, Ishikawa-Tobita-ptep, Ishikawa-Tobita-anp,Ishikawa-Tajima-Tobita-ptep, Ishikawa-Oda,  Ishikawa-Oda-Nishiwaki}. Modern understanding of  wave-packet sizes emcompasses the following feature:  
\begin{itemize}
\item Wave-packet sizes differ between the initial and final states, exhibiting a wide diversity.
\item The transition amplitudes evaluated under the unmodified interaction Hamiltonian $H_{int}$ differ fundamentally from those calculated using plain waves with $e^{-\epsilon |t|}H_{int}$, and explicitly depend on the initial and final wave packets. 
\end{itemize}

Wave-packet formalim introduces   a new scale of length through  the wave-packet size.  One length is    $\sim 10^{-15}$ meter for nucleus and    $\sim 10^{-10}$ for atoms. These are  sizes of bound states and expressed by  parameters in the Lagrangian. Another  length  is  the size of  continuum energy levels. This  is characterized  by a mean free path of particles in matter and can become large in dilute system.  This  
 is  not a parameter of the Lagrangian  but governs the absolute probability.  Because  the spatial size  can be as large as $10^5$ M or larger, this induces quantum phenomena  in macroscopic scales.  Wave-packet sizes  have universal features that  depend on the environment.   Systematic studies of the wave packet sizes based on the principles of quantum mechanics   are necessary  for  understnding quantum phenomena in wide area.  Extending previous  analysis on the sizes   
\cite{Ishikawa-Shimomura, Ishikawa-Tobita-ptp},  we present systematic studies of the wave packets.

The wave-packet sizes of  initial and final states in the interaction picture of scattering processes serve as essential physical  parameters and have been studied extensively  \cite{Ishikawa-Shimomura, Ishikawa-Tobita-ptp, Ishikawa-Oda,  Ishikawa-Oda-Nishiwaki,Ishikawa-Nishiwaki-Jinnouchi-Oda}.  In previous  literature,  other sizes  defined from a superposition of stationary continuum states of a total Hamiltonian have been  used. These states do neither represent  initial and final states seperately  nor  express the scattering of isolate states for short-range  potentials. This difficulty   is caused by the  non-orthogonality of  stationary continuum states with different energies  in short-range potentials as originally noted by Landau  \cite{landau}.  Wave packets constructed from such eigenstates exhibit  time-dependent norms \cite{ishikawa_1,ishikawa_2}, and fail to represent isolate states.  Likewise,  formalisms relying on the adiabatically switched  interaction $ H_{int} e^{-\epsilon |t|}$ neglect wave-packet overlap in  asymptotic regions  \cite{ Goldberger,newton,Sasakawa}.  Clarifying these distinctions resolves past ambiguities surrounding   wave- packet spatial scales.

This paper   presents a unified  wave-packet formalism applicable to both  $P_{I}$ and $P_{II}$ across diverse physical systems. While  wave-packet sizes are challenging  to measure directly, they  are intimately connected with  environmental conditions and  span a broad hierarchy  from microscopic to macroscopic scales. By analyzing   the structural properties of matters,  we clarify and estimate these physical  sizes.   
Specifically, we offer new  perspectives of  (1) quantum transition of large wave packets, (2)  the wave-packets of undetected particles, (3) two class of states and  their sizes  (4) transition of wave packet  in natural phenomena.

The paper is organized in the following manner. In the second section, the principles of quantum mechanics and general properties of one-particle wave functions are summarized. These are deeply connected with subtle problems such as  non-orthogonality of continuum stationary states, the probability interpretation of the wave function, classical-quantum correspondence, and others.  The sizes of the incident particles are found  in Section 3. The sizes of the outgoing particles observed with detectors  are found  in Section 4.  The wave packets play important roles also in natural phenomena, and the sizes  are studied in Section 5. The summary is given in Section 6. Appendix is devoted to equations used  in the main text and wave packets in uniform electric and magnetic fields.

%%%%%%%%%%%%%%%%%%%%%%%%%%%%%%%%%%%%%%%%%%%%%%%%%%%%%%%%%%%%%%%%%%%%%%%%%%%%%%%
%%%%%%%%%%%%%%%%%%%%%%%%%%%%%%%%%%%%%%%%%%%%%%%%%%%%%%%%%%%%%%%%%%%%%%%%%%%%%%%%%
\section{Wave packets in the quantum mechanics}
%%%%%%%%%%%%%%%%%%%%%%%%%%%%%%%%%%%%%%%%%%%%%%%%%%%%%%%%%%%%%%%%%%%%%%%%%%%%%%%
%%%%%%%%%%%%%%%%%%%%%%%%%%%%%%%%%%%%%%%%%%%%%%%%%%%%%%%%%%%%%%%%%%%%%%%%%%%%%%
%First we describe fundamentals  of wave-packets, and clarify various subtle points for succeeding sections.  

%The wave packets  in natural situations or in  experiments.  
 Particles in matter are either bound to  atoms or move freely. These are expressed by bound states of discrete energy or by scattering states of continuous energy. 
The states   in the continuum energies   follow  fundamental principles but vary  drastically  depending  on the situations.  Thypical sizes of charged particles and  photons are presented.

%%%%%%%%%%%%%%%%%%%%%%%%%%%%%%%%%%%%%%%%%%%%%%
\subsection{Principles of the quantum mechanics}
%%%%%%%%%%%%%%%%%%%%%%%%%%%%%%%%%%%%%%%%%%%%%%%%%
The wave-packet formalism follows  four  principles. Three of these principles---(1) the superposition principles, (2) commutation relations, and (3) the Schr\"{o}dinger equation-- map physical phenomena on to a mathematical space.  Physical space is represented by a complex vector space in which  a vector expresses a state, while   position $q_i$ and momentum $p_j$ are operators  satisfying the canonical commutation relations 
%\begin{eqnarray}
  $  [q_i,p_j]= i\hbar \delta_{ij}$. 
%    \end{eqnarray}
The state satisfies the Schr\"{o}dinger equation,
%\begin{eqnarray}
 $i \hbar \frac{\partial}{\partial t} \psi(t)=H \psi(t)  \label{Schroedinger_0}$.
%\end{eqnarray}

Now, the fourth principle introduces a probability framework  connecting  vector spaces with transition  processes.    
The  scalar product of states $\alpha$ and $\beta$, $\langle \psi_{\alpha}| \psi_{\beta} \rangle$, is a complex number, and the square of its modulus represents the probability that $\beta$ is included in $\alpha$, ( or the probability of their transition), for normalized states $\langle \alpha| \alpha \rangle=\langle \beta|\beta \rangle =1$. The probability satisfies $0 \leq P_{\alpha ,\beta} \leq 1$. If $P_{\alpha,\beta}=0$ the transition does not occur, whereas a process always  occurs if the corresponding probability is unity. A transition  $\alpha \rightarrow \beta_0$  always occus if $P_{\alpha \beta_0}=1$. In that case, a transition to other state does not occur because  $P_{\alpha,\beta'} =0$ for $ \beta' \neq \beta_0$. This  
 transition then  reveals a classical motion. As  will be shown later,  this  ensures a stability of a detection process. Fourth principle is  more general than the  Born's probability interpretation. \cite{ishikawa_3}.

Physical quantities derived from probability principle  include $\Gamma$ and $P^{(d)}$ in Eq.$(1)$. To account for their computations, the formalism holds  unique mathematical relations  that  have not appeared in  previous  formalisms.  
%%%%%%%%%%%%%%%%%%%%%%%%%%%%%%%%%%%%%%%%%%%%%%%%%%
\subsection{One-particle states of wave packets}
%%%%%%%%%%%%%%% %%%%%%%%%%%%%%%%%%%%%%%%%%%%%%
One particle state in the vacuum is a fundamental building block of matter, specified by mass, spin, and  momentum $|m,s,\vec p \rangle $, and normalization  
\begin{eqnarray}
    \langle m_1,s_1, \vec p_1| m_2,s_2,\vec p_2 \rangle=\delta_{m_1,m_2} \delta_{s_1,s_2} (2\pi)^3 \delta (\vec p_1-\vec p_2)
\end{eqnarray}
using the Dirac delta function.  The square of the delta function  is formally  expressed as  
\begin{eqnarray}
\delta(\vec p_1-\vec p_2)^2= \delta(\vec 0) \delta(\vec p_1-\vec p_2),
\end{eqnarray}  
where $\delta(0)$ is divergent.  The approximation $\delta(\vec 0)=(2\pi)^3 V$ with  integration   volume $V$ is commonly used. However, this is an approximation, and the  next-order term in $\frac{1}{V}$ is unknown for  this state.  Despite the fact that the scalar products of plane waves are treated using  the Dirac delta functions, the divergence $\delta(\vec 0)$ in its square can not be avoided.   \cite{schwartz}
Because the  probability for each of these plane-wave states to exist is not  unity,  the transition probability suffers ambiguity.   This  deficit is resolved by using  wave packets.  Normalized states  make transitions with finite probabilities. Each particle is thus characterized by mass, spin, momentum, and wave packet size.    

  A   wave packet satisfying 
\begin{eqnarray}
    & &|\langle \Psi(t) |\Psi (t)\rangle|^2 =|\langle \Psi(0)|\Psi(0) \rangle|^2 =1 
\end{eqnarray}
and governed by the  the free  wave equation  $i \hbar \frac{\partial}{\partial t} \psi(t)=H_0 \psi(t)$ for a free Hamiltonian $H_0$     is expressed as a  superposition of plane waves:
\begin{eqnarray}
\Psi(\vec x,t) = \int d{\vec p} \langle {\vec x}|{\vec p}\rangle \langle t,{\vec p}| {\vec P}_0, {\vec X}_0, T_0 \rangle.  \label{wave-packet}
\end{eqnarray}
Here a plane wave of a momentum $\vec p$ is given  by
\begin{eqnarray}
\langle \vec x| \vec p \rangle= \frac{1}{(2 \pi \hbar)^{3/2}}e^{i {\frac{\vec p}{\hbar}}\vec x},
\end{eqnarray}
  and the  weight factor is given  by
\begin{eqnarray}
\langle t, {\vec p}| {\vec P}_0, {\vec X}_0, T_0  \rangle= N e^{\xi(\vec p,\vec P_0,\vec X_0)} \label{wave-packet_{1_1}}. 
\end{eqnarray}
Fig.1. shows a wave packet $\Psi(\vec x,t)$,  plane waves $\langle \vec x| \vec p \rangle $, and a weight $\langle t, {\vec p}| {\vec P}_0, {\vec X}_0, T_0  \rangle $. A weight is shown in blue color and plotted in a vertical direction. This has a peak  energy $E(\vec P_0)$ and   a peak momentum  $\vec P_0$ at a center position ${\vec X}_0$ at an initial time $T_0$. An exponential factor  has a form  
 \begin{eqnarray}
\xi(\vec p,\vec P_0,\vec X_0) =-i\frac{E(\vec p)}{\hbar}(t-T_0)-i{\frac{\vec p}{\hbar}{\vec X_0}}-\gamma[{\vec p}-{\vec P_0}] \label{wave-packet_{shape}}
\end{eqnarray}
 where $E(\vec p)$ is determined from the wave equation and  $\gamma[
\vec p-\vec P_0]$ is real and  positive semi-definite.  The normalization factor is given by
\begin{eqnarray}
& &|N|^2=\left(\int d^3p e^{-2\gamma[{\vec p}-{\vec P_0}] }\right)^{-1}.
\end{eqnarray}
\begin{figure}[t]
\includegraphics[width=.4\textwidth, bb=0 0 541 532]{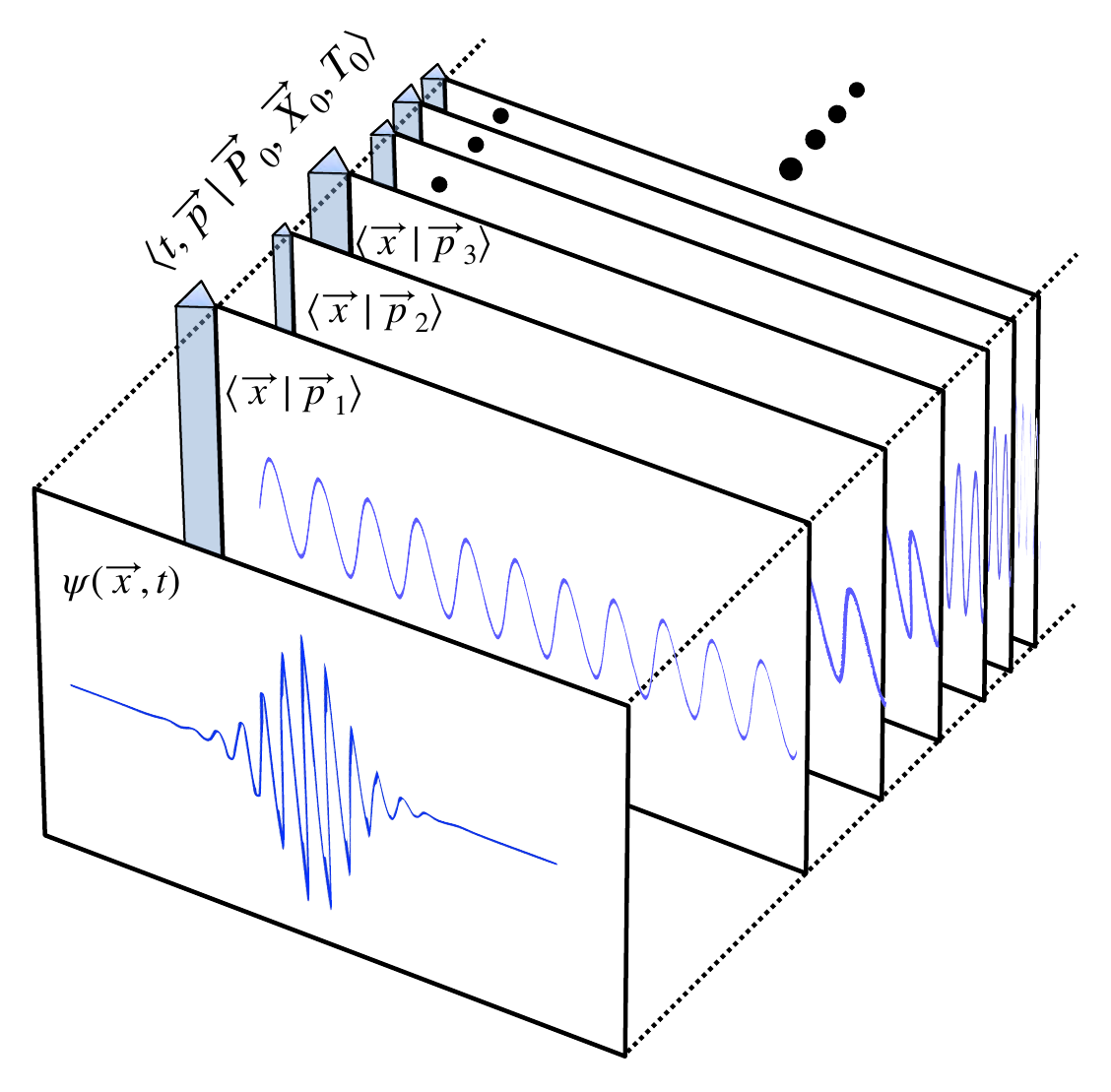}
\caption{  A wave packet $\Psi({\vec x},t)$  is a superposition of plane waves with different momenta ${\vec p}_i$, $ \langle x|\vec p_i \rangle $,  weighted by a distribution function, and is localized around a center position ${\vec X}_0$ and momentum ${\vec P}_0$.  The  weight is plotted along  the vertical directions. }
\label{fig:superposition of plane waves}
\end{figure}
%%%%%%%%%%%%%%%%%%%%%%%%%%%%%%%%%%%%%%%%%%%%%%%%%%%%%%%%%
An explicit form of $\gamma[{\vec p}-{\vec P}_0]$ will be presented in the next subsection.

Wave packets  satisfy  normalization and completeness  conditions,  
\begin{eqnarray}
& &\langle {\vec P}_0,{\vec X}_0,T_0| {\vec P}_0,{\vec X}_0,T_0 \rangle =1, \label{normalization}\\
& & \int \frac{d^3P_0 d^3X_0 }{(2\pi \hbar)^3} |{\vec P}_0, {\vec X}_0,T_0 \rangle \langle {\vec P}_0, {\vec X}_0,T_0 |=1 \label{completeness},
\end{eqnarray}
over the variables  ${\vec P}_0$  and ${\vec X}_0$  
 \cite{Ishikawa-Shimomura}.  A normalized vector $|\Psi\rangle$  is expressed as
\begin{eqnarray}
|\Psi \rangle= \int \frac{d^3P_0 d^3X_0 }{(2\pi \hbar)^3}  |{\vec P}_0, {\vec X}_0,T_0 \rangle \langle {\vec P}_0, {\vec X}_0,T_0 |\Psi \rangle,
\end{eqnarray}
and the scalar product of two states is expressed as
\begin{eqnarray}
\langle \Phi|\Psi \rangle= \int \frac{d^3P_0 d^3X_0 }{(2\pi \hbar)^3} \langle \Phi|  {\vec P}_0, {\vec X}_0,T_0 \rangle \langle {\vec P}_0, {\vec X}_0,T_0 |\Psi \rangle.
\end{eqnarray}
The square of the norm 
\begin{eqnarray}
\langle \Phi|\Phi \rangle= \int \frac{d^3P_0 d^3X_0 }{(2\pi \hbar)^3} | \langle \Phi|  {\vec P}_0, {\vec X}_0,T_0 \rangle|^2,
\end{eqnarray}
vanishes if and only if  $ \langle \Phi|  {\vec P}_0, {\vec X}_0,T_0 \rangle=0$. 
Thus, wave packets with arbitrary values of position and momentum form a complete set.  Moreover,  matrixes  $\langle \alpha | \vec x | \beta \rangle $, $\langle \alpha | \vec p | \beta \rangle $,  $\langle \alpha | H(\vec x,\vec p) | \beta \rangle $, and others satisfy   $ \langle \alpha | O | \beta \rangle =(\langle \beta | O  | \alpha  \rangle)^*  $ for normalized states $|\alpha \rangle$ and $|\beta \rangle $.  \footnote{ For normalized functions,  using the integration by part,  $ \int_{-\infty}^{\infty} dx \alpha(x)^{*}( {\frac{\partial }{\partial x }})^2 \beta(x) = (\int_{-\infty}^{\infty} dx \beta(x)^{*}( {\frac{\partial }{\partial x }})^2 \alpha(x))^{*}$. }

%%%%%%%%%%%%%%%%%%%%%%%%%%%%%%%%%%% 
\subsubsection{Gaussian wave packets}
%%%%%%%%%%%%%%%%%%%%%%%%%%%%%%%%%%%

  The simplest   weight factor in Eq.$( \ref{wave-packet_{shape}})$ is    given by a  Gaussian function    
\begin{eqnarray}
\gamma[{\vec p}-{\vec P_0}]=\frac{\sigma}{2}({\vec p}-{\vec P_0})^2,
\end{eqnarray}
where   $\sigma$ is a width parameter. 
Then      
 \begin{eqnarray}
& &\langle t, {\vec p}| {\vec P}_0, {\vec X}_0, T_0  \rangle= N_3 \sigma^{3/2} e^{\xi_G(\vec p,0)},  \label{wave-packet_1} \nonumber\\
& &\xi_G(\vec p,0) =-i\frac{E(\vec p)}{\hbar}(t-T_0)-i{\frac{\vec p}{\hbar}{\vec X_0}}-\frac{\sigma}{2}({\vec p}-{\vec P_0})^2,\nonumber \\
\end{eqnarray}
where $N_3=(\pi \sigma)^{-3/4}$  and  $1/\sigma$ represents the variables  in the momentum distribution. The wave function in the coordinate space is 
\begin{eqnarray}
 & &\langle t,{\vec x}| {\vec P}_0, {\vec X}_0, T_0  \rangle=    \label{wave-packet_2}  
N_3 {\left( \frac{\sigma}{2\pi}\right)^{3/2}}\int d{\vec p} e^{\xi_G(\vec p, \vec x)}, \nonumber \\
& &\xi_G(\vec p, \vec x)=  -i\frac{E(\vec p)}{\hbar}(t-T_0)+i\frac{{\vec p}}{\hbar}({\vec x}-{\vec X_0})-\frac{\sigma}{2}(\vec p-{\vec P_0})^2. \nonumber \\
\end{eqnarray}
Integrating  over the momentum for   a large  $t-T_0 $ using the   stationary phase approximation around ${\vec p}={\vec P}_0$ yields,
\begin{eqnarray}
& &\langle t,{\vec x}| {\vec P}_0, {\vec X}_0, T_0  \rangle \nonumber \\
&=& N_3 e^{-\frac{1}{2 \sigma \hbar^2}\left( \vec x-{\vec X}_0 -{\vec v}_0(t-T_0)\right)^2 -i\frac{E({\vec P}_0)}{\hbar}(t-T_0)+i\frac{{\vec P}_0}{\hbar}({\vec x}-{\vec X}_0)}   \label{wave-packet_3}, \nonumber  \\
 & &v_0^i = \frac{\partial}{\partial p_i}E(p)|p=P_0, 
\end{eqnarray}
in the region where wave-packet spreading  is negligible.
At the initial time $t=T_0$, the wave function resembles  a plane wave in the  spatial region $\frac{|{\vec x}-{\vec X}_0|}{  \sqrt \sigma} \leq 1$ and vanishes at large $\frac{|{\vec x}-{\vec X}_0|}{\sqrt \sigma} \gg 1$. 
At  later time $t >T_0$,the wave in  Eq.$(\ref{wave-packet_3})$, retains its initial shape and width, centered at  $\vec x={\vec X}_0 +{\vec v}_0(t-T_0)$. The  overall  phase at the center can be  rewritten  as
\begin{eqnarray}
& & \frac{1}{\hbar}\left( -E({\vec P}_0)(t-T_0)+{\vec P}_0({\vec x}-{\vec X}_0)\right)  \nonumber \\
 & &  =- \frac{1}{\hbar}\left( E({\vec P}_0)  - {\vec P}_0  {\vec v}_0 \right) (t-T_0), \label{frequency_moving}
\end{eqnarray}
where the right-hand side corresponds to the   phase in the moving frame of  a constant group velocity ${\vec v}_0$.  At $t-T_0 \rightarrow \infty$,  the wave function is finite at an area of infinite distance around $ {\vec X}_0 +{\vec v}_0(t-T_0) $. 
 The function Eq.$(\ref{wave-packet_3})$ remains  normalized at all  times.

A product of wave packets with different configurations possesses  the same  property and does not vanish at $t=\pm \infty $. The product of wave packets and its integral also remain finite.  
The overlap matrix element with   wave packets of configuration  $(\vec P_1,\vec X_1,T_1)$ of size $\sigma_1 $ and $(\vec P_2, \vec X_2,T_2)$ of size  $ \sigma_2$    is given by
\begin{eqnarray}
& &\langle {\vec P}_2, {\vec X}_2, T_2 ; \sigma_2|{\vec P}_1, {\vec X}_1, T_1; \sigma_1 \rangle \nonumber \\
%& &= \int d \vec x   \langle {\vec P}_2, {\vec X}_2, T_2 ; \sigma_2|t, \vec x \rangle \langle t, \vec x | {\vec P}_1, {\vec X}_1, T_1; \sigma_1 \rangle \nonumber \\
%& &=N_3(\sigma_2) N_3(\sigma_1) \int d{\vec x}e^{-\frac{1}{2 \sigma_2 \hbar^2}\left( \vec x-{\vec X}_2 -{\vec v}_2(t-T_2)\right)^2 +i\frac{E({\vec P}_2)}{\hbar}(t-T_2)-%i\frac{{\vec P}_2}{\hbar}({\vec x}-{\vec X}_2)} \nonumber \\ & &\times  e^{-\frac{1}{2 \sigma_1 \hbar^2}\left( \vec x-{\vec X}_1 -{\vec v}_1(t-T_1)\right)^2 
%-i\frac{E({\vec P}_1)}{\hbar}(t-T_1)+i\frac{{\vec P}_1}{\hbar}({\vec x}-{\vec X}_1)} \nonumber \\
& &=\left(\frac{2 \sqrt {\sigma_1 \sigma_2}}{\sigma_1+\sigma_2}\right)^{3/2} 
e^{-\frac{\sigma_s}{2}(\vec p_1-\vec P_2)^2-\frac{(\vec X_1(t)-\vec X_2(t))^2}{2(\sigma_1+\sigma_2)}-i{\bar \vec X(t) }(\vec P_1-\vec P_2)},\label{overlap_wp} \nonumber \\
\end{eqnarray}
where
\begin{eqnarray}
  \vec X_i(t)=\vec X_i+ \vec v_i(t-T_i); \quad i=1,2.  
\end{eqnarray}
The square of the modulus satisfies
\begin{eqnarray}
| \langle {\vec P}_2, {\vec X}_2, T_2 ; \sigma_2|{\vec P}_1, {\vec X}_1, T_1; \sigma_1 \rangle|^2   \leq \left(\frac{2 \sqrt {\sigma_1 \sigma_2}}{\sigma_1+\sigma_2}\right)^{3} \leq 1, 
\end{eqnarray}
where the first equality when 
\begin{eqnarray}
    \vec P_2=\vec P_1, \vec X_2(t)=\vec X_1(t)
\end{eqnarray}
and the second equality holds when   
\begin{eqnarray}
    \sigma_2=\sigma_1.\label{classical}
\end{eqnarray}
 In this case,the transition probability  $1 \rightarrow 2 $ is  unity, and the probability  ot transitions  to other states vanishes.  Accordingly,  a wave packet  follows a particle trajectory of  classical motion.  The probability is less than the unity for $\sigma_2 \neq \sigma_1$. A quantum mechanical process obeys  a classical causality of motion  when its probability is  unity.  
%%%%%%%%%%%%%%%%%%%%%%%%%%%%%%%%%%%%%%%%%%%%%%%%%%%%%%
\subsection{Stationary states in external potentials}
%%%%%%%%%%%%%%%%%%%%%%%%%%%%%%%%%%%%%%%%%%%%%%%%%%%%%%

A non-relativistic particle  of mass $m$ with kinetic energy $H_0=\frac{{\vec p}^2}{2m}$ in  a potential $V(\vec x)$ is described by  the Schr\"{o}dinger equation
\begin{eqnarray}
& &i \hbar \frac{\partial}{\partial t} \Psi({\vec x},t )  =(H_0+V(x)) \Psi( {\vec x},t),  \label{Schroedinger} \nonumber \\
& &\Psi({\vec x},t)=\langle  {\vec x}| \Psi({\vec x},t )   \rangle.
\end{eqnarray}
The probability density and  probab ility current density,  defined by
\begin{eqnarray}
& &\rho(\vec x,t)=\psi^{\dagger}(\vec x,t) \psi(\vec x,t),  \nonumber \\
& &\vec j(\vec x,t)=\psi^{\dagger}(\vec x,t) \frac{\vec p}{2m}\psi(\vec x,t)-\left(\frac{\vec p}{2m}\psi(\vec x,t)^{\dagger} \right) \psi(\vec x,t)\nonumber \\
\end{eqnarray}
satisfy the continuity equation and conservation law:
\begin{eqnarray}
& &\frac{\partial }{\partial t  } \rho(\vec x,t)-\nabla  \vec j(\vec x,t)=0, \label{continiuty} \nonumber \\
& &\frac{\partial }{\partial t   } \int d \vec x  \rho(\vec x,t)-\int d \vec S   \vec j(\vec x,t)=0, \label{conservation} 
\end{eqnarray}
where the spatial integration of the divergence was converted into a surfact integral using   Gauss's theorem.

Stationary solutions are  energy eigenstates of the Hamiltonian  
\begin{eqnarray}
& & \Psi({\vec x},t ) =e^{iEt/{\hbar }} \langle {\vec x}| E \rangle, \nonumber \\
& & (H_0+V(\vec x))  \langle {\vec x}| E \rangle=E \langle {\vec x}| E \rangle, 
\end{eqnarray}
satisfying  $\frac{\partial }{\partial t  } \rho(\vec x,t)=0$.
 A stationary  state is either localized or extended. Localized states are described by normalized wave functions with discrete energy spectra, wher extended states are described by non-normalized wave functions with continuous energy spectra.

%%%%%%%%%%%%%%%%%%%%%%%%%%%%%%%%%%%%%%%%%%%%%%%%%%%%%% 
\subsubsection{Bound state}
%%%%%%%%%%%%%%%%%%%%%%%%%%%%%%%%%%%%%%%%%%%%
 Bound states in an attractive potential  $V(\vec x)$ have  discrete energy $E_l$ satisfying  
\begin{eqnarray}
& & (H_0+V(\vec x))  \langle {\vec x}| E_l \rangle=E_l \langle {\vec x}| E_l \rangle, \quad l=1,2,\cdots. \label{bound-state}
%& & \langle E_m, {\vec x} | E_l, {\vec x} \rangle  =\delta_{l,m},
\end{eqnarray}
 Multiplying $ \langle E_m |\vec x \rangle $ from the left and integrating over coordinates gives 
\begin{eqnarray}
& & \int d {\vec x} \langle E_m |\vec x \rangle [(H_0+V(\vec x) ) \langle {\vec x}| E_l \rangle]=E_l \langle E_m | E_l \rangle. \label{bound-state_s} 
%& & \langle E_m, {\vec x} | E_l, {\vec x} \rangle  =\delta_{l,m},\nonumber \\
\end{eqnarray} 
 Taking   the complex conjugate of Eq.$ \ref{bound-state} $ yields 
\begin{eqnarray}
& & \langle E_m|{\vec x} \rangle (H_0+V(\vec x))^{\dagger}  =E_m^{*} \langle E_m| {\vec x} \rangle,  \label{bound-state_c}
%& & \langle E_m, {\vec x} | E_l, {\vec x} \rangle  =\delta_{l,m},
\end{eqnarray}  
and $\langle {\vec x}| E_l \rangle $  is given by
\begin{eqnarray}
& & \int d {\vec x} [\langle E_m |\vec x \rangle [(H_0+V(\vec x) )^{\dagger} ]\langle {\vec x}| E_l \rangle=E_m^{*} \langle E_m | E_l \rangle. \label{bound-state_s2} \nonumber \\
%& & \langle E_m, {\vec x} | E_l, {\vec x} \rangle  =\delta_{l,m},
\end{eqnarray} 
 Left-hand sides of Eq.$(\ref{bound-state_s} )$ and Eq.$( \ref{bound-state_s2})$ agree because $ |\langle E_m |\vec x \rangle|_{x =\pm \infty}$ vanishes and the right-hand sides satisfy 
\begin{eqnarray}
(E_l-E_m^{*}) \int d {\vec x}\langle E_m|{\vec x} \rangle \langle {\vec x} |E_l \rangle=0. \label{selfajoint}
\end{eqnarray}
For $l=m$ or $E_l-E_m \neq 0  $, this implies
\begin{eqnarray}
& & E_l=E_l^{*} ,\nonumber \\
& &  \int d {\vec x}\langle E_m|{\vec x} \rangle \langle {\vec x} |E_l \rangle=0.
\end{eqnarray}
The inner product  is normalized with   Kronecker delta  $\delta_{lm}$:   
 \begin{eqnarray}
%& & (H_0+V(x))  | E_l, {\vec x}  \rangle=E_l | E_l, {\vec x}  \rangle \\
& & \int d {\vec x}\langle E_m|{\vec x} \rangle \langle {\vec x} |E_l \rangle  =\delta_{l,m} \label{orthonormal}
\end{eqnarray} 
 satisfying
\begin{eqnarray}
& &\delta_{lm}=1; l=m \nonumber \\
& &\delta_{lm}=0; l \neq m, 
\end{eqnarray}
where non- degeneracy is assumed for simplicity.   
Thus,   the Hamiltonian is self-adjoint in the subspace of bound states.   Equation.$(\ref{orthonormal})$ holds    at an arbitrary  time,
\begin{eqnarray}
\int d{\vec x} | \langle t, E_m,|  {\vec x}\rangle \langle {\vec x}| t,E_l  \rangle =\delta_{l,m} \label{norm_bound}
\end{eqnarray}
which is consistent with probability  conservation, where 
\begin{eqnarray}
| t,E_l  \rangle=e^{\frac{iE_l t}{\hbar}}   | E_l  \rangle.
\end{eqnarray}
A  bound state is localized  within the potential region, and  ts  characteristic  size lis governed  by spatial extent  of its  wave function. 
  
  Bound states in many-body systems are more intricate than single-particle states. Nevertheless, their wave functions remain in extent  and are well described by wave packets. 
For instance, the spatial size of an electron wave function in an hydrogen atom is approximately  $10^{-10}$ [m] and it becomes significantly smaller  for inner shell electrons in atoms with  larger atomic number.  
In contrast, the sizes of the nucleons or nucleus is around  $10^{-15}$ [m], which is roughly   $10^5$ times  smaller than an average atomic size. 
These spatial dimensions  naturally dictate  the characteristic sizes of the wave packets of detected particles.

 %%%%%%%%%%%%%%%%%%%%%%%%%%%%%%%%%%%%%
\subsubsection{Continuum states}
%%%%%%%%%%%%%%%%%%%%%%%%%%%%%%%%%%%%%%%
Wave functions of stationary states in continuum spectra are not normalized and are not expressed by $L^2$ functios.  Nevertheless, scattering states can be  expressed by non-stationary  $L^2$ functions  which belong to Hilbert space. Then   the transition probability Eq.$(\ref{absolute_probability})$ is computed following     the four principles. \cite{Ishikawa-Tobita-ptep,Ishikawa-Tobita-anp,Ishikawa-Tajima-Tobita-ptep}. Mathematical relations involved differ  from those of the standard plain-wave  formalism. 
\footnote{Self-adjoint extensions within the von-Neumann formalism  provide a rigorous  method for an observable expressed by operator's expecation value \cite{sdextension,sdextension_2}.  The   modified  space depends on the operator, and the method does not provide the probability ,Eq.$(\ref{absolute_probability})$.  The present wave-packet formalism relies on  normalized states  where   $H$  is manifestly  self-adjoint, in which  extensions  are not necessary.}    

 A wave function in short range potential  is characterized  by  a momentum $\vec k$ and satisfies
\begin{eqnarray}
& & (H_0+V(\vec x)) \langle {\vec x}  | E(\vec k), {\vec k} \rangle=E(\vec k) \langle {\vec x} | E(\vec k), {\vec k} \rangle, \label{scattering_states}
\end{eqnarray}
where  $E(\vec k)=\frac{p^2}{2m}$ for a non-relativistic particle.     These wave functions  behave as  free waves at  $|\vec x| \rightarrow \pm \infty$, and left-hand sides of Eq.$(\ref{bound-state_s} )$ and Eq.$( \ref{bound-state_s2})$ do not coincide.     Their scalar product does not   satisfy  Eq.$(\ref{selfajoint})$ and  is   written as 
\begin{eqnarray}
& & \langle E(\vec k'),{\vec k'}  | E(\vec k), {\vec k} \rangle=\delta({\vec k}-{\vec k'})+ \epsilon(\vec k,\vec k'), \label{product_scattering}  
\end{eqnarray}
where the last term in  right-hand side do not vanish for $E(\vec k) \neq E(\vec k')$. Accordingly, stationary states with different eigenvalues are non-orthogonal.   
The non-orthogonal term in Eq.$( \ref{product_scattering} )$ is much smaller than the first term for $\vec k = \vec k'$  \cite{landau} in many situations, and has been ignored at majority places.  Nevertheless, the non-orthogonality  is critical for studying rigorous probability, because a superposition of  continuum  states Eq.$(\ref{scattering_states})$  defined as
\begin{eqnarray}
& &\langle t, {\vec x}| {\vec P}_0, {\vec X}_0, T_0  \rangle= \int d {\vec k}  \alpha({\vec P}_0, {\vec X}_0; {\vec k} ) \langle \vec x |E(k),  {\vec k} \rangle e^{\frac{E(k) (t-T_0)}{i \hbar} }, \nonumber \\
 \label{wave_packet_s} 
\end{eqnarray}
where c-number functions $ \alpha({\vec P}_0, {\vec X}_0; {\vec k} )$ satisfy 
\begin{eqnarray}
\int d{\vec k} |\alpha({\vec P}_0, {\vec X}_0; {\vec k} )|^2=1,
\end{eqnarray}
has a time-dependent norm   
\begin{eqnarray}
& &\int d{\vec x} |\langle t,{\vec x}| {\vec P}_0, {\vec X}_0, T_0  \rangle|^2  =1+ \nonumber \\
& & \int d{\vec k} d{\vec k'}  \alpha^{*}({\vec P}_0, {\vec X}_0; {\vec k'} )  \alpha({\vec P}_0, {\vec X}_0; {\vec k} )  \epsilon(k,k')  e^{\frac{(E(k)-E(k')) (t-T_0)}{i \hbar} }  \label{norm_scattering} \nonumber \\
\end{eqnarray}
\cite{ishikawa_1} from the  second term in the right hand side.   Since the oscillating function vanishes by averaging over a time,  physical quantities defined with averaging procedure would represent observed quantities. However,the time dependent  norm implies that the state  does not represent an isolate state. 
From Eq.$(\ref{conservation})$, the state with time-dependent norm has the probability current flowing from or to outside of the system.  The finite current indicates that this state does not represent an isolate state. 
 Therefore, amplitudes of the wave packets that are constructed from these eigen states of total Hamiltonian do not represent the transition of the isolate state \cite{ishikawa_1,ishikawa_2}. 

The physical reasons why wave functions in continuum states fail to satisfy   orthogonality   were  clarified recently \cite{ishikawa_1,ishikawa_2} based on physical requirements  on integrations in the present formalism.  These insights provide  an important  link between  wave packets and physical processes and are briefly summarized below.   
The definite integral over the interval,$-\Lambda$ and $\Lambda$ representing the scalar product  of plane waves $\phi(k,x)=e^{ikx}$ approaches the Dirac delta function at $\Lambda \rightarrow \infty$,
\begin{eqnarray}
& &  \int_{-\Lambda}^{\Lambda} dx \phi(k_2,x)^{*} \phi(k_1,x)= \frac{e^{-i(k_2-k_1) {\Lambda} }-e^{i(k_2-k_1)\Lambda }}{-i(k_2-k_1)} \nonumber \\
& &=\frac{ -2i \sin  (k_2-k_1) {\Lambda}}{-i(k_2-k_1)} \rightarrow 2 \pi \delta(k_2-k_1). \label{dirac_delta}
\end{eqnarray}
{\bf Scaling hypothesis}

For  large $\Lambda$, the scattering physics  within the region $ (-\Lambda/2, \Lambda/2) $ must be  equivalent to that within  $(-\Lambda, \Lambda)$. This  hypothesis is expressed as   
 \begin{eqnarray}
\frac{e^{-i(k_2-k_1) {\Lambda} }-e^{i  (k_2-k_1)\Lambda}}{-i(k_2-k_1)}=\frac{e^{-i(k_2-k_1) {\Lambda/2} }-e^{i  (k_2-k_1)\Lambda/2}}{-i(k_2-k_1)}. \label{scaling}
\end{eqnarray}
    Combining Eq.$(\ref{scaling})$ with the trigonometric  identity, 
\begin{eqnarray}
& &\frac{e^{-i(k_2-k_1) {\Lambda} }-e^{i  (k_2-k_1)\Lambda}}{-i(k_2-k_1)}=2\cos((k_2-k_1)\Lambda/2)\nonumber \\
& &\times \frac{e^{-i(k_2-k_1) {\Lambda/2} }-e^{i  (k_2-k_1)\Lambda/2}}{-i(k_2-k_1)},
\end{eqnarray}
we find  that    in the scaling region as  $ \Lambda \rightarrow  \infty$,   
\begin{eqnarray}
2 \cos ((k_2-k_1)\Lambda/2) \rightarrow 1.
\end{eqnarray}
This  condition is solved  by 
\begin{eqnarray}
(k_2-k_1)\Lambda/2 \rightarrow \frac{\pi}{3} sign(k_1-k_2)+2\pi n
\end{eqnarray}
where  $n$ is an integer. \footnote{Under this condition,  the large $\Lambda$ limit of the phase factor of the scalar product is not random.} At these momenta,   
\begin{eqnarray}
 \sin((k_2-k_1)\Lambda/2)\frac{e^{-i(k_2-k_1) {\Lambda/2} }-e^{i(k_2-k_1)\Lambda/2 }}{-i(k_2-k_1)} \nonumber \\
=  sign(k_2-k_1) \sin(\frac{\pi}{3})   \frac{e^{-i(k_2-k_1) {\Lambda/2} }-e^{i(k_2-k_1)\Lambda/2 }}{-i(k_2-k_1)}.
\end{eqnarray}
%\footnote{Under this condition,  the large $\Lambda$ limit of the phase is not random and the value becomes different from a previous result \cite{bar}.}
%$\delta(k_2-k_1)$
The values at these momenta characterize  the functional properties within the scaling region specified by Eq.$(\ref{scaling})$  even though  $\delta(k_2-k_1)$ is a distribution rather than  normal function in Eq.$(\ref{dirac_delta})$.  Applying the same limit procedure yields 
\begin{eqnarray}
& &\lim_{\Lambda \rightarrow \infty} \frac{e^{ - i(k_2-k_1) \Lambda}-1 }{ -i(k_2-k_1) } \rightarrow \lim_{\Lambda  \rightarrow \infty} \nonumber \\
& &e^{-i(k_2-k_1)(\Lambda/2)} \frac{e^{  -i(k_2-k_1) \Lambda/2}- e^{  i(k_2-k_1) \Lambda/2}}{- i(k_2-k_1)}  \nonumber \\
& & = \left( \cos (\pi/3)  -i \sin(\pi/3)  sign(k_2-k_1) \right)   2 \pi \delta(k_2-k_1). \label{dirac_delta_3} \nonumber \\
\end{eqnarray}
We thus arrive at the following resul : 

{\bf Theorem 1}. In the limit  $\lim_{\Lambda \rightarrow \infty}$,
  
\begin{eqnarray}
& & \frac{e^{ - i(k_2-k_1) \Lambda}-1 }{ -i(k_2-k_1) } \rightarrow  \pi (1+i \sqrt 3 sign(k_1-k_2))\delta(k_2-k_1),  \label{delta_m} \nonumber \\
& &\frac{e^{i (k_2-k_1) \Lambda}-1 }{-i(k_2-k_1)} \rightarrow -\pi (1-i \sqrt 3 sign(k_1-k_2))\delta(k_2-k_1). \nonumber \\
 \end{eqnarray}
The first  term in  the second line of  Eq.$(\ref{dirac_delta_3})$  is continuous, whereas the second term is discontinuous in $k_2-k_1$ at $k_2-k_1=0$. The integral  of its  product with any continuous function over the momentum difference $k_2-k_1$ vanishes. Hence, the last term may be omitted  in Eq.$(\ref{dirac_delta_3})$. The integrals defined under  the scaling hypothesis  are used throughout this paper. \footnote{ An expression of the left-hand side, $\frac{\cos (k_2-k_1) \Lambda -1}{-i(k_2-k_1)} +\frac{\sin(k_2-k_1) \Lambda}{k_2-k_1}$, is not proportional to $\delta(k_2-k_1)$ if $ \cos((k_2-k_1) \Lambda) \rightarrow 0 $ at $\Lambda \rightarrow \infty$ and is inappropriate.  }.
 Note that these momenta differ from those defined in a closed system  satisfying  periodic boundary conditions.
Similarly, for an integral  over a large interval $\Lambda$ and a lower limit $x_0$, the dependence of its functional property  on the interval is expressed as:  
\begin{eqnarray}
& & \frac{e^{ - i(k_2-k_1)  (\Lambda+x_0)}- e^{ - i(k_2-k_1) x_0}}{ -i(k_2-k_1)}=e^{-i(k_2-k_1)(x_0+\Lambda/2)} \nonumber \\
& &\times \frac{e^{  -i(k_2-k_1) \Lambda/2}- e^{  i(k_2-k_1) \Lambda/2}}{- i(k_2-k_1)} \rightarrow e^{-i(k_2-k_1)x_0}\pi \delta(k_2-k_1). \label{dirac_delta_2} \nonumber \\
\end{eqnarray}
%  The last term on the right-hand side of Eq.$(\ref{delta_m})$ should not be discarded 
These formulae guarantee  that the delta function is well-defined via the definite integral. Next, we apply Theorem 1 to continuum states in  short-range potentials.

 %%%%%%%%%%%%%%%%%%%%%%%%%%%%%%%%%%%%%%%%%%%%
{\bf Short-range potential }
%%%%%%%%%%%%%%%%%%%%%%%%%%%%%%%%%%%%%%%%%%%%

For a general short-range potential, $V(x)$, satisfying $V(x)=0$ at $x<x_0$, and $x_0'< x$, a wave function is expressed in term of the scattering coefficients $R(k)$ and $T(k)$ as,
\begin{eqnarray}
\psi(k,x)&=&e^{ikx}+R(k) e^{-ikx} ;x<x_0 \label{delta_potential_3}\\
&=& T(k) e^{ikx} ;x_0'<x,\nonumber 
\end{eqnarray}
by scattering coefficients, $R(k)$ and $T(k)$.

{\bf Theorem 2}.

The overlap integral is decomposed into an to an energy conserving term and an energy-non-conserving term  as,  
\begin{eqnarray}
& &\int dx \psi(k_2,x)^{*} \psi(k_1,x) \rightarrow 2  \pi \delta({k_1-k_2}) \nonumber\\
%&=&i\left( (T(k_2)^{*} T(k_1)e^{i (k_1-k_2) x_2}- e^{i (k_1-k_2) x_1}+e^{-i(k_1-k_2)x_1} R(k_2)^{*}R(k_1)\right)\frac{-1}{k_1-k_2} \nonumber \\
%& &-i{\left(R(k_1)e^{-i(k_1+k_2)x_1}- R(k_2)^{*} e^{i(k_1+k_2)x_1}\right)}\frac{1}{k_1+k_2}\nonumber\\
& &+{\left(R(k_1)+ R(k_2)^{*} \right)}  \pi \delta(k_1+k_2)-\Delta,\label{scalar_product} 
\end{eqnarray}
where  
\begin{eqnarray}
& &\Delta=i\left( \left(T(k_2)^{*} T(k_1)-1\right) + R(k_2)^{*}R(k_1)\right)\frac{1}{k_2-k_1} \nonumber \\
& & +i{\left(R(k_1)- R(k_2)^{*}\right)}\frac{1}{k_1+k_2}. \label{non-conserving} 
\end{eqnarray}
The first  and second terms in the right-hand side of Eq.$(\ref{scalar_product})$  vanish at $E_{k_1} \neq E_{k_2} $, whereas the third term does not vanish and represents energy non-conserving term.
From formulae  Eqs.$(\ref{scalar_product})$ and $(\ref{non-conserving})$,   the overlap integral is expressed by    the scattering coefficients, $R(k)$ and $T(k)$.  

In most short range potentials, Theorem 2 implies that the off-diagonal term does not vanish, ($ \Delta \neq 0$). Intuitively, the non-orthogonality of stationary states with different energies can be understandable from the spatialy behavior of wave functions: while  they behave  as free waves in the asymptotic  region, they deviate substantially within the  potential region, preventing  the scalar products of dstates with different energies from vanishing. 

This non-orthogonality of continuum states  might seem to contradict  the standard theorem  that the scalar products of eigen vectors corresponding to distinct eigenvalues of a Hermitian matrices are always orthogonal. However, the former applies to the infinite-dimensionnal spaces, whereas  the latter holds for finite-dimensional spaces. If  orthogonality were  preserved when taking the  infinite-dimensional limit, the two results would agree. The disagreement indicates that the limit is non-uniform.   Details on these limiting processess have confirmed  a breakdown of associativity in state-operator products, $ \langle \psi_1|(H|\psi_2 \rangle) \neq (\langle \psi_1|H)| \psi_2 \rangle$. While associativity for product of three operators  $ (AB)C = A(BC)$ hold strictly for   finite- dimensional matrices  and is customarily  assumed in quantum mechanics, its breakdown causes  quantum mechanical calculations to  fail in this context. 

{\bf Wave-packet construction}

Wave packets constructed from waves Eq.$( \ref{delta_potential_3})$ conserve the norm  with $\Delta = 0$ and represent isolated states.  
As demonstrated in \cite{ishikawa_2}, exact  orthogonality is satisfied   in systems with a free field, a uniform electric field, a uniform magnetic field, a harmonic oscillator, or a periodic potential.  
Wave packets constructed in these specific systems represent trully isolated states.  In the systems of short range potentials, wave packets constructed  with free waves describe the physical particles.  These evolve uniquely according to time-dependent   Schr\"{o}dinger equation Eq.$(\ref{Schroedinger})$. \cite{Reed-Simon,Kato}

{\bf Theorem 3}.

The normalized solution  with a short-range interaction is described by the perturbative solution  in the time-dependent   Schr\"{o}dinger equation Eq.$(\ref{Schroedinger})$ in $L^2$ space   (  $ \int dx |f(x)|^2 =1 $), which is complete and extension is unnecessary.

%%%%%%%%%%%%%%%%%%%%%%%%%%%%%%%
\subsection{A wave packet of continuum states: particle of a finite mean free path }
%%%%%%%%%%%%%%%%%%%%%%%%%%%%%%%%%%%%%%%%%%%%%%%%%%%%%%
The size of wave-packet within  continuum states  is governed by  the  particle's mean free path.

.

%%%%%%%%%%%%%%%
%%%%%%%%%%%%%%%%%%%%%%%%%%%%%%%%%%%%%%%%%%%%
\subsubsection{Perturbative  solution} 
%%%%%%%%%%%%%%%%%%%%%%%%%%%%%%%%%%%%%%%%%%%%%%%%%%%%%%%%%%
We  uncover a connection  between  the wave-packet size in the continuum states and a mean free path  using   Schr\"{o}dinger equation Eq.$(\ref{Schroedinger})$ and continuity equation Eq.$(\ref{continiuty})$. 
%Schroedinger equation Eq.$(\ref{Schroedinger})$ is solved with a Green function.% 
%interaction picture 
%\begin{eqnarray} 
%& &\psi(\vec p,t)=U(t,t_0) \psi_0(\vec p ,t_0),~~U(t,t_0)=1+\int_{t_0}
%^t dt' H_{int}(t')+\cdots \\
%& &U(t,t_0)^{\dagger} U(t,t_0)=1
%\end{eqnarray}
%satisfies 
%\begin{eqnarray}
%\langle \psi(\vec p, t)| \psi(\vec p,t) \rangle=\langle \psi_0(\vec p,t_0)| \psi_0(\vec p,t_0) \rangle.
%\end{eqnarray}
%Accordingly, a wave packet 
%\begin{eqnarray}
%\psi_{wp}(p_0,t)= \int d \vec k a(\vec k, \vec p_0) \psi(\vec k,t)
%\end{eqnarray}
 %has a constant norm and represents an isolated state. 
A solution of Eq.$(\ref{Schroedinger})$ is expressed  with a series 
\begin{eqnarray}
& &\psi(t,\vec x)= \phi(t, \vec x)\nonumber \\
& &+\int dt' d {\vec x'}[ G(t,\vec x;t',\vec x')V(\vec x')+\cdots] \phi(t', \vec x') \label{time_d}
\end{eqnarray}
using  the Green's function  defined by  
\begin{eqnarray}
\left( i \hbar\frac{\partial}{\partial t}-\frac{{\vec p}\,^2}{2m}\right) G(t,\vec x;t',\vec x')= \delta(t-t')\delta( \vec x-\vec x').
\end{eqnarray}
$\phi(\vec x,t)$ is a wave packet of  satisfying   
%\begin{eqnarray}
$\int d {\vec x}  \phi(\vec x, t)^{*} \phi(\vec x,t) =\int d {\vec x}  \phi(\vec x,t_0)^{*}  \phi(\vec x,t_0)$ .
%\end{eqnarray}
Because the norm of $\psi(\vec x,t)$ is given from Eq.$(\ref{time_d} )$ by 
\begin{eqnarray}
\int d {\vec x} \psi(\vec x, t)^{*}  \psi(\vec x,t) =\int d {\vec x}  \phi(\vec x,t)^{*}  \phi(\vec x,t)
\end{eqnarray}
 $\psi(\vec x,t)$  satisfies  
\begin{eqnarray}
\int d{\vec x}  \psi(\vec x, t)^{*}   \psi(\vec x,t) =  \int d{\vec x}  \psi(\vec x, t_0)^{*}   \psi(\vec x,t_0) 
\end{eqnarray}
and represents an isolated state.

For  short range potential 
\begin{eqnarray}
V= g \delta (\vec x),
\end{eqnarray}
 the first order correction to the wave function  is expressed by  Green's function   
 \begin{eqnarray}
 G(t,\vec x;t' \vec x')=\left(\frac{2\pi\hbar |t-t'|}{m}\right)^{-1/2} e^{i\frac{m{(\vec x-\vec x')}^2}{2 \hbar|t-t'|}}
 \end{eqnarray}
 in Eq.$( \ref{time_d})$ as
 \begin{eqnarray}
 %& &\int dt' d{\vec x}' G(t,\vec x;t',\vec x')V(\vec x') \phi(t', \vec x') = 
 & &N_3 \int dt'  \left(\frac{2\pi\hbar |t-t'|}{m}\right)^{-1/2}  e^{i\frac{m{(\vec x)}^2}{2 \hbar|t-t'|}} \nonumber \\
 & & \times e^{-\frac{1}{2 \sigma \hbar^2}( -{\vec X}_0 -{\vec v}_0(t'-T_0))^2 -i\frac{E({\vec P}_0)}{\hbar}(t'-T_0)+i\frac{{\vec P}_0}{\hbar}(-{\vec X}_0)}. 
 \end{eqnarray}
The wave function expands over time and is approximated  at a large time $t$ with    stationary phase satisfying  
\begin{eqnarray}
& &\frac{\partial }{\partial t'}\left[\frac{m(\vec x)^2}{2 |t-t'|}-E(\vec P_0)t' \right] =0,
\end{eqnarray}
which is sloved by 
\begin{eqnarray}
    \frac{m}{2} \left(\frac{\vec x}{t}\right)^2 =E(\vec P_0)
\end{eqnarray}
and the size at $t$ is given by
\begin{eqnarray}
  \sqrt{\frac{2E(P_0)}{m}} t=v_0 t. \label{size_t}
\end{eqnarray}
When the state is detected at time $t$, the wave function   is reset at the same position. That  evolves from the detected one, and the size follows Eq.$( \ref{size_t})$.  When the state is not detected,   a transition of the state occurs in matters.  A mean life time is finite even though the detection is not made. Average  size at mean life time $\tau$  represents  a size of wave packet
\begin{eqnarray}
    \sqrt \sigma =v_0 \tau \label{mean_free_path_0}.
\end{eqnarray}

%%%%%%%%%%%%%%%%%%%%%%%%%%%%%%%%%%%%%%%%%%%%%%%%%%%%%%%%%%%%%%%%%%%%%%
\subsubsection{Many short-range potentials  }
%%%%%%%%%%%%%%%%%%%%%%%%%%%%%%%%%%%%%%%%%%%%%%%%%%%%%%%%%%%%%%%%%%%%%%%%%
 A wave function of a particle    in a system of  many short-range potentials 
 \begin{eqnarray}
V= \sum_l  g_l \delta (\vec x-\vec X_l),
\end{eqnarray}
where $g_l$ and $\vec X_l$ are the coupling strength and position of $l$th potentials  is a superposition of waves
\begin{eqnarray}
\psi= \sum_l \psi_l(\vec x-\vec X_l).
\end{eqnarray}
Positions $(\vec X_l)$ are randomly distributed, and a  total scattering cross section  is approximated well by a sum of each cross section.  A  mean free path of a particle   is given by the classical formula 
 \begin{eqnarray}
L_{\rm mfp}=\frac{1}{\sigma_c \rho} \label{mean_free},
\end{eqnarray}
where $\sigma_c$ is the cross section and $\rho$ is the density of atoms.   A  free wave within  $L_{\rm mfp}$   is  expressed by a Gaussian wave packet of  size $\sqrt \sigma=L_{\rm mfp}$. 
The  mean free path and  wave-packet size  are shown in Figure 2.

% the Schr\"{o}dinger  equation 
\begin{figure}[t]
\includegraphics[width=.4\textwidth, bb=0 0 541 532]{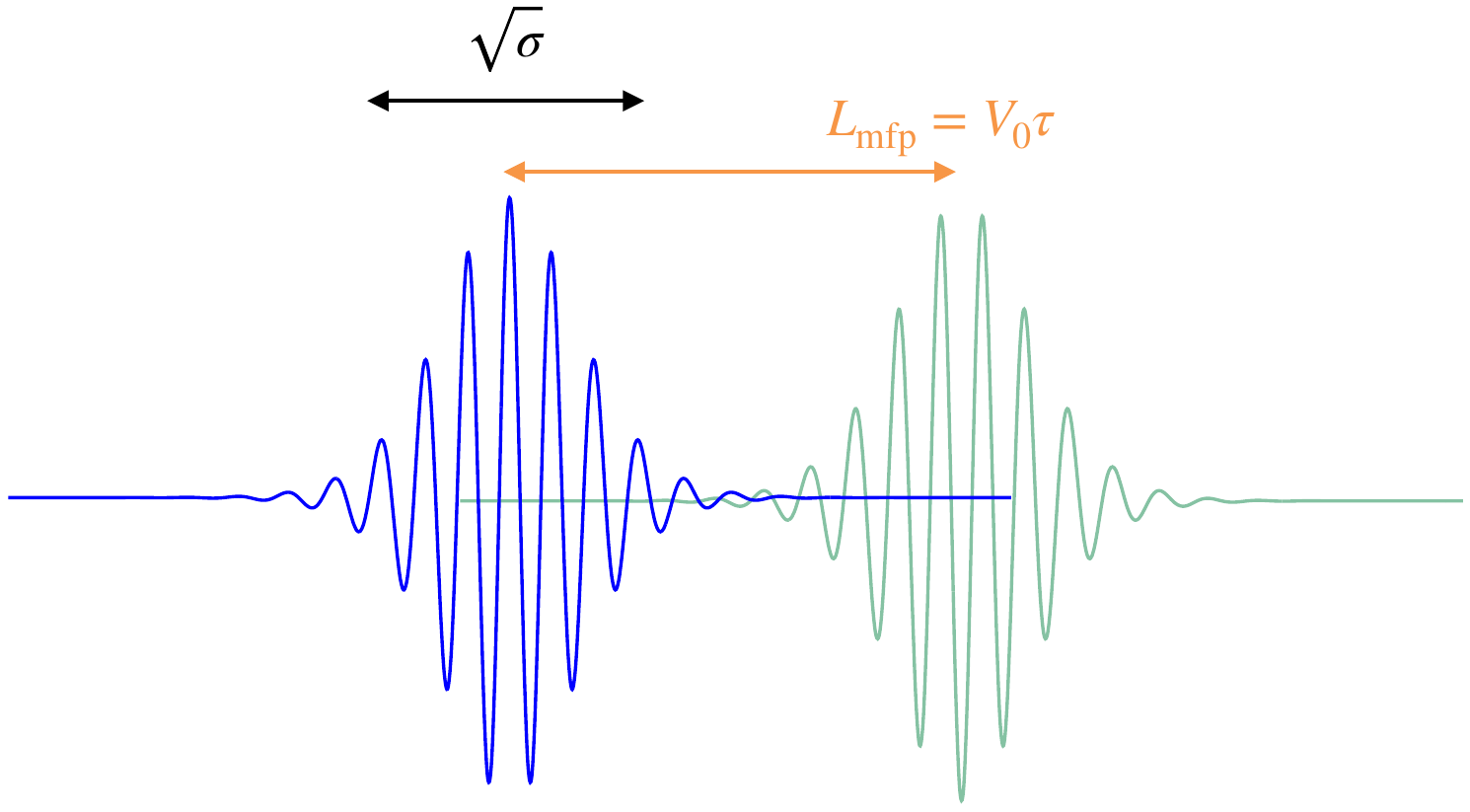}
\caption{  A size of wave packet $\Psi({\vec x},t)$  is a coherent length of the wave in the vacuum or in the dilute gas, which is proportional to the mean free path in matter.   }
\label{fig:superposition of plane waves}
\end{figure}
%%%%%%%%%%%%%%%%%%%%%%%%%%%%%%%%%%%%%%%%%%%
\subsubsection{ wave packet in space with  a boundary }
%%%%%%%%%%%%%%%%%%%%%%%%%%%%%%%%%%%%%%%%%%
We show  that  the connection  between coherence lengths and wave-packet sizes  is preserved   in space  of a boundary,  where matter is present in $z<0$ and absent in $z>0$.
In $z<0$, a particle's norm decreases with time due to the scattering with matter with a finite relaxation time, $\tau$, and the wave function is expressed as 
\begin{eqnarray}
\psi_{m}(p_0,t)&=& N_3 e^{-\frac{t-T_0}{2\tau_m}} e^{-i\frac{E({\vec P}_0)}{\hbar}(t-T_0)+i\frac{{\vec P}_0}{\hbar}({\vec x}-{\vec X}_0)}, \label{wp_damping}
\end{eqnarray} 
wheareas  in $z>0$ 
\begin{eqnarray}
\psi_{v}(p_0,t)&=&  N_3  e^{-\frac{1}{2 \sigma \hbar^2}( \vec x-{\vec X}_0 -{\vec v}_0(t-T_0))^2 -i\frac{E({\vec P}_0)}{\hbar}(t-T_0)+i\frac{{\vec P}_0}{\hbar}({\vec x}-{\vec X}_0).}\label{wp_damping} \nonumber \\
\end{eqnarray}
The continuity  Eq.$(\ref{conservation} )$ at  the boundary $z=0$  at $t=T_0, \vec x= \vec X_0$ derives  
\begin{eqnarray}
 \frac{1}{\tau}=\frac{v_0}{\sqrt \sigma},   
\end{eqnarray}
when  the expectation  values at   $z \geq 0$, and $z \leq 0$ are  substituted. This is identical to Eq.$(\ref{mean_free_path_0} )$.

Next we study wave packets expressed  with complex enery and momentum in the same space.  The wave packet size and life time in the former region is $( \sigma_1, \tau_1)$ and $( \sigma_2, \tau_2)$  in another  region.  
%A continuity of the wave functions at the boundary  is written using 
%\begin{eqnarray}
%\frac{1}{\psi} \left(i \hbar \frac{\partial}{\partial t}-\frac{({-i \hbar \frac{\partial}{\partial x}})^2}{2m}\right)\psi.
 %   \end{eqnarray}
 For wave packets of a complex energy of imaginary part $i \frac{1}{\tau} $, and a complex momentum of   imaginary part  $i \frac{1}{\sqrt \sigma} \vec n$, where $\vec n$ is a unit vector in 
 the direction of the momentum,   a continuity of the wave functions at the boundary 
\begin{eqnarray}
\frac{1}{\psi} \left(i \hbar \frac{\partial}{\partial t}-\frac{({-i \hbar \frac{\partial}{\partial x}})^2}{2m}\right)\psi
    \end{eqnarray}
    gives  equality  
\begin{eqnarray}
& &E+i \frac{1}{\tau_1} -\frac{\left(\vec p- \frac{i}{\sqrt \sigma_1}\vec n\right)^2}{2m} =E+i \frac{1}{\tau_2} -\frac{\left(\vec p- \frac{i}{\sqrt \sigma_2}\vec n\right)^2}{2m}. \nonumber \\
\end{eqnarray}
For  a relativistic particle 
\begin{eqnarray}
& &\left(E+i \frac{1}{\tau_1} \right)^2-\left(\vec p-i \frac{1}{\sqrt \sigma_1}\vec n\right)^2-m^2 \nonumber \\
&=&\left(E+i \frac{1}{\tau_2} \right)^2-\left(\vec p-i \frac{1}{\sqrt \sigma_2}\vec n \right)^2-m^2.
\end{eqnarray}
From  real parts  the energies are given and  from the imaginary parts for $\frac{1}{\tau} << E$, we have
\begin{eqnarray}
\frac{1}{\tau_1}+v \frac{1}{\sqrt \sigma_1}=\frac{1}{\tau_2}+v \frac{1}{\sqrt \sigma_2},    \end{eqnarray}
in agreement with Eq.$(\ref{mean_free_path_0})$. If one side is the vacuum, the life time there is infinite. The size of the wave packet in the other side  is determined from the life time and the wave packet size,
   \begin{eqnarray}
v \frac{1}{\sqrt \sigma_v}=\frac{1}{\tau_m}+v \frac{1}{\sqrt \sigma_m}.    \end{eqnarray} 
This is an extended form of Eq.$(\ref{mean_free_path_0})$.
%%%%%%%%%%%%%%%%%%%%%%%%%%%%%%%%%%%%%%%%%%%%

%%%%%%%%%%%%%%%%%%%%%%%%%%%%%%%%%%%%%%%%%%%%%%%%%%%%%%%%%%%%%%%%%%%%%%%%%%%%%%%%%
%%%%%%%%%%%%%%%%%%%%%%%%%%%%%%%%%%%%%%%%%%%%%%%%%%%%%%%%%%%
\subsection{Waves in many-body states.}
%%%%%%%%%%%%%%%%%%%%%%%%%%%%%%%%%%%%%%%%%%%%%%%%%%%%%%%%%%%
Inside of ordinary matter there are nearly infinite number of atoms. A particle propagating  this region  is affected by many-body effects.  
A many-body state composed of a particle $a$  of a momentum $\vec p_a$ at a position $\vec X_a$,  a particle $b$  of a momentum $\vec p_b$ at a position $\vec X_b$, and many particles $A$'s of  momenta  $\vec P_l$ at  positions $\vec X_l$
   is expressed using creation operators as 
\begin{eqnarray}
| \Psi \rangle=a^{\dagger}({\vec p_a},{\vec X}_a)b^{\dagger}({\vec p_b},{\vec X_b})\prod_{l}A^{\dagger}(\vec P_l,\vec X_l)|0 \rangle. 
\end{eqnarray}
where $a^{\dagger}({\vec p_a},{\vec X}_a) $  is a creation operator of a momentum ${\vec p}_a$ and a position ${\vec X}_a$. The other operators $b^{\dagger}({\vec p_b},{\vec X_b}) $, and $A^{\dagger}(\vec P_l,\vec X_l)$ are creation operators. 
This  system is described by a free Hamiltonian $H_0$ and an interaction Hamiltonian $H_{int}$,  which  are  sums of  terms of  $a$, $b$, and $A$,  
\begin{eqnarray}
H_0&=&H_0(a)+H_0(b)+H_0(A) \nonumber \\
H_{int}&=&H_{int}(a,b)+H_{int}(a,A)+H_{int}(b.A)+H_{int}(A), \nonumber \\
\end{eqnarray}
and  an many-body equation
\begin{eqnarray}
i \hbar \frac{\partial}{\partial t} |\Psi \rangle=(H_0+H_{int} )|\Psi \rangle. \label{many-body sch}
\end{eqnarray}
 
For bound states,  spatial sizes of wave functions represent coherence lengths. Continuum states are described by extended wave functions, and have finite mean free path due to interactions with other particles in matter. 
Coherence lengths depend on their physical processes with detectors, and classified into two cases.
 
 (1) $a$ and $b$ are observed but $A$ are not observed as in Fig.3 (1).  $a$ and $b$ evolve according to the total Hamiltonian and make change coherently. The states loose coherence when its transition probability 
exceeds the unity. A mean free path shows a coherence length. The coherence length of $a$ and $b$ are estimated from total probability to all possible final configurations of $A$.

 (2) When particles in $A$,  $a$, and $b$  are observed as in Fig.3 (2),  the wave functions are entangled and the transition probability becomes different. 
 Transition probability of state $A$ to  a certain final configuration is computed directly without summing over the final states. 
An entanglement of the wave functions appears in this probability.   
  \begin{figure}[t]
\includegraphics[width=0.5\textwidth, bb=0 0 715 249]{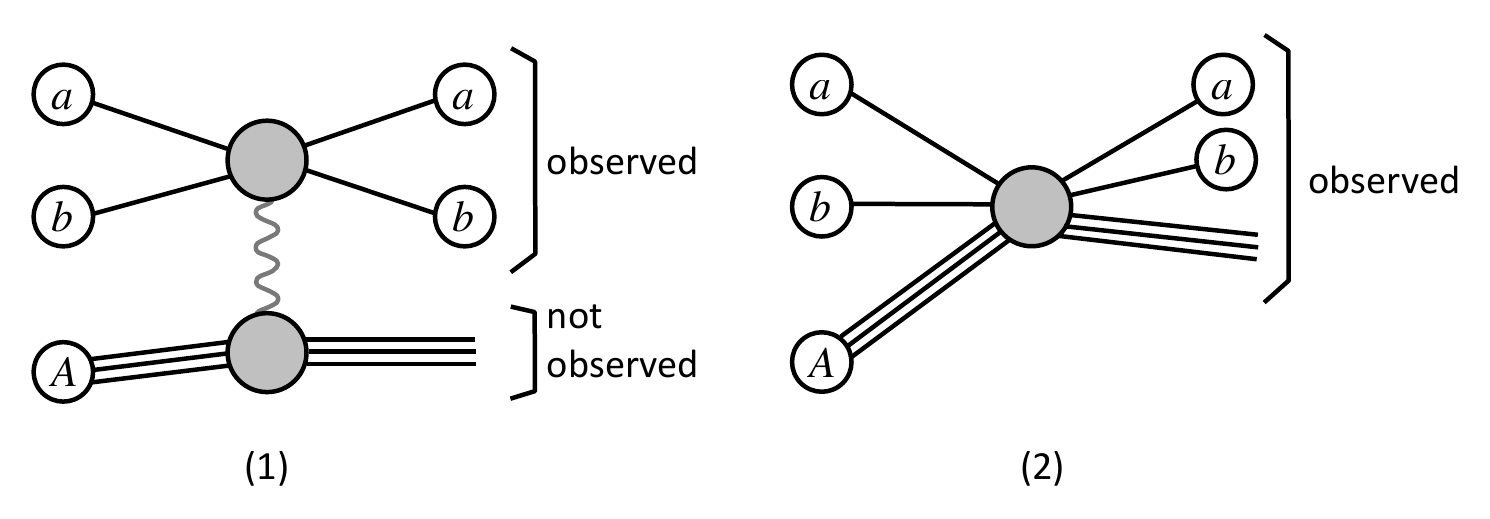}\caption{  Particles a,  b, and A collide and a and b in final states  are observed but A are not observed, in (1). In (2), a,b,and A are observed. }
\label{fig:superposition of plane waves}
\end{figure}

%%%%%%%%%%%%%%%%%%%%%%%%
%%%%%%%%%%%%%%%%%%%%%%%%%%%%%%%%%%%%%%%%%
\subsection{Motion of wave packets   }
%%%%%%%%%%%%%%%%%%%%%%%%%%%%%%%%%%%%%%%%%%
Free wave packets  are modified, and reveal unique behaviors       in external potentials.
%%%%%%%%%%%%%%%%%%%%%%%%%%%%%%%%%%%%%%%%%
\subsubsection{Semiclassical motion   }
%%%%%%%%%%%%%%%%%%%%%%%%%%%%%%%%%%%%%%%%%%
Motion of a wave packet in slowly varying potential, V, can be analyzed with a semiclassical approximation. Parameters of wave packets are varied slowly then.  We study a potential $V=V_0$.  As a sum of the kinetic energy and the potential energy is kept constant, a momentum in  the potentia region,  ${\vec p}_1$, and a momentum in a vacuum region,  ${\vec p}_2$ are connected. Decomposing  them into  central values and deviations,    ${\vec p}_1 +\delta {\vec p}_1 $ and ${\vec p}_2 +\delta {\vec p}_2 $, we have 
\begin{eqnarray}
E({\vec p}_1 +\delta {\vec p}_1)+V_0=E({\vec p}_2+\delta {\vec p}_2), \label{Semiclassical_energy}
\end{eqnarray}
from the conservation of  total energy. Next we expand   both sides in $\delta {\vec p}_i,i=1,2$ , and we have 
\begin{eqnarray}
& &E({\vec p}_1)+V_0+ \vec v_1  \delta {\vec p}_1  =E({\vec p}_2 ) + \vec v_2  \delta {\vec p}_2 \\
& &\vec v_1=\frac{\partial E({\vec p}_1)}{\partial \vec  p_1} ,\vec v_2= \frac{\partial E({\vec p}_2)}{\partial \vec p_2}, 
\end{eqnarray}
where $\vec v_i, i=1,2$ are the wave packet velocities, and 
\begin{eqnarray}
& &E({\vec p}_1)+V_0  =E({\vec p}_2 ) \label{central_energy}\\
& & \vec v_1 \delta {\vec p}_1 =\vec v_2  \delta {\vec p}_2 \label{width_change}
\end{eqnarray}
Eq.$(\ref{central_energy})$ indicates  a conservation of  the central energy and Eq.$(\ref{width_change})$ indicates  that a deviation  of the momentum from the central value is inversely proportional to the velocity. Accordingly a spatial  extension of  the wave $\delta x^l$  is expressed by the uncertain relation from the momentum variation  in the direction to the velocity $\delta p_i^l (i=1,2)$ as
\begin{eqnarray}
 \delta x_i^l=\frac{\hbar}{\delta p_i^l}.
\end{eqnarray}
Substituting to Eq.$(\ref{width_change})$, we have   an identity  
\begin{eqnarray}
\frac{ \delta x_1^l}{v_1}=\frac{ \delta x_2^l}{v_2}. \label{crossing_time} 
\end{eqnarray}
The left-hand side is an elapsed time of particle 1 that passes an interaction  point, and the right-hand side is the one of particle 2.  Lengths $\delta x_i^l$ can be considered as the wave packet sizes.   
%%%%%%%%%%%%%%%%%%%%%%%%%%%%%%%%%%%%%%%%%%%%%%%%%%%%%
\subsubsection{Uniform force }
%%%%%%%%%%%%%%%%%%%%%%%%%%%%%%%%%%%%%%%%%%%%%%%%%%%%
 Wave packets in a uniform force in z-direction  are constructed with   Airy function,
\begin{eqnarray}
& &H \phi_E(z)=[\frac{p_z^2}{2m}+mgz] \phi_E(z) \\
& &\phi_E(z)= A_E(\xi-\xi_1), \xi=\frac{z}{c},~ \xi_1=\frac{E}{cmg}, c=(\frac{1}{2m^2g})^{1/3} \nonumber \\
& &A_E(\xi)=\frac{1}{\pi} \int_0^{\infty}du \cos(\frac{u^3}{3}+\xi u).
\end{eqnarray}
 Airy functions satisfy orthogonality 
\begin{eqnarray}
& &\int_{-\infty}^{\infty} dz \phi_E^{*}(z-Z_1) \phi_{E'}(z-Z_2)  \nonumber \\
& &= c\delta(\frac{E'-E}{cmg} +\frac{Z_2-Z_1}{c}). 
%\end{eqnarray}
%\begin{eqnarray}
%& &\tilde \xi_1=\frac{E}{cmg} +\frac{Z_1}{c},~\tilde \xi_1'=\frac{E'}{cmg} +\frac{Z_2}{c}
\end{eqnarray}
\cite{landau}.
%\newpage
Wave packets defined by
\begin{eqnarray}
& &| E_1, Z_1,T_1;\sigma \rangle=N_1\int d E e^{ iET_1} e^{-\frac{\sigma}{2} (E-E_1)^2} \phi_E(z-Z_1)  \nonumber \\
%\end{eqnarray}
%\begin{eqnarray}
& & N_1^2=(c^2 mg  \frac{\pi}{\sqrt{2\sigma_1}} )^{-1},  
\end{eqnarray}
have   scalar product
\begin{eqnarray}
& & \langle E_1,Z_1,T_1 ; \sigma_1| E_2,Z_2,T_2;\sigma_2 \rangle \nonumber \\
& & = \frac{2 \sqrt{\sigma_1\sigma_2}}{\sigma_1+\sigma_2}    e^{ -\frac{(T_1-T_2)^2}{(\sigma_1+\sigma_2)}-\frac{\sigma_1\sigma_2}{(\sigma_1+\sigma_2)}(E_1-E_2+\frac{\Delta Z}{mb})^2 }. 
\end{eqnarray}
The magnitude  becomes maximum at energy  $ E_2=E_1+\frac{\Delta Z}{mb}$ consistent with  Eq.$(\ref{Semiclassical_energy})$. At this  energy the probability  becomes maximum   at size
\begin{eqnarray}
\sigma_2=(T_1-T_2)^2+\sqrt{(T_1-T_2)^4+\sigma_1^2 }.
\end{eqnarray}
At large $|T_1-T_2|$, the wave packet expands with the velocity of light. 
\begin{eqnarray}
\sqrt{\sigma_2}=\sqrt 2|T_1-T_2|.
\end{eqnarray}

\subsubsection{Uniform magnetic field}
In a uniform (weak) magnetic field, Hamiltonian and eigenfunctions are 
\begin{eqnarray}
& &H=\frac{(p_x-eBy)^2+p_y^2}{2m}  \nonumber \\
& &H \psi(x,y)=E \psi, \psi=e^{ik_x x}\phi_l(k_x,y),E_l=  \omega(l+1/2) \nonumber \\
\end{eqnarray}
(1) Wave packet 1
\begin{eqnarray}
& &|E_l,P_x,X,T_1 \rangle=N_1\int d k_x e^{iE_lT_1-\frac{\sigma}{2}(k_x-P_x)^2+ik_x(x-X)} \nonumber \\
& &\times \phi_l(k_x,y), \nonumber  \\
& &\langle E_{l^1} ,P_x^1,X^1,T_1 | E_{l^2},P_x^2,X^2,T_2 \rangle=N_1N_2(2 \pi) \delta_{l_1,l_2} \nonumber \\
& &e^{-iE_{l^1}(T_1-T_2)}e^{ -\frac{1}{2(\sigma_1+\sigma_2)}(X_1-X_2)^2-\frac{\sigma_1 \sigma_2}{2(\sigma_1+\sigma_2)}(P_x^1-P_x^2)^2}e^{i \phi_0} \nonumber \\
\end{eqnarray} 
where $\phi_0$ is a constant phase.
The energies agree exactly and positions and momenta approximately agree.  

(II)Wave packet 2
\begin{eqnarray}
& &|E_l,k_x,X,T_1 \rangle=\int d E e^{iE T_1-\frac{\sigma}{2}(E-E_0)^2+ik_x(x-X)}\phi_l (k_x,y), \nonumber \\
& &\langle E_{l^1} ,k_x^1,X^1,T_1 | E_{l^2},k_x^2,X^2,T_2 \rangle = N (2 \pi) \delta_{l_1,l_2} \nonumber \\
& & \times e^{ik_x^1(X_1-X_2)} \sqrt{ \frac{2 \pi}{\sigma_1+\sigma_2}} e^{- \frac{1}{\sigma_1+\sigma_2}(T_1-T_2)^2-\frac{\sigma_1\sigma_2}{\sigma_1+\sigma_2}(E_1-E_2)^2 +i \phi } .   \nonumber \\
\end{eqnarray} 
The energies agree approximately at $T_1-T_2 \sim 0$. 
Symmetric gauge is studied in Appendix.

%%%%%%%%%%%%%%%%%%%%%%%%%%%%%%%%%%%%%%%%%%%%%%%%%%%%%%%%%%%%%%%%%%%%%%%
\subsubsection{Wave packet of intermediate state: decay and scattering}
%%%%%%%%%%%%%%%%%%%%%%%%%%%%%%%%%%%%%%%%%%%%%%%%%%%%%%%%%%%%%%%%%%%%
Contraly to the wave packets in initial and final states,   the sizes of wave packets in intermediate states are indirectly determined. For a stable intermediate state,  an overlap integral of the initial wave functions with intermediate states becomes maximum when they agree from Eq.$(\ref{classical} )$.    This result can be understood using semi-classical picture. For a case that a length of a target wave function is much shorter than an average mean free path of the initial state, an overlapping  period $\delta \tau$ 
is expressed using the coherence length of the initial particle, ${\delta x}_i$ and its velocity.  The period is also expressed using the length of the intermediate particles, ${\delta x}_m$ and its velocity. Conversely, the coherence length  are connected  with  their velocities by Eq.$(\ref{wave-packet_3})$ and Eq.$(\ref{crossing_time})$ as
\begin{eqnarray} 
& &{{\delta x}_i \over v_i}={{\delta x}_m \over v_m }=\delta \tau,\\
& &{\delta x}_m={v_m \over v_i}{\delta x}_i.\label{crossing_time_{inter}} 
\end{eqnarray}
Further study will be given in Section 4. The wave packet size has universal properties and  will be estimated  in various situations. 

%Formulae for given initial and final states in \cite{ishikawa_1,ishikawa_2} are useful to determine sizits size reflects the sizes of the initial state.

%%%%%%%%%%%%%%%%%%%%%%%%%%%%%%%
\subsection{ Charged particle in matter    }
%%%%%%%%%%%%%%%%%%%%%%%%%%%%%%%%%%%%%%%%%%%%%%%%%%%%%%
A  particle propagating in matter receives many-body effects successively from atoms in matter.   The cross section of a charge particle with an atom  is governed by  the  long-range electromagnetic interaction and the short range strong interaction.  The electromagnetic interaction  is main sources for a charged particle to lose its coherence and energy.  The strong interaction is short range and gives weaker effects.

Approximating the cross section of  wave packet  with those  of plane wave,  we estimate the mean free path using the formula of Eq.$(\ref{mean_free} )$.
Collision processes such as Rutherford scattering, Bethe-Bloch process, Bremsstrahlung, and others, make  a particle to lose its energy. The average length of losing 
 energy  is expressed  using  an energy loss rate per a unit of length  $x$ in  a matter's number  density $\rho_m$,  
\begin{eqnarray}
& &L_\text{charge}={1 \over {\frac{1}{E} \frac{dE}{ \rho_m d x} \times \rho_m }}=\frac{E}{ \frac{dE}{dx}}, \label{energy_loss}
\end{eqnarray}
where dimension is   
\begin{eqnarray}
& &[x]=m
\end{eqnarray}  
The particle keeps the majority of  energy  below the  length Eq.$( \ref{energy_loss} )$. $ L_\text{charge}$  represents a coherent length in term of the energy. 
%We will show later that $l_{charge} \approx  0.5$ meters for $1$ GeV proton in Fe.

Coulomb  interaction between charged particles is long-range and the cross section is expressed by Rutherford formula.  The Rutherford cross section is  divergent in a forward direction and regularized due 
to   charge screening effect with a cutoff parameter $\log \Lambda$. The cross section for a charged particle of the charge $e$ and the energy $E$ is expressed as 
\begin{eqnarray}
& &\sigma_{Ru}=4 \pi \left(\frac{e^2}{ 4 \pi E }\right)^2 \log \Lambda \label{Rutherford}\\
& &E=\frac{mv^2}{2},
\end{eqnarray}   
where a standard value  is $\log \Lambda =10$. Assuming the parameters: 
charge of the scatterer is $|e|$,  the density of scatterers of mass number $N$, $\rho= \frac{6 \times 10^{23}}{N \times 10^{-6}}$ [1$/m^3$] to Eq.$(\ref{mean_free}, )$, we have the lengths for  energy $1$ GeV,$10$ MeV, $50$ keV. and $1$ keV, and $N= 55$. 
\\
\\
 Mean free path due to Rutherford scattering  is given as, 
\\
\\
%\begin{table}[htbp]
%\begin{table}
%\centering
%\caption{ Mean free path by Rutherford scattering} }
\begin{tabular}{|c|c|c|c|c|}
\hline
 Energy ($E$)   &  1~GeV  & 10~ MeV & 50~ keV & 1~keV   \\
 \hline
  $L_{mfp}$  [m]     & $ \sim 10^3     $   & $ \sim 0.1$ & $\sim 2.5 \times10^{-6}  $  &  $\sim 10^{-9}$  \\
 \hline
 \end{tabular}.
%\end{table}
\\
\\
The length  in the low energy  is much shorter than the value in the intermediate energy. 
In Figure 4, the energy loss rate of an electron of energy $E$ in Lead is shown. At an energy higher than $10$ MeV, The Bremsstrahlung is dominant in the energy lower than $10$ MeV. The rate from ionization increases at low energy. $X_0$ is given by
\begin{eqnarray}
 X_0(\rm Pb)=6.37 \,[\text{g}/{\text cm}^2]    
\end{eqnarray}
for Lead.
\begin{figure}[t]
\includegraphics[width=.6\textwidth, bb=0 0 541 532]{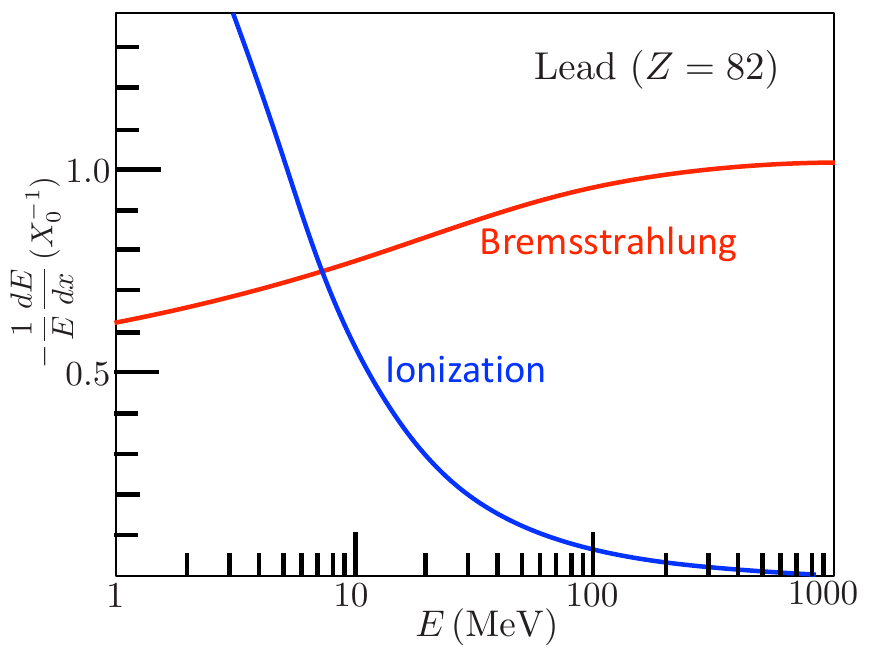}
\caption{  An energy loss rate  of an electron  in Lead by ionization or Bremsstrahlung is shown as a function of energy. In low energy, ionization is a dominant process. Simplified and redrawn from \cite{particle_data_1}}
\label{fig:energy loss rate of charged particle}
\end{figure}
%A   mean free path of a particle emitted from a box  of size $L$ in each direction filled with matter.   
 %In a box of $L > L_\text{charge} $, each  particle produced in matter loses its coherence in every distance $L_\text{charge}$. A  particle that loses the coherence before  reaching %the boundary is irrelevant.  A particle that reachs the boundary before losing  the coherence is  emitted with the same  wave packet size. 
%  In a case  $L <  L_\text{charge} $, the size of wave packets is given by  Eq.$( \ref{size_mean})$. The emitted wave packets keep their sizes in the out side. 

%Situations  that the particles are confined in the inside are    not studied in the present paper \cite{breene}.

%%%%%%%%%%%%%%%%%%%%%%%%%%%%%%%%%%%%%%%%%%%%%%%%%%%%%%%%%
%\subsection{Photon  }
%%%%%%%%%%%%%%%%%%%%%%%%%%%%%%%%%%%%%%%%%%%%%%%%%%%%%%%

%For a photon, an optical length can be used to express the size of the wave packet. 
%The strong interaction is stronger than the electromagnetic interaction but the force range is short. Consequently the cross section varies slowly with the energy, and the magnitude is %much smaller than that of the electromagnetic interactions.  The mean free path is not so short even in low energy. 

%Elementary processes of the photon in matter is photoelectric effect, Compton scattering, and $e \bar e$ annihilation. The size of the photon wave packet in the final state is  %determined by this length Eq.$(\ref{mean_free_path})$.  
% \cite{Ishikawa-Tobita-ptp}.

%%%%%%%%%%%%%%%%%%%%%%%%%%%%%%%%
%%%%%%%%%%%%%%%%%%%%%%%%%%%%%%%
\section{Wave packets of in-coming waves of scatterings}
%%%%%%%%%%%%%%%%%%%%%%%%%%%%%%%%%%%%%%%%%%%%%%%%%%%%%%%%%%%%
%%%%%%%%%%%%%%%%%%%%%%%%%%%%%%%%%%%%%%%%%%%%%%%%%%%%%%%%%%

%A wave packet  at $t=T_0$ evolving in a vacuum at $t >T_0$ is expressed by a wave function Eq.$(\ref{wave-packet_3})$.  Based on this expression, 

  We  apply 
{\bf Theorem 3}
and study properties of wave packet of proton,   pion,  muon,   neutrino, and others.
% following partly a method of  \cite{Ishikawa-Shimomura,Ishikawa-Tobita-ptp,Ishikawa-Tobita}.
 %{\bf pulse size }
%A pulse of particle or wave is composed of many particles or waves. A  size of each particle or wave is less than or equal to the pulse size.    
Because these have different properties, the sizes of wave packets depend on their species.  
%Particles in targets are normally  bound states of discrete energy spectra of microscopic sizes. Energy levels and other parameters of the wave functions are determined %from microscopic parameters such as particle masses and coupling strengths in the Lagrangian. These are different from those of continuum states.  

In a system of a many-body interaction $H_{int}$, an initial state makes transitions to others according to the absolute probability.    Transition amplitudes  are expressed by a matrix element of a unitary S-matrix between initial and final states.  The states  satisfy  normalization conditions when   behave as particle and described by wave packets, which are characterized by finite sizes. The size   in beam is determined by its mean free paths in matter.     
Wave packets of the initial states are analyzed  using the formulae of the previous section.
%%%%%%%%%%%%%%%%%%%%%%%%%%%%%%%%%%%%%%%%%%%%%%%%%%
\subsection{Scattering matrix and probability formula}
%%%%%%%%%%%%%%%%%%%%%%%%%%%%%%%%%%%%%%%%%%%%%%%%%%%

Using a  many-body state evolving  according to  Eq.$(\ref{many-body sch})$ from  an initial state $|\Psi_i \rangle $, the scattering matrix to observe   a final state $| \Psi_f \rangle $ is expressed by Dyson series
\begin{eqnarray}
& &S=\langle  \Psi_f| U(T) |\Psi_i \rangle  \nonumber \\
& &U(T)=1+\frac{1}{i \hbar} \int_0^T dt_1 H_{int}(t_1) \nonumber \\
& &+\frac{1}{i \hbar} \int_0^T dt_1  \frac{1}{i \hbar} \int_0^{t_1} dt_2 H_{int}(t_1)    H_{int}(t_2)+  \cdots  . \label{Dyson}
\end{eqnarray}
For explicit calcultions,  Theorem 3 is extended. 
 
{\bf Theorem 4}.

The transition probability $P(T)$ of  normalized initial and final states $|\Psi_i \rangle $ and  $| \Psi_f \rangle$ with a short-range interaction is described by the S-matrix with wave packets in the interaction picture Eq.$(\ref{Dyson})$. This space is complete and applicable to general systems.

   An initial state is composed of incident particles or waves that are propagating in vacuum or dilute matter and targets which are at rest normally. 
These are described by Gaussian wave packets in this paper. 
Gaussian wave packet is parametrized by one parameter $\sigma$ in Eq.$(\ref{wave-packet_1})$ and provides a systematic expression of  rigorous amplitude  and  probability satisfying  all the requirements for physical quantity, as can be seen partly  in Appendix 1.   

% Importance of wave packets were addressed in \cite{ } and the amplitudes  of a fixed  %position were extensively analyzed in  \cite{araki}.  
  This formalism   has been  applied  to neutrino processes  and  radiative transitions  in  \cite{Ishikawa-Tobita-ptep,Ishikawa-Tobita-anp,Ishikawa-Tajima-Tobita-ptep}, and 
  to other  QED and strong interaction processes  in \cite{Asahara,Ishikawa-Jinnouchi,Ushioda,threshold}.
  The absolute probability $P(T)$ in Eq.$(\ref{absolute_probability} )$ for these processes have been obtained.   $\Gamma$ is independent of the wave-packet sizes and agrees with the known quantity obtained from the plain-wave formalism, whereas   $P^{(d)}$ depends on  the wave-packet sizes. Processes $P_{I}$ and $P_{II}$ have been studied.  
     
%\bibitem{Asahara} A.~Asahara, K.~Ishikawa, T.~Shimomura, and T.~Yabuki,
%Prog. Theor. Phys. \textbf{113}, 385(2005); T.~Yabuki and K.~Ishikawa,
%Prog. Theor. Phys. \textbf{108}, 347(2002).

%\bibitem{Wilson-OPE} K.~Wilson, in Proceedings of the Fifth International Symposium on %Electron and Photon Interactions at High Energies, Ithaca, New York, 1971,~p.115~(1971).
% edited by N. B. Mistry (Cornell Univ. Press, Ithaca, New York, 1971 

%\bibitem{Ishikawa-Jinnouchi} K.~Ishikawa, O.~Jinnouchi, A.~Kubota, T.~Sloan, T. %H.~Tatsuishi, and R. ~Ushioda, `` On experimental confirmation of the correction to  %Fermi's golden rule''
%	Prog. Theor. Exp. Physics.  \textbf{2019}, 033B02,(2019).
%\bibitem{Ushioda} R.~Ushioda, O.~Jinnouchi, K.~Ishikawa and T.~Sloan ,`` Search for the %correction to Fermi's golden rule in positron annihilation '' 	Prog. Theor. Exp. Physics.  %\textbf{2020}, April, (2020). 
%\bibitem{particle_data_2} S.Navas et al.(Particle Data Group).Phys. Rev. D 110,030001 
%(2024) and 2025 update, Sec. 34.Passage of Particles , Fig. 34.15
%\bibitem{threshold} K.~Ishikawa, O.~Jinnochi, K.~Nishiwaki,  and K.~Oda, Euo.Phys. J.C %83(2023) 978  http://dx.do.org/10.1140/epj/s10052-0123-12077-7.

%{ \bf Demonstrate wave-packet amplitude  $S(T)$ and the probability $P(T)$ obtained  in a previous paper.} 

%to shown to depend on the wave packet sizes

Incident particles are prepared with an apparatus and   have    finite coherence lengths.  These govern   the wave-packet sizes  for the initial states.

 %%%%%%%%%%%%%%%%%%%%%%%%%%%%%%%%%%%%
 \subsection{ Proton}
 %%%%%%%%%%%%%%%%%%%%%%%%%%%%%%
Coherence length of a proton is governed by its reactions with atoms in matter.   Coulomb force  between charges  and a force from an  electric field  act as  static forces.      
 The proton loses energy   by ionization of matter and  Bremsstrahlung.  The coherence lengths due to these will be found  not short. A mechanism specific to  high energy accelerator is studied next.    During acceralation in high energy accelerator,  protons rotate many times and form  bunches due to periodic motions. Each  proton in the bunch is  interacting  with other protons through long range Coulomb forces and forms unique correlated states.   
We will see that the proton's coherence length  in this situation is short.
%%%%%%%%%%%%%%%%%%%%%%%%%%%%%%%%%%%%%%%%%%%%%%%%%%%%%%%%%
\subsubsection{ Single proton in matter }
%%%%%%%%%%%%%%%%%%%%%%%%%%%%%%%%%%%%%%%%%%%%%%%%%%%%%%%%%%
A proton is stable and  used as an incident particle in many experiments. A proton  interacts with surrounding atoms by various interactions. 

(1) Scattering due to the Coulomb interaction is expressed by Rutherford scattering Eq.$( \ref{Rutherford})$. Because the cross section is large in low energy, the mean free path is the same.  The values are presented in Table for $|e|=1$ and $N=50$. 

(2) Strong interaction of a proton with a nucleus is short-range and the cross section is around $\frac{1}{m_{\pi}^2}$ [mb]. An mean free path for a density $n=N_A$, where $N_A$ is the Avogadro number,  per $\rm cm^3$ is given by
\begin{eqnarray}
 l_{{\rm mfp},s}=\frac{1}{n_A \sigma} \sim 1.5 [\rm m].   \label{mfp_strong}
\end{eqnarray}
This value is much longer than the  values  presented in the Table.  

(3) The proton loses energy by ionizing atoms and Bremsstrahlung. An average coherence length of the proton in matter can be computed with 
the transition probabilities. Matter effects are so complicated that the estimations are made here with measured energy loss values. From Data summarized in the particle data summary \cite{particle-data}, we estimate the values hereafter. An average time interval for the proton to hold its coherence in matter is an average relaxation time, which is an inverse of the energy loss rate, Eq.$(\ref{energy_loss})$.        
 %\begin{eqnarray}
%L_\text{proton}={1 \over {\frac{1}{E} \frac{dE}{  d x} \times \rho }}.
%\end{eqnarray}
The proton's energy loss rate at the momentum, $1$GeV/c, for several 
metals such as Pb, Fe, and
others are 
\begin{eqnarray}
-{d E \over  d x} \sim 1-2  ~[ \text{GeV} \text{m}^{-1} ],   
\end{eqnarray}
%\begin{eqnarray}
%-{d E \over \rho_m d x}=1-2 ~[\text{MeVg}^{-1}\text{cm}^2],   
%\end{eqnarray}
hence an average coherence length of the proton of $1~\text{GeV}/c$ 
is estimated for the matter of density $\rho_m$  as  

 \begin{eqnarray}
L_\text{proton}&=& {1 ~[\text{GeV}] \over (1-2) ~[\text{GeV }^{-1}  \text{ m}^{-1}]}  \label{mfp-proton-i1} \\
 & \sim & (0.5-1.0) [\text{m}] \nonumber
\end{eqnarray}
%
%\begin{eqnarray}
%L_\text{proton}&=& {1 ~[\text{GeV}] \over (1-2)\times
% 10 ~[\text{MeV g}^{-1} \text{cm}^2 \text{g cm}^{-3}]} \frac{\rho_m}{{\rho_m}_0} \label{mfp-proton-i1} \\
% & \sim & (0.50-1.0) ~[\text{cm}],  \nonumber
%\end{eqnarray}
At  energy,  $0.2~\text{GeV}/c$ ( $0.2~\text{MeV}/c$ )
%an energy loss rate of aproton is about $10~\text{MeVg}^{-1}\text{cm}^2$ and 
the average coherence length  is  
\begin{eqnarray}
L_\text{proton} \sim 0.10 ~[\text{m}]. ( 10^{-4}~[\text{m}]) \label{mfp-proton-i_2}
\end{eqnarray}
At  lower  energy, the length becomes shorter than Eq.$(\ref{mfp-proton-i_2} )$.
When these particles are emitted from matter to the vacuum or dilute gas, they propagate freely. The wave is expressed by Eq.$(\ref{wave-packet_3})$.

When a proton is accelerated, its width is varied following Eq.$(\ref{width_change})$.  
Accordingly  a particle of a size, $L_{\rm before}$,  becomes to have   a new  value, $L_{\rm after}$  at a velocity $v_{\rm after}$,
\begin{eqnarray}
L_\text{after}= L_\text{before}\times{v_\text{after} \over
 v_\text{before}}.\label{ratio_length}
\end{eqnarray}
A velocity is bounded
by the light velocity $c$,
and the velocity ratio from $1~\text{GeV}/c$ to $10~\text{GeV}/c$
 is about $1.2$ and that from $0.2~\text{GeV}/c$ to $10~\text{GeV}/c$
 is about five. Hence from Eq.$(\ref{ratio_length})$, the proton 
 of $10~\text{GeV}/c$  has the mean 
free path
\begin{eqnarray}
L_\text{proton}\sim 0.40 - 1.0~[\text{m}].
\label{mfp-proton-i2}
\end{eqnarray}   
in vacuum or in an accelerator. 

The strong interaction between hadrons is short range and does not give large effect to the energy loss rate. The mean free paths of losing energy is roughly the same as naive estimations,  Eq.$(\ref{mfp_strong})$. 

At  low energy,  the coherence length given in Eqs.$(\ref{mfp-proton-i_2})$ is short, and is appropriate for the size of proton wave packet. This value is in agreement with the value considered in  1960's, \cite{Goldberger}     
%%%%%%%%%%%%%%%%%%%%%%%%%%%%%%%%%%%%%%%%%%%%%%%%%%%%%%%%%%%%%%%
\subsubsection{Protons in a beam bunch of  high energy accelerators }
%%%%%%%%%%%%%%%%%%%%%%%%%%%%%%%%%%%%%%%%%%%%%%%%%%%%%%%%%%%%%
In recent high energy experiment such as LHC at CERN,  a proton in beam has a peculiar coherence  of wave packets. The beam is composed of bunches of semi-macroscopic number of protons around $10^{11}$. Protons are correlated by  Coulomb force like electrons in metals or nano-particles.  
The proton's coherence length in the bunch  is affected by various factors.  Beams are affected by strong electric and magnetic fields long time.  Typical lengths of the fields are  macroscopic. Consequently,  the macroscopic coherence may disappear and the coherence length may be provided from a length of Coulomb wave function.   In the following, we estimate the proton coherence length of  Coulomb gas. 

 Protons in a bunch interact with other protons with Coulomb force of strength proportional to  a dimensionless coupling constant.
    This many-body proton state  is  Coulomb gas of protons.  A coherence length  is  obtained with the length of Coulomb wave function expressed by proton mass
\begin{eqnarray}
a_\text{proton}  = \frac{4 \pi \epsilon_0 \hbar^2}{m_p c^2} =2.89 \times 10^{-14} [\text {m}]. \label{proton_C}
\end{eqnarray}   
This gives the momentum variation
\begin{eqnarray}
& &\Delta p_\text{proton}=\frac{\hbar c}{a_\text{proton} }=6.9 \,\text{MeV}.
\end{eqnarray} 
This value is much larger than the momentum spreading of a  proton in matter described by Eq.$(\ref{mfp-proton-i2})$. It would be interesting to observe their physical effects.

%%%%%%%%%%%%%%%%%%%%%%%%%%%%%%%%%%%%%
\subsection{Pion }
%%%%%%%%%%%%%%%%%%%%%%%%%%%%%%%%%%%%%
\subsubsection{Pion production}
Pions are produced by collisions of a proton or a nucleus with nucleus in targets and live for  finite time.  Charged pions decay by the weak interaction and  neutral pions decay by the electromagnetic interaction. An average life time of the charged one at rest is $\sim 10^{-8}$ seconds and that of the neutral one is $\sim 10^{-16}$ seconds.   As these times are much longer than a time scale of the strong interaction,  these values are approximated with the infinity, when pion's coherence property due to the strong interaction is studied.  Accordingly, we study coherence properties of a stable pion  hereafter. This is governed by properties of proton, of nucleus, and of collision processes.  A proton in matter interacts with atoms and has a finite coherence length. Nucleus has intrinsic sizes of order $10^{-15}$~m and is expressed by  certain wave functions of baound states.  Their sizes are slightly larger than nucleus size. So we use in
the present paper the value $10^{-15}$~[m] for the nucleus size.

\subsubsection{Pion coherence length }

 Owing to the short range nature of the strong interaction, the  pion pruduction takes place  inside  of a proton or nucleus.  Coherence length of the  produced pion in the proton collision  with nucleus is governed by the  proton's coherence length and target size. There are two lengths for the proton coherence length, Eq.$(\ref{mfp-proton-i2})$ in  the normal matter and  Eq.$(\ref{proton_C})$ in the extreme high energy beam.

In the  normal matter, a proton has much longer  mean free path  than a length of a target wave function. The coherence length of the pion, ${\delta x}_\text{pion}$, is expressed by an overlap time $\delta \tau$, the coherence length of a proton, ${\delta
x}_\text{proton}$,  and their velocities by Eq.$(\ref{wave-packet_3})$, Eq.$(\ref{crossing_time})$, and Eq.$(\ref{crossing_time_{inter}})$.
%\begin{eqnarray}
%& &{{\delta x}_i \over v_i}={{\delta x}_f \over v_f }=\delta \tau,\\
%& &{\delta x}_f={v_f \over v_i}{\delta x}_i\approx {\delta x}_i .
%\nonumber
%\end{eqnarray}
Consequently  the pion's coherence
of $1~\text{GeV}/c$ or larger 
momentum is given by Eq.$(\ref{mfp-proton-i2})$ as, 
\begin{eqnarray}
L_\text{pion}\sim  0.40 -1.0~[\text{m}].
\label{mfp-pion_i}
\end{eqnarray} 
In the  high energy beam in LHC, the pion coherence length  determined by  the proton coherence length  Eq.$(\ref{proton_C})$  is  
  \begin{eqnarray}
L_\text{pion}\sim 10^{-14} ~[\text{m}].
\label{mfp-pion_c}
\end{eqnarray} 
This  is  much shorter than Eq.$( \ref{mfp-pion_i} )$.    The beam properties of experiments are necessary to find coherence  lengths.  It is  important to find which length should be used because two lengths are very different.
  We use Eq.~$(\ref{mfp-pion_i})$ or Eq.$( \ref{mfp-pion_c})$ in latter sections.  

Pions keep the same coherence lengths  while propagating  freely in the vacuum.   In
a dilute gas, the interaction with matter is negligible and pions freely propagate with the same coherence lengths.    
The size of  pion wave packet in the initial or final states thus derived is different from  an average length of  pion production   area.   
By Fourier transforming the  pion production amplitude, a position dependence of the pion can be computed.  This length represents a spatial size of interaction region and is not related to the size of pion wave packet.       

%%%%%%%%%%%%%%%%%%%%%%%%%%%%%%%%%%%%%%
\subsection{Muon }
%%%%%%%%%%%%%%%%%%%%%%%%%%%%%%%%%%%%%%
A muon is produced by a pion decay. Its  coherence length is determined by the coherence length of the pion and decay processes. 
%%%%%%%%%%%%%%%%%%%%%%%%%%%%%%%%%%%%%
\subsubsection{Decay of pion}
%%%%%%%%%%%%%%%%%%%%%%%%%%%%%%%%%%%%
Coherence length of a muon produced by a pion decay is connected with that of a  pion by a
ratio of velocities by Eq.$(\ref{wave-packet_3})$,  Eq.$(\ref{crossing_time})$ , and Eq.$(\ref{crossing_time_{inter}})$ as
 \begin{eqnarray}
{{\delta x}_\text{pion} \over v_\text{pion}}={{\delta x}_\text{muon} \over v_\text{muon}},
\end{eqnarray}
and is expressed as
\begin{eqnarray}
\delta x_\text{muon}={v_\text{muon} \over v_\text{pion}}\times \delta x_\text{pion}.
\label{meanfp-muon-ratioi}
\end{eqnarray}
For  relativistic particles,  velocities are light velocity and 
the velocity ratio is unity.
 
For an  initial pion of  momentum spreading, ${\Delta
p}_\text{pion}$,
the final
muon has also a momentum spreading, ${\Delta p}_\text{muon }$,  
\begin{eqnarray}
{\Delta p}_\text{muon}={\Delta p}_\text{pion}+O\left({\hbar \over \delta x}_i\right).
\end{eqnarray}
%%%%%%%%%%%%%%%%%%%%%%%%%%%%%%%%%%%%%%%%%%%%%%%%%%%%%%%%%
%\subsection{Muon coherence length}
%%%%%%%%%%%%%%%%%%%%%%%%%%%%%%%%%%%%%%%%%%%%%%%%%%%%%%%
Combining Eq.~(\ref{meanfp-muon-ratioi})  with Eq.~(\ref{mfp-pion_i}) 
we have the coherence length of muon 
\begin{eqnarray}
L_\text{muon}\sim  0.40 - 1.00~[\text{m}],
\label{mfp-muoni}
\end{eqnarray}  
and combining Eq.~(\ref{meanfp-muon-ratioi})  with Eq.~(\ref{mfp-pion_c}) 
we have the coherence length of muon 
\begin{eqnarray}
L_\text{muon}\sim  10^{-16} ~[\text{m}].
\label{mfp-muon_c}
\end{eqnarray}  
%%%%%%%%%%%%%%%%%%%%%%%%%%%%%%%%%%%%%
\subsection{Neutrino  } 
%%%%%%%%%%%%%%%%%%%%%%%%%%%%%%%%%%
Neutrinos  interact with matter so  weakly that  their detection are made indirectly by analyzing  processes specific to neutrino.      Consequently, the sizes of neutrino wave packets reveal  unusual   properties.  
Neutrinos are produced by transitions of nucleus or particles due to  the weak interactions of  tiny probabilities. The  time scales of the neutrino  reactions are  much longer than those of the strong or electromagnetic interactions.

%Neutrinos can be used
%as new means to study inside of the Earth, Sun, Moon, and other stars inside of which are not observed directly by %ordinary means such as lights, electrons, or protons. 
%%%%%%%%%%%%%%%%%%%%%%%%%%%%%%%%%%%%%%%%%%%%%%%%%%%%%%%%%%%%%%%%%%%
\subsubsection{Neutrino wave packets and  oscillations}
%%%%%%%%%%%%%%%%%%%%%%%%%%%%%%%%%%%%%%%%%%%%%%%%%%%%%%%%%%%%%%%%%
Neutrino oscillation experiments observe   
%~\cite{SK-Atom,SK-Solar,SNO-NC,KamLAND-Reactor,Borexino,K2K} 
neutrinos from the sun, accelerator, reactors, and atmosphere   \cite{particle-data}.  This  neutrino has a finite coherent length  and propagates almost freely. For a solar neutrino produced by the beta decay of a nucleus, the size is determined by the nucleus size. Similarly a neutrino produced in reaction 
\begin{eqnarray}
{}^7{\rm Be}+e^-\to{}^7{\rm Li}+\nu_e
%% & &Be^7+e \rightarrow Li^7 +\nu   % comment out by jinnouchi
%& &p+p \rightarrow p+n +\bar e+ \bar \nu
\end{eqnarray}
has a unique size if the wave packet size of $\text{Be}$ is finite.  The size of parent nucleus has been estimated by 
\cite{Krauss,Loeb,Kim_Versner,Nussinov,Kiers} from $10^{-6}$ [m] to  $10^{-10}$ [m].  For  neutrino processes in super novae, the value in the core is estimated as $10^{-16}$ [m],  and the value in surface is estimated as $10^{-11}$ [m] \cite{Anada_1,Anada_2}.  The estimations  are spreading widely.

As was proved in \cite{Kayser, Stodolsky}, thermally equilibrium quantities are determined by the transition rates, which are not  affected by the wave-packet sizess.  Non-equilibrium quantities, however,  are affected.  Because, observed oscillation patterns in the long distance are expressed by combinations of production and detection processes, these are approximated well with naive formulae based on the transition rates with small wave packet effect\cite{Giunti,Nussinov,Kiers,Lipkin} in long distance experiments so far. However,the neutrino detection patterns in short distances are affected by $P^{(d)}$  \cite{Ishikawa-Tobita-ptep,Ishikawa-Tobita-anp} that depends the wave packet sizes. 
%%%%%%%%%%%%%%%%%%%%%%%%%%%%%%%%%%%%%%%%%%%%%%%%%%%%%%%%%%%
\subsubsection{Wave packets in neutrino reactions }
%%%%%%%%%%%%%%%%%%%%%%%%%%%%%%%%%%%%%%%%%%%%%%%%%%%%%%%%%%%
For coherence length of a neutrino produced in decay of a particle, the size of parent particles plays an important role.
 A nucleus is a main source and interacts with other nucleus in matter with electromagnetic and strong interactions.  Its mean free path   given by  the Rutherford  cross section $ \sigma_{N}(Ru)$ and nucleus density is 
\begin{eqnarray}
L_{N}=\frac{1}{\sigma_{N}(Ru) n_N}.
\end{eqnarray}
 A neutrino produced from a decay of  nucleus at rest  of average life time $\tau$ is extended over  time by $\tau$.  The neutrino wave spreads in a spherically symmetric manner afterward with the light  velocity, and reaches average size $c \tau $. The wave-packet size  is   $l=c \tau$. Substituting the average life time, we have a very large size for the neutrino.  Solar neutrinos and cosmic neutrinos are the same.

A neutrino  from the pion decay   in high-energy experiments and atmospheric experiments have slightly different coherence lengths.  Charged pions have  average life time at rest is $\tau_{\pi} \sim 10^{-8}$ [s] and neutrinos from the decay at rest  has a coherence length 
\begin{eqnarray}
L_c=c \tau \sim 3 \times 10~ [\text{m}]. 
\end{eqnarray}
A coherence length of the neutrino from pion decays in flight becomes longer by a Lorentz factor. For a pion of an energy $E_{\pi}$,
\begin{eqnarray}
L_c=\tau c =\tau_0 c \times \frac{E_{\pi}}{m_{\pi}} 
\end{eqnarray}

Neutron has average  life time  at rest  $\tau_{neutron} \sim 15$ minutes, and the size  of the  neutrino  from the neutron  at rest is  
\begin{eqnarray}
L_c= 3 \times 10^8 \times 15 \times 60 \sim2.7 \times 10^{10} [\text{m}].
\end{eqnarray}    
 A neutrino wave packet size from nucleus beta decay is determined by the avera life time also. 
 The size of neutrino wave packet $\sigma_{\nu}$ for the initial states are macroscopic length.
%%%%%%%%%%%%%%%%%%%%%%%%%%%%%%%%%%%%%%%%%%%%%%%%%%%%
\subsubsection{Field theory and neutrino wave packets}
%%%%%%%%%%%%%%%%%%%%%%%%%%%%%%%%%%%%%%%%%%%%%%%%%%%%%%%%%%%%%%%%%%%%%%%%%%%
 The probability $P(T)$  with interaction $ H_{int}$    reveals the long range correlations, and  includes   $P^{(d)}$ that is not expressed by the golden rule \cite{Ishikawa-Tobita-anp}. The new term gives unusual behaviors that depend on the wave-packet sizes to  the transition probability of elementary processes \cite{Ishikawa-Shimomura, Ishikawa-Tobita-ptp, Ishikawa-Oda, Ishikawa-Oda-Nishiwaki}.

It is worthwhile to compare the present  packets sizes  with others   in the literature.  One uses a
size defined with the one-particle momentum distribution of the final states \cite{Akhmedov}.  This size  represents a spatial length  of interaction  region, and is not related  to the size of initial or final states \cite{Giunti_2,Giunti_3,Beuthe}  
An average length of  interaction region  is different from  the length of  initial state or final state. Another  length  obtained from a density matrix of many-body states \cite{Jones} under the interaction $e^{-\epsilon |t|}H_{int}$  represents an effective length of statistical ensemble derived from the eigen states of the total Hamiltonian. This coherence length is translated from a momentum and a energy distribution of the energy eigen states, and provides a particle property of the transition probability. This size represents a spatial size of the stationary states, and  is not related with the size of wave packets of an initial beam of this section or of a next section. These lengths do not represent the sizes of the initial and final states.   Difference will be evident in the next section.   
%%%%%%%%%%%%%%%%%%%%%%%%% geometric picture %%%%%%%%%%%%%%%%%%%%%%%%%%%%%%%%%%%%%%%%
%\begin{figure}[t]
% \includegraphics[scale=1.2]{coherence1.eps}
%\caption{The geometry of the neutrino interference experiment. The
% neutrino is observed by the detector at T and produced at $t_1$ or $t_2$.}
%\label{fig:geo}
%\end{figure}
%%%%%%%%%%%%%%%%%%%%%%%%% geometric picture %%%%%%%%%%%%%%%%%%%%%%%%%%%%%%%%%%%%%%%%

%%%%%%%%%%%%%%%%%%%%%%%%%%%%%%%%%%%%%%%%%%%%%%%%%%%%%%%%%
\subsection{Electron and positron}
%%%%%%%%%%%%%%%%%%%%%%%%%%%%%%%%%%%%%%%%%%%%%%%%%%%%%%%
Electrons are  stable and exist in ordibnary matters.  Isolated  electrons can be used as  incident particles, after  extracted from matter.   Its  coherence  depends on its processes in matter. 
A  band electron in   metal is eigenstate under the peridic potential of the ions and does not interact with periodic ions but with impurities and disorders. A finite relaxation time  measured from the electric resistance of matter is estimated as $10^{-7}$ [m] \cite{Ishikawa-Jinnouchi, Ushioda}.

For  electrons in nano-particles and the bulk, the coherence length is around atomic distance. 
Electron  are either bound states or continuum states. The size  in a  hydrogen atom  is  given by Bohr's radius,
\begin{eqnarray}
a_0= \frac{4 \pi \epsilon_0 \hbar^2}{m_e c^2} =0.53 \times 10^{-10}  [\text{m}] \label{atom_r}.
\end{eqnarray}   
For two  electron system,  wave functions decrease rapidly in the short distance described by  a slope 
\begin{eqnarray}
\frac{1}{a_0},
\end{eqnarray} 
and  this length gives the  electron coherence length in electron gas.
The  momentum variation is 
\begin{eqnarray}
& &\Delta p_e=\frac{\hbar c}{a_0}=3.77  \text{keV}. 
\end{eqnarray} 
This value  is equivalent to the size of bound states. On the other hand,  the relaxation length in metal caused by impurities is longer. At low temperature, free electrons near Fermi energy have  relaxation times due to scattering with impurities at low temperature \label{atom_r}. The corresponding length is around $10^{-7}$ [m] and much longer than Bohr's radius $a_0$.   A naive length between two electrons in metal also is also around  $a_0$.

A positron is the electron's antiparticle and   produced by 
particle collisions or beta decays. In beta decays, the average life time is very large and the energy width is very narrow. The wave packet property of the positron emitted from a nucleus at rest in matter is determined by  a coherence length of the parent wave function initially. 
 The spatial sizes of the parent wave functions are approximately $10^{-15}-10^{-14}$ [m].
 A positron emitted from the nucleus at rest have the nuclear length at the initial instant of time. The wave function expands afterward following a free equation. Similarly, gamma rays emitted from these objects have these lengths at the initial instant of time, and expand afterward following a free equation as will be  studied later in Section 5-A. Its coherent length at a position of distance $l$ from the source is given by the distance $l$.   

A positron  interacts with other charge by Coulomb interaction in matter.
Its mean free path of the propagating positron in solid matter is equivalent to the one of the electron. These are determined by Rutherford cross section, and are much shorter than the above $l$. As a positron is anti-particle of an electron, they annihilate and become two photons.  A positron in matter has a finite average life time. Its value depends on electron density and is around nano-second in solid.  
High energy electrons  will be studied in a succeeding section.
 
%%%%%%%%%%%%%%%%%%%%%%%%%%%%%%%%%%%%%%%%%%%%%%%%%%%%%%%%%
\subsection{Photon }
%%%%%%%%%%%%%%%%%%%%%%%%%%%%%%%%%%%%%%%%%%%%%%%%%%%%%%%
In classical electromagnetism, the electromagnetic wave is expressed by a real number and  satisfies the Maxwell equation. Electromagnetic fields commute with each other and are direct observables.  In quantum  theory,  electromagnetic fields are operators in a many-body space and are not commutative. The many-body states satisfy the many-body Schr\"{o}dinger equation. Photons are  emitted from radiative transitions of bound states such as molecule, atom, nucleus, hadrons,  and others,  are quantum waves of finite extensions.  These are wave packets and their properties are determined by the parents and decay dynamics. These sizes  are  approximately $10^{-15}-10^{-14}$ [m] in nucleus and particles,  $10^{-12}-10^{-10}$ [m] in atoms, larger in molecules and other larger systems. Photons emitted from these objects have these lengths at the initial instant of time, and expand afterward following a free equation as will be  studied later in Section 5-A. Its coherent length at a position of the distance $l$ from the source is given by the distance $l$.  

Photon's interactions with matter are governed by Thomson, Rayleigh, and  Compton scatterings, and photo-electric effect. Cross sections of these processes are 
  the following.

Cross section of  a photon and electron scattering in low energy,   Thomson scattering, is expressed by electron radius $r_e$,
\begin{eqnarray}
    & &\sigma_{\rm Thomson}=\frac{4 \pi}{3} r_e^2 \nonumber \\
    & &r_e=\frac{e^2}{4 \pi \epsilon_0 m_e c^2}.
\end{eqnarray}

The cross section of a photon and an atom(  molecule) scattering,  Rayleigh scattering, is proportional to $\lambda^{-4}$.
\begin{eqnarray}
    \sigma_{\rm Rayleigh}=\frac{8 \pi}{3}\left(\frac{\pi \alpha}{\epsilon_0}\right)^2 \lambda^{-4}
\end{eqnarray}
where $\alpha$ is a polarizability and $\lambda$ is wave length.

Differential cross section of a X-ray or gamma ray with a electron, Compton scattering, is given by 
\begin{eqnarray}
& &    \frac{d \sigma_{\rm Compton}}{d\Omega}=\frac{r_e^2}{2}\left(\frac{k_0'}{k_0 }\right)^2\left( \frac{k_0}{k_0'}+\frac{k_0'}{k_0}-\sin^2 \theta\right) \nonumber \\
& &k_0'=\frac{m k_0}{m+k_0(1-\cos \theta)}
\end{eqnarray}
Cross section of  photo-electric effect of an atom of atomic number $Z$ is given by
\begin{eqnarray}
    \sigma_\text{Photo-electric}=\frac{16}{3}\sqrt 2 \pi r_e^2 \alpha^4\frac{Z^5}{k^{3.5}},
\end{eqnarray}
which is proportional to $\frac{1}{k^{3.5}}$ and becomes large in low energy.  The photo-electric cross section is as large as $10$ Mb for Lead(Z=82) at $10$ eV.  

In Figure 5, the photon cross sections in Lead are shown. The photo-electric effect, Compton effect, and pair annihilation give about the same contribution at energy around $1$ MeV, and 
the photo electric effect gives dominant contribution at lower energies.

\begin{figure}[t]
\includegraphics[width=.8\textwidth, bb=0 0 541 532]{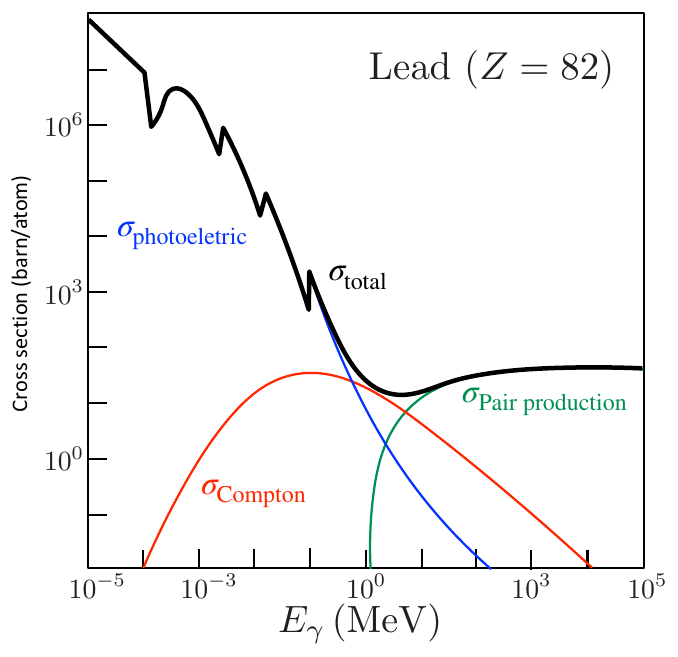}
\caption{   Cross sections of processes of a photon with Lead are shown. At low energy, photo-electric effect is dominant, and Compton scattering and pair annihilation become about the same at intermediate energy.Simplified and redrawn from \cite{particle_data_2}  }
\label{fig:cross section of a photon}
\end{figure}%%%%%%%%%%%%%%%%%%%%%%%%%%%%%%%%%%%%%%%%%%%%%%%%%%%%%%%%
\subsection{Neutron}
%%%%%%%%%%%%%%%%%%%%%%%%%%%%%%%%%%%%%%%%%%%%%%%%%%%%%%

A neutron is slightly heavier than a proton and decays to a proton, an electron, and a neutrino by the weak interaction. The neutron is the proton's iso-spin partner and composed of three quarks and electrically neutral but has 
  a finite magnetic moment. The  strong  interaction of neutron with  nucleus or other hadrons are almost equivalent to  that of the proton.  Accordingly,  the neutron  is produced in the same manner as the proton.    Neutron is electrically  neutral and the electromagnetic interactions  is through magnetic moment. Consequently, the energy loss rate of a neutron in matter is different from a proton. A wave packet size is  expressed with a mean free path.   
%%%%%%%%%%%%%%%%%%%%%%%%%%%%%%%%%%%%%%%%%%%%%%%%%%%%%%%%%
\subsection{Particles in targets}
%%%%%%%%%%%%%%%%%%%%%%%%%%%%%%%%%%%%%%%%%%%%%%%%%%%%%%%%%%%%%%%
\subsubsection{Solid state}
 The targets using  solid   are composed of many atoms. These  atoms  are bound at periodical positions in crystals. Electrons are either bound to atoms or move freely. The electrons in the latter are expressed by waves of  a finite relaxation time and coherence length.   The wave functions of bound atoms and valence electrons are normalized with fixed  center positions.   The size of wave packet agrees with the size of the bound states. 

\subsubsection{Gas}
 For gas targets, atoms are not bound but move freely. These are expressed by wave packets of the sizes  determined by the mean free paths, and depend on the density and other parameters of circumstance.    

%%%%%%%%%%%%%%%%%%%%%%%%%%%%%%%%%%%%%%%%%%%%%%%%%%%%%%%%%%%%%
%{ \bf KOKOMADE}
%%%%%%%%%%%%%%%%%%%%%%%%%%%%%%%%%%%%%%%%%%%%%%%%%%%%%%%%%%%
\section{Wave packets of out-going waves of scatterings }
%%%%%%%%%%%%%%%%%%%%%%%%%%%%%%%%%%%%%%%%%%%%%%%%%%%%%%%%%%%%
%%%%%%%%%%%%%%%%%%%%%%%%%%%%%%%%%%%%%%%%%%%%%%%%%%%%%%%%%%%%%%%
The final states in the scattering  represent particles observed by detectors. Microscopic states in the detection processe provide the wave-packet sizes.  
Conversly these sizes are determined by  microscopic states in the detection processes. 
  The sizes are analyzed using the relations of Section 2

 Out-going waves  interact with matter waves  in the detector.  The  amplitude of this process  is expressed by a scalar product of the out-going state  with the state in the detector as shown in  a following subsection. A square of the modulus  becomes maximum and   the proceses reveal classical behavior when two wave-packet sizes  coincide.  Conversly  the classical causality in the detection processes is guranteed when   the wave-packet sizes of the out-going states coincide with the coherence lengths of the microscopic states in detector.        
  This is understandable from the fact that observed signals  created from atoms or other units of matter in the detector  represent  the scattering probability then.  Accordingly, a spatial size of a unit of microscopic state where quantum mechanical processes take place determines the size of wave packet of the out-going wave.  This  wave-packet size is  different generally from sizes of  in-coming waves of the scattering.

 The wave-packet amplitude defined in this manner  is  different from the amplitude derived from  the flux of eigen states of the total Hamiltonian. The latter method  uses a ratio of fluxes of the outgoing wave with that of incoming waves following a classical formula. The classical  expression is valid for a measument  made locally in space-time position. This differs from the real measured values  obtained  from the scalar product of the initial and fianl states, and  neither includes the final states nor depends on a detection process.  In fact, it is not known how  to compute the cross section that implements  final states from this method. Moreover, a computation of an interference term or a correlation between the initial and final states is unknown. Considering these behaviors, the cross section formula of using the eigen states of the total Hamiltonian is different from ours. This is not studied in the present paper.  
%%%%%%%%%%%%%%%%%%%%%%%%%%%%%%%%%%%%%%%%%%%%%%%%%%%%%%%%%%%%%
\subsection{ Probability of the processes in the detector assumes   $P_{\alpha \beta}=1$}
%%%%%%%%%%%%%%%%%%%%%%%%%%%%%%%%%%%%%%%%%%%%%%%%%%%%%%%%%%%%
The wave packet sizes of the final states reflect  the  processes  in the detector. The transition amplitude of  a scattered particle in a state $f_\text{final}(\vec x)$ with detector   is proportional to an overlap integral of  $f_\text{final}(\vec x)$ with  a  wave function of the detecting particle in the detector $g_\text{detect}(\vec x)$     
\begin{eqnarray}
    \int d \vec x f_\text{final}(\vec x) g_\text{detect} (\vec x)
    \Psi_N (\vec x), \label{integral_d}  
\end{eqnarray}
where $\Psi_N(\vec x)$ represents the wave function of undetected particles. It is convinient to express  undetected states with  plane waves 
\begin{eqnarray}
  \Psi_N (\vec x) =F_N e^{i \sum_l \vec p_l \vec x}. 
\end{eqnarray}
Then the magnitude of integral Eq.$(\ref{integral_d})$  
is proportional to the product of wave packets as given by Eq.$(\ref{overlap_wp})$, and the numerical factor becomes the unity only for 
\begin{eqnarray}
f_\text{final}(\vec x,t) =g_\text{detect}( \vec x,t)
\end{eqnarray}
from Eq.$(\ref{two_wave_packets})$  in the Appendix. Then the transition process in the detector  follows  a   motion that satisfies classical causality and excludes other process.  In this situation, the wave packet of the final state in the scattering is in agreement with the wave packet in the detector. 
\begin{eqnarray}
\sigma_{\text final}=\sigma_{\text detect}  \label{wave_ps} 
\end{eqnarray}
This  ensures the classical causality of the detection. Experiments of assuming the clasical causality of detection process are represented by this wave packet.

The one particle  is either in a bound state or in a continuum state. In the former case, the size is an atomic scale.  In the latter case, the size varies and depends on the situation. The size is large for propagating  waves in dilute systems, and becomes huge in exceptional situations.    The measured probabilities depend on the environments. The wave-packet sizes 
 of particles in the final states are evaluated in this section.

\subsection{Proton }
Wave-packet size of  a proton in a final state depends on  detection processes and   detectors.   

\subsubsection{In solid}
 In standard experiments, a detection of proton  is made with  a signal generated  by the detector. A microscopic system that produces the signal is an atom in the solid, which  is expressed by bound states. The  interaction between the proton and atoms are  Bethe-Bloch processes or Bremsstrahlung in the energy regions of ordinary high-energy  experiments \cite{particle-data}.  These are specified by  two lengths.  One is the spatial size of atoms and another is the mean free path of the proton in the material. 
  % Because identifications of particle species are made and momenta are evaluated from these radiations, which are emitted from excited states of atoms, the size is governed %by the electron's wave functions. The overlap of their wave functions becomes the unity, and the size of the atom represents the size of the proton wave packet. 

The sizes  of the electron wave functions  are between the sizes of nucleus and atoms. Average size is a size of the inner core of the atom, and  is close to sizes of the nucleus. The size of the inner core for $Z=50$ is
\begin{eqnarray}
L_i=\frac{1}{50} R_\text{Bohr}, \quad R_\text{Bohr}=0.529 \times 10^{-10} [\text{m}].\label{mfp_finalp}
\end{eqnarray}
This length  represents an average  size of wave packets  of  charged particles in the final states, 
\begin{eqnarray}
    \sqrt{ \sigma_p}=L_i,
\end{eqnarray}
 from Eq.$(\ref{wave_ps})$.  The size determined by the strong interaction is around $10^{-15}$ [m].  
The wave packet size of proton is short  when that is detected by bound states of atom in solid. 

\subsubsection{Dilute gas}
In dilute gas, density is low and the size becomes different. An atom is moving almost freely in space and  described by a wave of large mean free path. When this atom  interacts with the  proton,  the proton  wave-packet is  large.   This is much larger in the dilute gas than in the solid. 

%Even if a proton is not detected but makes a transition, the events should occur according to the corresponding probability.   The wave %packet sizes vary and can be much larger. 

%%%%%%%%%%%%%%%%%%%%%%%%%%%%%%%%%%%%%
\subsection{Pion and muon}
%%%%%%%%%%%%%%%%%%%%%%%%%%%%%%%%%%%%%

Detection processes of pions and muons by Bethe-Bloch processes with the atom in solid are almost equivalent to those of protons.  Consequently, the sizes of wave packets of pion and muon are almost the same as the size of proton Eq.$(\ref{mfp_finalp})$.
 Cross section of the strong interaction is small of the order $20$ mb.  Accordingly, processes that have origins only  in the strong interaction, which is possible for pions, wave packets are small.   They are around $10^{-15}$ [m].   
  In this situation, the wave packet size of the muon in the final state is not related with that of the pion. 

%%%%%%%%%%%%%%%%%%%%%%%%%%%%%%%%%%%%%
\subsection{Neutrino  }
%%%%%%%%%%%%%%%%%%%%%%%%%%%%%%%%%%

A wave packet size of the out-going neutrino  based on  the probability principle  is very different from that of the in-coming neutrino, and  determined by  the detection processes.
A neutrino interacts extremely weakly with matter and reactions rarely occur. The out-going neutrino   propagates freely untill detected.  This process was described in  Sub-section 4-A, and  the wave-packet size agrees to the wave-packet size in the detector, %It is hard to have large statistics in neutrino reactions. If the wave packet effect is large in the neutrino identifications, small number of the events does not matter, once the neutrino events are identified. 
with which neutrino identification is made.     Accordingly, the size of neutrino wave packet is determined from a size of the wave function of microscopic state that a neutrino interacts with from  Eq.$( \ref{wave_ps})$. The wave packet sizes are given by the sizes of the nucleus or electrons. The nucleus has a size of the order $L_{N}=10^{-15}$[m] and an
electron's wave function has a size of order $L_{e}=10^{-10}$[m].  The size of the wave packet is $10^{-10}$~[m] in the former, or $10^{-15}$[m] in the latter.

A muon neutrino has reactions  
\begin{eqnarray}
& &\nu_{\mu}+e^{-} \rightarrow e^{-}+\nu_{\mu}, 
\label{numu-leptonic}\\
& &\nu_{\mu}+e^{-} \rightarrow \mu^{-}+\nu_{e}, 
\label{numu-lcharged}\\ 
& &\nu_{\mu}+A \rightarrow \mu^{-}+(A+1)+X,
\label{numu-hcharged}\\ 
& &\nu_{\mu}+A \rightarrow \nu_{\mu} +A  +X,
\label{numu-hneutral}
\end{eqnarray}
in the detectors, where $A$ stands for nucleus of atomic number $A$. A  size of the neutrino wave packet   in processes $(\ref{numu-leptonic})$ and
$(\ref{numu-lcharged})$ is given by the size of atoms  
\begin{eqnarray}
\sqrt \sigma_{\nu}\sim10^{-10}~[\text m]
\label{mfp-neutrino-e}
\end{eqnarray}
 and  $\sqrt \sigma_{\nu}$ in processes $(\ref{numu-hcharged})$ and
$(\ref{numu-hneutral})$ is  is given by the size of nucleus 
\begin{eqnarray}
\sqrt \sigma_{\nu} \sim10^{-15}[\text m].
\label{mfp-neutrino-N}
\end{eqnarray}
Lengths Eq.$(\ref{mfp-neutrino-e})$ and Eq.$(\ref{mfp-neutrino-N})$ are so short that may not influence observed  neutrino oscillation pattern. We will see that these lengths are sufficient to observe neutrino oscillations in the long distance. However,  observed  pattern and strength of a muon neutrino in short distance, or in intermediate distance is affected.  A distance between a position where the neutrino is produced and a position where the neutrino is detected is up to few hundred Meters in the former cases and longer than the few hundred kilometers in the latter ones.  The time differences of the neutrino arrival to the detectors is estimated. 

Neutrinos propagate with nearly the velocity of light. A velocity of  the
 neutrino of energy  $1~\text{GeV}/c^2$ and  the mass 
$0.1~\text{eV}/c^2$   is
\begin{eqnarray}
& &v/c=1-2\epsilon, \\
& &\epsilon=\left({m_{\nu}c^2 \over E_{\nu}}\right)^2=5\times 10^{-21}\nonumber.
\end{eqnarray} 
The  distance $l$ that this neutrino propagates is slightly smaller that the distance $l_0$ that the light propagate in the same time,   
\begin{eqnarray}
l=l_0(1-\epsilon)=l_0-\delta l,\quad\delta l= l_0\times \epsilon.
\end{eqnarray}
This difference of distance,
$\delta l$, becomes
\begin{eqnarray}
& &\delta l\sim 5\times 10^{-19}~[\text{m}]; ~l_0=100~[\text{m}], \\ 
& &\delta l \sim 5\times 10^{-18}~[\text{m}]; ~l_0=1000~[\text{m}],
\end{eqnarray}
which is much shorter than the nucleus size, and the lengths Eq.$(\ref{mfp-neutrino-e})$ and $(\ref{mfp-neutrino-N})$.

%%%%%%%%%%%%%%%%%%%%%%%%% geometric picture %%%%%%%%%%%%%%%%%%%%%%%%%%%%%%%%%%%%%%%%
%\begin{figure}[t]
% \includegraphics[scale=1.2]{coherence1.eps}
%\caption{The geometry of the neutrino interference experiment. The
% neutrino is observed by the detector at T and produced at $t_1$ or $t_2$.}
%\label{fig:geo}
%\end{figure}
%%%%%%%%%%%%%%%%%%%%%%%%% geometric picture %%%%%%%%%%%%%%%%%%%%%%%%%%%%%%%%%%%%%%%%

Interactions of an electron neutrino  in detectors are 
\begin{eqnarray}
& &\nu_{e}+e^{-} \rightarrow e^{-}+\nu_{e}, 
\label{nue-leptonic}\\ 
& &\nu_{e}+A \rightarrow e^{-}+(A+1) +X,
\label{nue-hcharged}\\ 
& &\nu_{e}+A \rightarrow e  +A  +X.
\label{nue-hneutral}
\end{eqnarray}
A neutrino 
wave packet  $\sqrt \sigma_{\nu}$ 
is of order   $10^{-10}$[m] in processes $(\ref{nue-leptonic})$ and  of order   $10^{-15}$[m] in processes $(\ref{nue-hcharged})$ and
$(\ref{nue-hneutral})$. They are treated in
the same way as the neutrino from the pion  decay.

Low energy neutrino  from the sun or reactors must be
treated separately and will be studied in a next paper.

%%%%%%%%%%%%%%%%%%%%%%%%%%%%%%%%%%%%%%%%%%%% Section 3 %%%%%%%%%%%%%%%%%%%%%%%%%%%%%%%%%%%%%%%%%%%%%%%%%%%
%%%%%%%%%%%%%%%%%%%%%%%%%%%%%%%%%%%%%%%%%%%%%%%%%%%%%%%%%%%%%%%%%%%%%%%%%%%%%%%%%%%%%%%%%%%%%%%%%%%%%%%%%%
%%%%%%%%%%%%%%%%%%%%%%%%%%%%%%%%%%%%%%%%%%%%%%%%%%%%%%%%%
\subsection{Electron and positron }
%%%%%%%%%%%%%%%%%%%%%%%%%%%%%%%%%%%%%%%%%%%%%%%%%%%%%%%
Wave packets of the electron and positron in high energy is different from that in low energy.  
An electron injected to the detector with high energy interacts with atoms by Bremsstrahlung and Bethe-Bloch processes. Electron and positron emit radiations when accelerated. The Bremsstrahlung is caused by the nucleus electric field, which is strong near the nucleus. The reaction takes place in this region, and its spatial size determines the  wave-packet size.

Bethe-Bloch processes are the ionization of atoms by charge particles characterized by the wave functions of electrons. 
The size of the wave function is the Bohr radius $R_{Bohr}$ in the hydrogen atom. That are smaller in heavier elements or larger for molecules or semiconductors. As an electron is scattered by these states in the detector, the size of the wave packet is determined by its size \cite{Ishikawa-Jinnouchi, Ushioda}. The size of the wave packet in a cascade reaction should be larger than the atomic size.   
%%%%%%%%%%%%%%%%%%%%%%%%%%%%%%%%%%%%%%%%%%%% Section 3 %%%%%%%%%%%%%%%%%%%%%%%%%%%%%%%%%%%%%%%%%%%%%%%%%%%
%%%%%%%%%%%%%%%%%%%%%%%%%%%%%%%%%%%%%%%%%%%%%%%%%%%%%%%%%
%\subsection{Gamma rays }ermined by a spatial size of a physical system which the photon interacts with. A size of  area of the Compton scattering  varies with its energy from an atomic size to a size of an inner core. The  area  of pair creation is from the size of the inner core to the size of nucleus. The sizes are approximately $10^{-15}-10^{-14}$ [m] for the nucleus and particles,  $10^{-12}-10^{-10}$ [m] for atoms. 

The wave-packet size  also depends on detectors. 

NaI(Tl) scintillator:

The size of ion wave functions in I is $2.06 \times 10^{-10}$[m]. This size leads a momentum width $\frac{c \hbar}{\Delta X}=\frac{2 \times 10^5 \times 10^{-15}}{2.06 \times 10^{-10}}=1$ keV.  An energy resolution of the NaI(Tl) detector due to statistical origin is much larger. The wave packet effect is smaller than the statistical uncertainty. 

Germanium detector (HPGD) : 

The size of the electrons in Ge atom is about the same as that of I, but
a size  of  electron wave functions of semiconductor is larger than the atomic size. An effective Bohr radius is enhanced by an effective mass $m^*$ and dielectric constant $\epsilon^*$ in matter and given by
\begin{eqnarray}
a_B^*=a_B \frac{m_0}{m^*} \epsilon^*
\end{eqnarray}
Substituting the transverse mass  $0.08\, m_0 $ and $\epsilon^*=16 $, 
\begin{eqnarray}
    a_B^* \approx 200 \, a_B= 100\times 10^{-10} [\text m]
\end{eqnarray}
A wave packet size is approximately $33$ times  
of atomic size. This leads a momentum width around  $0.02$ keV. 
  The size of the electron wave function is larger for molecules and other larger systems.  Electrons wave functions in detectors composed of these systems  are large. and momentum widths are small. 

  Photomultiplier :
 The primary process of the photon in a photomultiplier is the photo-electric effect. The sizes of the electrons determine the wave packet sizes.    

  Calorimeter :

  If a photon interacts with physical system composed of large number of atoms, the size of wave packet is large.

%%%%%%%%%%%%%%%%%%%%%%%%%%%%%%%%%%%%%%%%%%%%%%%%
\subsection{Neutron}
%%%%%%%%%%%%%%%%%%%%%%%%%%%%%%%%%%%%%%%%%%%%%%
A neutron interacts with nucleus in matter by either the  strong interaction or  electromagnetic interaction through a magnetic moment.  Neutron nucleus  cross section varies rapidly with energy in energy regions of excited states. Their geometrical sizes follow the atomic number, and are around $10^{-15}$[m].  For processes of electromagnetic interaction through a magnetic moment, a size can be much larger. 

%%%%%%%%%%%%%%%%%%%%%%%%%%%%%%%%%%%%%%%%%%%%%
\subsection{Quarks}
 %%%%%%%%%%%%%%%%%%%%%%%%%%%%%%%%%%%%%%%%%%%%%%%%%%%%%%%%%%%%%%%%%%%%%%%%%%%%%%%
Quarks are constituent particles of hadrons that are confined in short distances due to a confining interaction, and have no asymptotic states. Quarks are produced in hadron collisions, and exist for short period of time. Quarks do not behave like isolated particles,  but are observed as hadron jets.   Formation and detection processes of quarks are non-perturbative quantum processes which are difficult to analyze analytically. It is hard to express precisely this process, but the processes are considered coherent \cite{particle-data}. Then these are described by wave packets. A spatial size of quark jet is characterized by a confining scale. Accordingly, jets have finite spatial size and are described by wave packets. A size of quark wave packet is a hadronization length, of the order of a confinement scale,  $\frac{1}{500-1000}$ MeV$^{-1}$. This is much shorter than the length of electromagnetic interactions of particles in matter, MeV$^{-1}$, hence. an energy width of a quark wave packet becomes few hundred MeV.   This value is higher than the sizes of hadron wave packets.     
%%%%%%%%%%%%%%%%%%%%%%%%%%%%%%%%%%%%%%%%%%%%%%%%%%%%%%%%%%%
%%%%%%%%%%%%%%%%%%%%%%%%%%%%%%%%%%%%%%%%%%%%%%%%%%%%%%%
\section{Wave packets of natural phenomena}
%%%%%%%%%%%%%%%%%%%%%%%%%%%%%%%%%%%%%%%%%%%%%%%%%%%%%%%%%%%%
%%%%%%%%%%%%%%%%%%%%%%%%%%%%%%%%%%%%%%%%%%%%%%%%%%%%%%%%%%%%%
In  transition processes of natural environments, no measurements are made, but particle collisions  follow  the transition probabilities  which are governed by the initial and final states. These particles have finite coherence lengths due to interactions with  matters in the enviorment, and  are expressed with wave packets.  In a system of an extremely low matter density. the wave-packet sizes  can be much larger than those in ordinary matter. 
A time interval between  initial and final states also varies widely, and a spreading of the wave packet may not be negligible.  Transitions of states under extraordinary conditions may exhibit unusual behaviors. 

A wave packet loses coherence  at a finite $T$ if probability $P(T)$  to others exceed the unity, i.e.,  decoherence effect. In the following, we study the situation where the 
decoherence effect is negligible.   

%%%%%%%%%%%%%%%%%%%%%%%%%%%%%%%%%%%%%%%%%%%%%
\subsection{Spreading of  wave packets}
%%%%%%%%%%%%%%%%%%%%%%%%%%%%%%%%%%%%%%%%%%
We use a unit $c=1$ and $\hbar=1$ for simplicity.   Because the velocity of a wave of  momentum ${\vec p}$ and energy $E(p)=\sqrt{p^2+m^2}$,
\begin{eqnarray}
(\vec v)^i=\frac{ \partial E(\vec p) }{\partial p_i}=\frac{p_i}{E(\vec p)}
\end{eqnarray}
varies with  momentum,   a wave packet spreads after long period of time.  The spreading of  wave packet is negligible  in the standard experiments, but rellevant in many natural phenomena of large time interval.     

A phase factor of the integrand and
the integral over the momentum in Eq.$(\ref{wave-packet_2})$ at a large $|t-T_0|$ is evaluated with the stationary phase approximation and becomes different from that of a small $|t-T_0|$. 
A solution of   
\begin{eqnarray}
& &\frac{\partial }{\partial p_i} \xi_G(p)|_{p^i=p_{X}^i }=0 
\end{eqnarray}
 is expressed as 
\begin{eqnarray}
& &{\vec p}_{X}=m\frac{1}{\sqrt{(t-T_0)^2-({\vec x}-{\vec X}_0)^2}}( \vec x-{\vec X}_0)+O\left(\frac{1}{t-T_0}\right), \nonumber \\
\end{eqnarray}  
where the last term gives a next order contribution \cite{Ishikawa-Shimomura}.
Substituting the stationary momentum into Eq.$(\ref{wave-packet_2})$, we have  a wave function   
\begin{eqnarray}
 & &\langle t,{\vec x}| {\vec P}_0, {\vec X}_0, T_0  \rangle=N_3\left(\frac{1}{ i \frac{\gamma_L}{\sigma}+1}\right)^{1/2}\left( \frac{1}{i \frac{\gamma_T}{\sigma}+1}\right) \nonumber \\
 & &\times e^{-\frac{1}{2}\sigma({\vec P}_X-{\vec P}_0)^2 +i \phi_0 +O\left(\frac{1}{t-T_0}\right)},     \label{wave_expand}
\end{eqnarray}
where
\begin{eqnarray}
& &\gamma_L= \frac{m^2|t-T_0|}{E({\vec P}_{X,0})^3}, ~ \gamma_T= \frac{|t-T_0|}{E({\vec P}_{X,0})}  \label{packet_sizes} \\  
& &\phi_0=-(t-T_0)E({\vec P}_X)+ {\vec P}_X \cdot ({\vec x}-{\vec X}_0).
\end{eqnarray}
 From Eq.$( \ref{wave_expand})$, the wave packet has a peak at ${\vec p}_X ={\vec p}_0$.  Eq.$(\ref{wave_expand})$ can be rewritten into a Gaussian  function of
  $\vec x$  of a moving center ${\vec x}_0$ ,
\begin{eqnarray}
& &N_3\left(\frac{1}{ i \frac{\gamma_L}{\sigma}+1}\right)^{1/2}\left( \frac{1}{i \frac{\gamma_T}{\sigma}+1}\right)e^{-\frac{1}{2} \sigma \left( \frac{1}{\gamma_T^2}({\vec x}-{\vec x}_0)_T^2+  \frac{1}{\gamma_L^2}({\vec x}-{\vec x}_0)_L^2\right)} \nonumber \\
& &\times e^{  -i \frac{m^2}{E(P_X)}(t-T_0)}  \label{wave_expand_x}\\
%\end{eqnarray} 
%around the center position  
%\begin{eqnarray}
& &{\vec x}_0={\vec X}_0+(t-T_0) \frac{\vec P_0}{E(P_0)}. \label{center-position}
\end{eqnarray}
Here  the position vector ${\vec x}-{\vec x}_0$ is decomposed into a  component   parallel to
 ${\vec P}_0$ and  another one in transverse directions.  This decomposition  is made for an arbitral  vector $\vec A$ as,   
\begin{eqnarray}
\vec A= {\vec A}_T +{\vec A}_L {\vec n}_L ,\quad {\vec n}_L=\frac{\vec P_0}{P_0},
\end{eqnarray}
where ${\vec A}_T $ is the transverse component and ${\vec A}_L$ is the longitudinal component. From the exponential factor of Eq.$( \ref{wave_expand_x})$, $\frac{\gamma_L}{\sqrt{\sigma}}$ in Eq.$(\ref{packet_sizes} )$ is a wave packet size in the longitudinal direction and $\frac{\gamma_T}{\sqrt{\sigma}}$ is a wave packet size in the transverse direction. The size in the transverse direction $\gamma_T$ increases in time interval $t-T_0$ regardless of the mass. However,the size in the longitudinal direction does not increase with time for $m=0$.  The wave packets expand with time in asymmetric manner. The transverse size always becomes $\infty$ at $t-T_0 \rightarrow \infty$. The phase factor in the amplitude is written as $\phi_0= - \frac{m^2}{E(P_X)}(t-T_0)$, and is small. An additional time dependence emerges from the position ${\vec x}_0 $ of Eq.$({\ref{center-position}})$.  \cite{Ishikawa-Shimomura}

Wave packets are  normalized at an arbitrary time, and a scalar product of wave packets is finite.  At $t -T_0 \rightarrow \infty$,  wave packets extend to infinitely large size and a probability density becomes infinitesimal.  Even though initial states are extended to large sizes, the transferred energy in a scattering is finite, as in the transition in short distance.  
This behavior is specific to the quantum scatterings.  In classical scattterings, an energy of the wave at one position is proportional to its strength there, and  becomes infinitesimal in infinitely large wave packets.  The finite amount of  the transffered energy is a natural consequnce of the probability principle.         

\subsection{Proton, pion, and muon }

Mean free path of a stable particle in a natural environment is determined by its collision with surrounding atoms. This process is the same as that in a detector, and the same formula is applied.  
In space of dilute matters, the size of wave packets of stable particles are large. Pions are produced in hadron or nucleus  collisions with finite coherence lengths and have coherence lengths determined by parents.       A muon produced from a decay of a pion  has the coherence length of the pion. 

%%%%%%%%%%%%%%%%%%%%%%%%%%%%%%%%%%%%%
\subsection{Neutrino  }
%%%%%%%%%%%%%%%%%%%%%%%%%%%%%%%%%%
The wave-packet size of the neutrino in the initial state of scatterings in natural environment  is determined by its production process.
A neutrino wave function produced from a  pion of the coherence length $L_{\pi}$ is broadened by the distance given by a finite time interval  $\tau_{\pi}$ and its velocity $c$ as
\begin{eqnarray}
L_{\nu, \pi}=L_{\pi} + c \tau_{\pi}.
\end{eqnarray}    
The coherence length of the neutrino produced in a muon decay is expressed by the coherence length and the average lifetime of the muon,
 \begin{eqnarray}
L_{\nu, \mu}=L_{\mu} + c \tau_{\mu}.
\end{eqnarray}    
In dilute matter, the size of wave packets of neutrinos in the initial states of reactions  are large.

The wave-packet size of the neutrino in the final state of scatterings  is treated in the equivalent manner as the previous section. The sizes are determined by the spatial sizes of the matter which  the neutrino interact with. However, a distance between the scattering and  detection positions can be extremely large, and the effects due to expansions may not be negligible.  
  
%%%%%%%%%%%%%%%%%%%%%%%%% geometric picture %%%%%%%%%%%%%%%%%%%%%%%%%%%%%%%%%%%%%%%%
%\begin{figure}[t]
% \includegraphics[scale=1.2]{coherence1.eps}
%\caption{The geometry of the neutrino interference experiment. The
% neutrino is observed by the detector at T and produced at $t_1$ or $t_2$.}
%\label{fig:geo}
%\end{figure}
%%%%%%%%%%%%%%%%%%%%%%%%% geometric picture %%%%%%%%%%%%%%%%%%%%%%%%%%%%%%%%%%%%%%%%

%%%%%%%%%%%%%%%%%%%%%%%%%%%%%%%%%%%%%%%%%%%% Section 3 %%%%%%%%%%%%%%%%%%%%%%%%%%%%%%%%%%%%%%%%%%%%%%%%%%%
%%%%%%%%%%%%%%%%%%%%%%%%%%%%%%%%%%%%%%%%%%%%%%%%%%%%%%%%%%%%%%%%%%%%%%%%%%%%%%%%%%%%%%%%%%%%%%%%%%%%%%%%%%

%%%%%%%%%%%%%%%%%%%%%%%%%%%%%%%%%%%%%%%%%%%%%%%%%%%%%%%%%
\subsection{Electron and  positron  }
%%%%%%%%%%%%%%%%%%%%%%%%%%%%%%%%%%%%%%%%%%%%%%%%%%%%%%%

An electron interacts with charged matter by Coulomb interaction, 
and the mean free path is determined by the Rutherford cross section and charge density.  The wave-packet size is extremely large in dilute matters.  A positron is the same.

%%%%%%%%%%%%%%%%%%%%%%%%%%%%%%%%%%%%%%%%%%%%%%%%%%%%%%%%%
\subsection{Photon and other massless waves }
%%%%%%%%%%%%%%%%%%%%%%%%%%%%%%%%%%%%%%%%%%%%%%%%%%%%%%%
\subsubsection{Long distance  propagation}
%%%%%%%%%%%%%%%%%%%%%%%%%%%%%%%%%%%%%%%%%
As massless waves propagate with the velocity of light, their wave packets spread in asymmetric way. In a direction parallel to its central momentum, the wave does not spread and keeps the shape. In a transverse direction, that spreads symmetrically with an average radius increasing in time. Velocities  of the spreading are from Eq.$(\ref{packet_sizes})$, 
\begin{eqnarray}
& &v_l=  0              \\
& &v_t=   \sqrt{\frac{2}{\sigma}}\frac{1}{E(P_0)}                     \nonumber
\end{eqnarray}
Accordingly the wave packet becomes pancake shape after some time. The velocity of spreading in a transverse direction is inversely proportional to the initial size.

%%%%%%%%%%%%%%%%%%%%%%%%%%%%%%%%%%%%%%%%%%%%%%%%%%%
\subsubsection{Light scattering }
%%%%%%%%%%%%%%%%%%%%%%%%%%%%%%%%%%%%%%%%%%%%%%%%%%%%%
 Light is produced when charged particles collide. The light propagates long distance almost freely in a space of dilute gas. Its quantum mechanical scattering reveals behavirs spefic to the probability principle differnt from the classical mechanical scattering. In classical physics, an energy of the light is proportional to the strength of the wave. If that has an energy $E_0$ at the distance $l_0$, then its  energy at a distance ,$l$, is give by
\begin{eqnarray}
E=\left(\frac{l_0}{l}\right)^2 E_0.
\end{eqnarray}  
For the initial energy $E_0=1$ [eV] , and the distances  $l_0=1$[m],  and $l=10^{18}$ [m],  the magnitude of energy is 
\begin{eqnarray}
E=10^{-36} \,[\text{eV}].\label{infinitesimal_e}
\end{eqnarray}
This is too small to excite an atom, and gives no effect in observations. 

In quantum mechanics, a wave packet of light created in a star of energy $E_0$ interacts with atoms even at a long distance with energy $E_0$.   Accordingly, 
the light of wave packet  produced by transitions of the wave packet of charged particles spreads while that  propagates long distance. When that  interacts with a target , its frequency looks different from $E_0$ from the formula, $(\ref{wave_expand_x} )$. The transition amplitude provided by integrating over the time reveals the same formula as that of no spreading.   Because the photon energy is determined by its frequency, that is much larger than Eq.$( \ref{infinitesimal_e})$  agrees with observations.     

%%%%%%%%%%%%%%%%%%%%%%%%%%%%%%%%%%%%%%%%%%%%%%%%%%%%%%%%%%
\subsubsection{ Rayleigh scattering} 
%%%%%%%%%%%%%%%%%%%%%%%%%%%%%%%%%%%%%%%%%%%%%%%%%%%%%%%%%
Rayleigh scattering is a scattering of a light of wave length $\lambda$ with an atom of the size $L_a$, in the region $\lambda \gg L_a$. The sun light is scattered by Rayleigh scattering due to atoms in the atmosphere at a high attitude. At the solar surface, the density of charged particles is low and the mean free path is long. The light has a large mean free path. In the space between the sun and the earth, the density is much lower and is like the vacuum.  The sun lights are expressed by large wave packets. 
The force in the atmosphere is the gravity of the earth and the pressure of the atmosphere. The former is approximately expressed by the linear potential. Eigen states are expressed by the Airy function. The force of the pressure gradient is in the opposite direction and balances the molecule. The wave packet is formed in the normal direction. There is no force in the tangential direction. The wave packet will become spherically symmetric after many reactions.

%%%%%%%%%%%%%%%%%%%%%%%%%%%%%%%%%%%%%%%%%%%%%%%%%%%
\subsubsection{Extinction and decoherence }
%%%%%%%%%%%%%%%%%%%%%%%%%%%%%%%%%%%%%%%%%%%%%%%%%
Wave packet sizes of gamma rays were determined by the average coherence lengths of the gamma rays. Another is the length that the probability of initial gamma ray to disappear in medium.   This is expressed with a penetration length. Scattering of a light with macroscopic objects of length $L$ with internal structure in the region $L \gg \lambda$ shows these behaviors. It appears in the scattering of light with macroscopic objects such as ices, droplet of water, aerosol, and others belong to the phenomenon.   Mie scattering is also an example in this category. Light form these scatterings have the short wave packet size determined by the penetration length.    
%%%%%%%%%%%%%%%%%%%%%%%%%%%%%%%%%%%%%%%%%%%%%%%%%%%%%%
\subsubsection{Cosmic microscopic background}
%%%%%%%%%%%%%%%%%%%%%%%%%%%%%%%%%%%%%%%%%%%%%%%%%%%
Photons in a plasma have a finite mean free path due to Thomson scattering given by Eq.$(\ref{mean_free})$. 
At a temperature higher than the ionization energy $T > T_c, kT_c=E_\text{ion}$, where $E_\text{ion}$ is the ionization energy, there are free electrons.  
%\begin{eqnarray} 
%n_e=  n_p= ,
%\end{eqnarray}

(I) Above the critical temperature:

Around the decoupling time of the early universe, the density of  photon,  electron, and proton are given by 
\begin{eqnarray}
& &n_p=n_e=4 \times 10^{17} [{\text m}^{-3}], \\
& &n_{\gamma}=10^9 \times n_p
\end{eqnarray}
Scattering cross section of Thomson scattering and Rutherford scattering are given by
\begin{eqnarray}
& &\sigma_{Ru}= 4 \pi\left(\frac{e^2}{4\pi\epsilon_0 mv^2}\right)^2 { \log \Lambda}=4.4 \times 10^{-17}[{\text m}^2]  \log \Lambda, \nonumber \\
& &\sigma_{Th}=\frac{8\pi r_e^2}{3}=0.6 \times 10^{-28} [{\text m}^2]
\end{eqnarray} 
where the Coulomb screening  factor, and the Thomson cross section are  
\begin{eqnarray}
& &\Lambda=\frac{\gamma \hbar v 4 \pi \epsilon_0}{e^2}, mv^2=kT, \nonumber \\
& &r_e=\frac{e^2}{4 \pi \epsilon_0 m_e c^2}.
\end{eqnarray}
Using a standard value  $\log \Lambda=10$, at $T=3000[{\text K}]$, the mean free path of electron due to Rutherford scattering and Thomson scattering are 
\begin{eqnarray}
& &l_{e,Ru}=5.7 \times 10^{-3} [{\text m}], \label{electron_mfp} \nonumber \\
& &l_{e,TH}= 5 [{\text m}].
\end{eqnarray}
The photon mean free path is given by
\begin{eqnarray}
l_{photon,Th}= 5 \times 10^9 [{\text m}].   
\end{eqnarray}
This length is equivalent to the distance of the Sun. Photon has an effective mass expressed by  plasma frequency
\begin{eqnarray}
& &\omega_P=\left(\frac{n e^2}{ \epsilon_0 m}\right)^{1/2}
\end{eqnarray}
in the presence of free electrons, and the frequency satisfies
\begin{eqnarray}
& &\omega^2=\omega_P^2+k^2c^2.
\end{eqnarray}
Photon of a frequency less than $\omega_P$ does not propagate. 

Free charged particle screen the electric field, by Debye screening. The screening length  is given by
\begin{eqnarray}
 \lambda_D=\left(\frac{\epsilon_0 kT}{ne^2}\right)^{1/2}   
\end{eqnarray} 

(II) Below the critical temperature: 

In a temperature below critical temperature, electrons and protons form hydrogen atoms. A photon interacts with a hydrogen with Rayleigh scattering, which is proportional to the fourth power in energy $E^4$. The cross section and the mean free path are
\begin{eqnarray}
& & \sigma_\text{Rayleigh,hydrogen}=\sigma_T\left(\frac{1215}{3000}\right)^4=0.027 \sigma_T(1215)  \nonumber \\
& & l_\text{photon,Rayleigh}=0.154 \times 10^{13} [\text{m}]
\end{eqnarray}
where $\sigma_T(1215)$ is the value of the wave length  $1215$ nano-meter.

Because free electrons are absent, a photon remains massless. The wave packets of photons of the size $l_{photon,Rayleigh}$ propagate. Mean free path of hydrogen atom is estimated with their scattering cross section    
\begin{eqnarray}
l_H=\frac{1}{\sigma_H n} = 2.5 \times 10^2 \text[\rm m],  
\end{eqnarray}
where $\sigma_H=10^{-20}$ $ [{\text m}^2]$ is substituted. The mean free path of hydrogen is much longer than the mean free path of electron Eq.$(\ref{electron_mfp})$.

%%%%%%%%%%%%%%%%%%%%%%%%%%%%%%%%%%%%%%%%%%%%%%%%% sub subsection %%%%%%%%%%%%%%%%%%%%%%%%%%%%%%%%%%%%%%%%%%%%%%%%
%%%%%%%%%%%%%%%%%%%%%%%%%%%%%%%%%%%%%%%%%%%%%%%%%%%
\subsubsection{Gravitational wave  }
%%%%%%%%%%%%%%%%%%%%%%%%%%%%%%%%%%%%%%%%%%%%%%%%%%%%%
Gravitational waves are described by wave packets in quantum mechanics. The wave packet of graviton produced in transitions of states are the same as the photon wave packets. The wave packet spreads in the perpendicular directions but does not in the parallel direction. When the wave is observed by an interaction with a detector, its frequency may shift from $E_0$ in $(\ref{wave_expand_x} )$. It turns the same after the integration over the position. The energy is determined by the frequency. 

As the gravitational interaction is extremely weak, an interaction of each wave packet with matter may be negligible.   The gravitational waves may be produced by transitions of macroscopic states such as a star and a galaxy. These waves are expressed by wave packets of the coherent state, and may reveal unique properties.  

%%%%%%%%%%%%%%%%%%%%%%%%%%%%%%%%%%%%%%%%%%%%%%%%%%%%%%%%%%%%%%%%%%%%%%%%%%%%%%%%%%%%%%%%%%%%%%%%%%%%%%%%%%%%%%%%%%%

%%%%%%%%%%%%%%%%%%%%%%%%%%%%%%%%%%%%%%%%%%%%%%%%%%%%%%%%%%%%%%%%%%%%%%%%%%%%%%%%%%%%%%%%%%%%%%%%%%%%%%%%%%%%%%%%%%%%5
\section{Summary and implications}
%%%%%%%%%%%%%%%%%%%%%%%%%%%%%%%%%%%%%%%%%%%%%%%%%%%%%%%%%%%%%%%%%%%%%%%%%%%%%%%%%%%%%%%%%%%%%%%%%%%%%%%%%%%%%%%%%
Unique  roles and detailed  properties of the wave packets were uncovered in the formalism based on the four principles.  The wave packets  represent the physical states of the probability of exisitence unity, i.e., particles.    These make   transitions  acccording to the probability  provided from the scalar product of states, which  resolves difficulties of  plane-wave formalism. 

 Gaussian wave packets characterized uniquely by sizes  represent all the specific features of the wave packets, and were  applied.  The size  is governed by the environment, and    plays substantial roles.

Wave packets express either bound states  or waves  propagating  in vacuum or in a dilute system with finite coherence lengths, and make transitions following the transition probabilities that depend on their sizes. The initial and final states are prepared independently at different spatial positions and their sizes are determined by their processes in the experimental apparatus prior to or after the transition. Accordingly sizes of the initial states are different from those of the final states. 
The sizes  of bound states are governed by microscopic constants   and those of propagating waves are by  coherence lengths, whereas the size in the latter cases varies widely. Because the sizes are not included in the Lagrangian, these are not fixed but  inevitable for  the transition probabilities and for the comparisons of  theory with experiments. These imply that the  correct understanding of transitions in natural phenomena are made with the wave-packet formalism.  

%Summary  

1. A wave packet is formed by its reactions  with matter in the environment, and propagates in dilute systems or the vacuum afterwards. As the fundamental interactions are local in space-time coordinates and preserve the causality and Lorentz invariance, wave packets have universal properties. Wave packets are classified into two categories depending upon their spectra. In the first category of discrete spectra, bound states determine sizes and are microscopic lengths expressed by fundamental parameters. In the second category of continuum spectrum, coherence lengths governed by environment determine the sizes and vary in a wide range. They  become extremely large in low density.  A number density of atoms is around $6 \times 10^{23}$ per $cm^3$ in the solid sate.  In the solar corona, for instance,  the value is $10^{10} $, and    the mean free path is larger by $10^{13}$ times of the former system. Consequently, the sizes of wave packets in the latter system is $10^{13}$ times larger.  
 
2. The size of wave packet in the incoming states depends on their interaction with matter in experimental setups.  For those  of  charged particles, electromagnetic interactions give dominant effects. The particle  loses energy during propagation in matter. The size of the proton wave packet in a beam is determined by  Eqs.$(\ref{mfp-proton-i1}) $,  $( \ref{mfp-proton-i2}))$, or $(\ref{proton_C})$. If a proton is extracted from matter by weak electric field, the size is small.  For higher field, the size is larger. The sizes of other charged particles are almost the same as the proton. The neutrino interaction is short-range and extremely weak. Its mean free path and the size of wave packet are large.

3. The size of wave packet of the outgoing state depends on its reactions  with matter in the detector.  For a charged particle, electromagnetic interactions give dominant effects. From a rate of losing  energies in the detector, size of a proton wave packet is determined by Eq.$(\ref{mfp_finalp})$, or larger sizes. The sizes of other charged particles are almost the same as the proton. For particles that is indirectly identified by the detectors, the size is determined by the wave-packet size of the particle in the detector from Subsection 4.A. The neutrino is such particle that is detected indirectly, and   probabilities to specific final states are measured by considering their reactions  with matter. The size of wave packet of the neutrino is determined by the size of the target wave functions in majority of detectors.  Accordingly, the size  is either nuclear size or atom size, which  are much smaller than that of the initial neutrino.  
Effects by wave packets appear in wide area of transitions in microscopic and macroscopic systems. Their magnitude becomes prominent in large wave packets. Light scatterings exhibit specific transitions of wave packets in wide area.  

4. The wave-packet size  in natural processes varies depending on the situations from microscopic  to macrosopic lengths deriving both processes of $P_I$ and $P_{II}$.  These are crucial for correct understandings of wide quantum  phenomena.      
%The absolute  probability is necessary for the correct understanding of physical  phenomena.    

  %%%%%%%%%%%%%%%%%%%%%%%%%%%%%%%%%%%%%%%%%%%%%%%%%%%%%%%%%%%%%%%%%%%%%%%%%%%
%\newpage
\section*{Acknowledgements}
One of the authors (K.I) thanks Drs.T. Shimomura, K.Nishiwaki,  K. Oda, M.Takesada, and H.Sakurai for useful discussions. 
\\
\section*{Declarations statesments}
Funding: 
This work was partially supported
by a Grant-in-Aid for Scientific Research(Grant No. 24340043 and No.26K00623 ) provided
by the Ministry of Education,
Science, Sports and Culture, Japan. \\
Availability of data and material: 
No data was used for the research described in the article.\\
Competing interests: 
Not applicacable\\
Author contributions: \\
K.Ishikawa: idea , calculations, and writing \\
O.Jinnouchi: idea and figure drawing, and writing.
%\section*{Appendix: Matrix elements of wave packets}

\section*{Appendix}
\subsection*{ Matrix elements of wave packets}

\subsubsection{  Integrals of the amplitudes over time }
An integral over a  time  for the plane  wave,  
\begin{eqnarray}
& &lim_{\epsilon \rightarrow 0} \int_{-\infty}^{\infty} dt e^{i (\Delta E t +i\epsilon |t|)  })=lim_{\epsilon \rightarrow 0}\frac{2 \epsilon}{(\Delta E)^2+\epsilon^2} \nonumber \\
& &= (2 \pi) \delta(\Delta E )  \label{time_integral_2}
\end{eqnarray}
is replaced to  an integral over a time  
 \begin{eqnarray}
   & &\int_0^T d t e^{i(\Delta E) t-\frac{(t-T_{int})^2}{2\sigma_t}} \nonumber \\
   & &=  \pi \sqrt{\sigma_t} {e^{-\frac{\sigma_t (\Delta E )^2}{2}} }  [sign(T_{int} )-sign(T_{int}-T)] \nonumber \\
   & & +\frac{1}{2}\frac{\sigma_t}{T-T_{int}+i(\Delta E ) \sigma_t}e^{-\frac{(T-T_{int})^2}{2\sigma_t}-i( \Delta E)(T-T_{int})} \nonumber  \\
& &- \frac{1}{2}\frac{\sigma_t}{-T_{int}+i(\Delta E ) \sigma_t}e^{-\frac{(-T_{int})^2}{2\sigma_t}-i(\Delta  E )(-T_{int})} \label{time_integral_g},
 \end{eqnarray}
for wave packet  at  a large $T$,  where $T_{int}$ is an intersection  time of the initial and final states and $\Delta E$ is an energy difference  modified from naive values.  

Squares of Eq.$(\ref{time_integral_2})$ and  Eq.$(\ref{time_integral_g})$ are proportional constants of the probabilities. The former is ill-defined but approximated with the time interval $T$ in the standard method, whereas the latter is  well-defined function of $T$.

\subsubsection{Free waves in a detector}
Matrix elements of wave packets $|\vec P_1,\vec X_1,\sigma_1 \rangle$, $|\vec P_2,\vec X_2,\sigma_2 \rangle$ are given by 
\begin{eqnarray}
M&=&\int d{\vec x} e^{ -i\sum_l \vec p_l \vec x +i \sum_l E_l t}    \langle P_2,X_2,\sigma_2|t,\vec x \rangle \langle t,\vec x| P_1,X_1, \sigma_1 \rangle \nonumber \\
 &=&(\frac{4\sigma_1\sigma_2}{(\sigma_1+\sigma_2)^2})^{-3/4} e^{-\frac{(\Delta X(t))^2}{2(\sigma_1+\sigma_2)}-\frac{\sigma_s}{2}(\delta p)^2 +i \phi_0},\label{two_wave_packets}
\end{eqnarray}
where $\sigma_s=\frac{\sigma_1 \sigma_2}{\sigma_1+\sigma_2}$,
\begin{eqnarray}
%& &\sigma_s=\frac{\sigma_1 \sigma_2}{\sigma_1+\sigma_2}, \\
& &\phi_0=\vec p_2 \vec X_2-E_2T_2-\vec P_1 \vec X_1+E_1T_1 \nonumber \\
& &+\sigma_s(\frac{\vec X_1(t)}{\sigma_1}+\frac{\vec X_2(t)}{\sigma_2})(\vec P_1-\vec P_2) \nonumber \\
& &-\sum_l (\vec p_l \vec X_{tot}-E_l t),\\
& &\delta X(t)= \vec X_1-\vec v_1(T_1-t)-\vec X_2+\vec v_2 (T_2-t), \nonumber\\
& &\delta P= \vec P_1-\vec P_2 +\sum_l {\vec p}_l.
\end{eqnarray}

   $|M|=1$ at  $\delta \vec X(t)=0, \delta \vec p_l=0$
 for
$ \sigma_1=\sigma_2$

 $|M| <  1 $ for $\sigma_1 \neq \sigma_2$,
%%%%%%%%%%%%%%%%%%%%%%%%%%%%%%%%%%%%%%%%%%%%%%%%%%%%%%%%%%%%%%	
\subsubsection{Overlap of wave packets in uniform force }
%%%%%%%%%%%%%%%%%%%%%%%%%%%%%%%%%%%%%%%%%%%%%%%%%%%%%%%%%%%
 Wave packets in a uniform force (weak)  by Airy function.
\begin{eqnarray}
& &H \phi_E(z)=[\frac{p_z^2}{2m}+mgz] \phi_E(z) \\
& &\phi_E(z)= A_E(\xi-\xi_1), \xi=\frac{z}{c},~ \xi_1=\frac{E}{cmg}, c=(\frac{1}{2m^2g})^{1/3} \nonumber \\
& &A_E(\xi)=\frac{1}{\pi} \int_0^{\infty}du \cos(\frac{u^3}{3}+\xi u)
\end{eqnarray}
Scalar  product of wave packets. 

(I) Wave packets of the same positions.
\begin{eqnarray}
& &\int_{-\infty}^{\infty} dz \phi_E^{*}(z) \phi_{E'}(z) = \int_{-\infty}^{\infty} c d \xi  A_E^{*}(\xi-\xi_1) A_{E'}(\xi-\xi_1')  \nonumber \\
%& &=c \int_{-\infty}^{\infty}  d \xi \int_0^{\infty}  \frac{1}{\pi} \int_0^{\infty} du \cos(\frac{u^3}{3}+ (\xi-\xi_1) u)     \frac{1}{\pi} \int_0^{\infty} dv \cos(\frac{v^3}{3}+ (\xi-\xi_1') v)  \nonumber \\
%& &=c \frac{1}{\pi^2}\int_{-\infty}^{\infty} d \xi\int_0^{\infty}  \frac{du dv}{2}[\cos( \frac{u^3+v^3}{3}+(\xi-\xi_1) u \nonumber \\
%& &+(\xi-\xi_1') v) +\cos( \frac{u^3-v^3}{3}+(\xi-\xi_1) u-(\xi-\xi_1') v)]\nonumber \\
%& &=c \frac{1}{\pi^2} \int_0^{\infty}  \frac{du dv}{2}[\cos( \frac{u^3+v^3}{3}+(-\xi_1) u+(-\xi_1') v)(2 \pi) \delta(u+v) \nonumber \\
%& &+\cos( \frac{u^3-v^3}{3}+(-\xi_1) u-(-\xi_1') v) (2\pi) \delta(u-v)] \nonumber\\
& &=c^2 mg \delta(E'-E).
\end{eqnarray}
%\newpage
(II) Wave packets of different positions.
\begin{eqnarray}
& &\int_{-\infty}^{\infty} dz \phi_E^{*}(z-Z_1) \phi_{E'}(z-Z_2) \nonumber \\
%= \int_{-\infty}^{\infty} c d \xi  A_E^{*}(\xi-\tilde \xi_1) A_{E'}(\xi-\tilde \xi_1')  \nonumber \\
& &={c}  \delta(\tilde \xi_1'-\tilde \xi_1)=c\delta(\frac{E'-E}{cmg} +\frac{Z_2-Z_1}{c}),
\end{eqnarray}
\begin{eqnarray}
\tilde \xi_1=\frac{E}{cmg} +\frac{Z_1}{c},~\tilde \xi_1'=\frac{E'}{cmg} +\frac{Z_2}{c}.
\end{eqnarray}
Wave packet is superposioin of differentr energies,
\begin{eqnarray}
& &| E_1, Z_1,T_1;\sigma \rangle=N_1 \int d E e^{ iET_1} e^{-\frac{\sigma}{2} (E-E_1)^2} \phi_E(z-Z_1),  \nonumber \\
\end{eqnarray}
where $N_1$ is obtained later. Their scalar product is
\begin{eqnarray}
& & \langle E_1,Z_1,T_1 ; \sigma_1| E_2,Z_2,T_2;\sigma_2 \rangle \nonumber \\
%& &={N_1N_2}\int dz\int d E e^{ -iE T_1} e^{-\frac{\sigma_1}{2} (E-E_1)^2} \phi_E(z-Z_1)^{*}  \nonumber \\
%& & \times \int d E'e^{iE' T_2} e^{-\frac{\sigma_2}{2} (E'-E_2)^2} \phi_{E'}(z-Z_2) \nonumber \\
%& &={N_1N_2}\int d E e^{ -iE T_1} e^{-\frac{\sigma_1}{2} (E-E_1)^2}  \int d E'e^{iE' T_2} e^{-\frac{\sigma_2}{2} (E'-E_2)^2} c \delta( [ \frac{E'-E}{cmg} +\frac{Z_2-Z_1}{c}  ]) \nonumber \\
%& &={N_1N_2}\int d E e^{ -iE T_1} e^{-\frac{\sigma_1}{2} (E-E_1)^2}  e^{i(E+\frac{\Delta Z}{mg}) T_2} e^{-\frac{\sigma_2}{2} (E+\frac{\Delta Z}{mb}-E_2)^2} c^2 mg  \nonumber \\
%& &={N_1N_2}c^2mg \int d E e^{ -iE T_1-\frac{\sigma_1}{2} (E-E_1)^2 + i(E+\frac{\Delta Z}{mg}) T_2 -\frac{\sigma_2}{2} (E+\frac{\Delta Z}{mb}-E_2)^2} \nonumber \\ 
& &={N_1N_2}c^2 mg \int d E e^{f(E)}={N_1N_2}c^2 mg  e^{f(E_0)} \sqrt{\frac{2 \pi}{\sigma_1+\sigma_2}},\nonumber \\
\end{eqnarray}
where $ \Delta Z=Z_1-Z_2 $ and 
\begin{eqnarray}
& &f(E)=-iE T_1-\frac{\sigma_1}{2} (E-E_1)^2 + i(E+\frac{\Delta Z}{mg}) T_2 \nonumber \\
& &-\frac{\sigma_2}{2} (E+\frac{\Delta Z}{mb}-E_2)^2  
\end{eqnarray} 

Normalization is  given by $E_1=E_2,Z_1=Z_2, \sigma_1=\sigma_2$
\begin{eqnarray}
& &\langle E_1,Z_1,T_1 ; \sigma_1| E_1,Z_1,T_1;\sigma_1 \rangle  =N_1^2c^2mg \sqrt{\frac{\pi}{\sigma_1}}=1 \nonumber \\
& &N_1=( c^2mg \sqrt{\frac{\pi}{\sigma_1}})^{-1/2}
\end{eqnarray}

%\newpage
Apply statioinary phase method and obtain stationary energy $E_0$, it follows that  
\begin{eqnarray}
& &f(E)=-iE T_1-\frac{\sigma_1}{2} (E-E_1)^2 + i(E+\frac{\Delta Z}{mg}) T_2 \nonumber \\
& &-\frac{\sigma_2}{2} (E+\frac{\Delta Z}{mb}-E_2)^2,  \\
& &f'(E_0)=0,  \nonumber \\
& &E_0=\frac{-i (T_1-T_2)+{\sigma_1} E_1 +{\sigma_2} (-\frac{\Delta Z}{mb}+E_2) }{\sigma_1+\sigma_2} 
%& &E_0-E_1=\frac{-i (T_1-T_2)-{\sigma_2} (E_1-E_2+\frac{\Delta Z}{mb})}{\sigma_1+\sigma_2}   \nonumber \\
%& &E_0-E_2+\frac{\Delta Z}{mb} =\frac{-i (T_1-T_2)  +{\sigma_1} (\frac{\Delta Z}{mb}-E_2+E_1) }{\sigma_1+\sigma_2} 
\end{eqnarray}
and
\begin{eqnarray}
& &f(E_0)
%=-iE_0 T_1-\frac{\sigma_1}{2} (E_0-E_1)^2 + i(E_0+\frac{\Delta Z}{mg}) T_2 \nonumber \\
%& &-\frac{\sigma_2}{2} (E_0+\frac{\Delta Z}{mb}-E_2)^2  \nonumber \\
%& &=i\frac{\Delta z}{mg}T_2 -i(T_1-T_2)\frac{-i (T_1-T_2)+{\sigma_1} E_1 +{\sigma_2} (-\frac{\Delta Z}{mb}+E_2) }{\sigma_1+\sigma_2} \nonumber \\
%& &-\frac{\sigma_1}{2} (\frac{-i (T_1-T_2)-{\sigma_2} (E_1-E_2+\frac{\Delta Z}{mb})}{\sigma_1+\sigma_2})^2 \nonumber \\
%& &-\frac{\sigma_2}{2}( \frac{-i (T_1-T_2)  +{\sigma_1} (\frac{\Delta Z}{mb}-E_2+E_1) }{\sigma_1+\sigma_2} )^2 \nonumber \\
=i[ f_{\phi} ] +f_R,  
\end{eqnarray}
where $f_{\phi}$ is a constant and
\begin{eqnarray}
%& &f_R=-\frac{(T_1-T_2)^2}{\sigma_1+\sigma_2}-\frac{\sigma_1}{2(\sigma_1+\sigma_2)^2}[-(T_1-T_2)^2+\sigma_2^2(E_1-E_2+\frac{\Delta Z}{mb})^2  ]\nonumber \\
%& &-\frac{\sigma_2}{2(\sigma_1+\sigma_2)^2}[-(T_1-T_2)^2+\sigma_1^2(E_1-E_2+\frac{\Delta Z}{mb})^2  ]\nonumber \\
& &f_R=-\frac{(T_1-T_2)^2}{2(\sigma_1+\sigma_2)}-\frac{\sigma_1\sigma_2}{2(\sigma_1+\sigma_2)}(E_1-E_2+\frac{\Delta Z}{mb})^2. \nonumber \\
 \end{eqnarray}
Overlap of the wave packets and its square  is given by
\begin{eqnarray}
& & \langle E_1,Z_1,T_1 ; \sigma_1| E_2,Z_2,T_2;\sigma_2 \rangle =e^{i f_{\phi}} \sqrt{ \frac{2\sqrt{ \sigma_1 \sigma_2}}{\sigma_1 +\sigma_2}} \nonumber \\
& &\times e^{ -\frac{(T_1-T_2)^2}{2(\sigma_1+\sigma_2)}-\frac{\sigma_1\sigma_2}{2(\sigma_1+\sigma_2)}(E_1-E_2+\frac{\Delta Z}{mb})^2 } \nonumber \\
& &  \\
& & |\langle E_1,Z_1,T_1 ; \sigma_1| E_2,Z_2,T_2;\sigma_2 \rangle|^2 \nonumber \\
& & =( {\frac{2\sqrt{ \sigma_1\sigma_2}}{\sigma_1+\sigma_2}} )   e^{ -\frac{(T_1-T_2)^2}{(\sigma_1+\sigma_2)}-\frac{\sigma_1\sigma_2}{(\sigma_1+\sigma_2)}(E_1-E_2+\frac{\Delta Z}{mb})^2 }. \nonumber \\
\end{eqnarray}
%Decoherence Properties

%Fixed  $\sigma$,
%\begin{eqnarray}
%& & |\langle E_1,Z_1,T_1 ; \sigma_1| E_2,Z_2,T_2;\sigma_2 \rangle|^2 \rightarrow 0; |T_1-T_2| \rightarrow \infty , |E_1-E_2+\frac{\Delta Z}
%{mb}| \rightarrow \infty
%\end{eqnarray}

%%%%%%%%%%%%%%%%%%%%%%%%%%%%%%%%%%%%%%%%%%%%%%%%
\subsubsection{Uniform magnetic field}
%%%%%%%%%%%%%%%%%%%%%%%%%%%%%%%%%%%%%%%%%%%%%%
 Eigenstate of Hamiltonian of a charge particle  in a uniform  magnetic field is  Landau level $\phi_l(k_x,y)$ 
\begin{eqnarray}
& &H=\frac{(p_x-eBy)^2+p_y^2}{2m}  \nonumber \\
& &H \psi(x,y)=E \psi, \psi=e^{ik_x x}\phi_l(k_x,y),E_l=  \omega(l+1/2) \nonumber \\
\end{eqnarray}
(1) Superposition of the continious $ k_x$ is given by 
\begin{eqnarray}
& &|E_l,P_x,X,T_1 \rangle=\int d k_x e^{iE_lT_1-\frac{\sigma}{2}(k_x-P_x)^2+ik_x(x-X)}\phi_l(k_x,y) \nonumber \\
\end{eqnarray}
and their overlap is
\begin{eqnarray}
& &\langle E_{l^1} ,P_x^1,X^1,T_1 | E_{l^2},P_x^2,X^2,T_2 \rangle 
%& &=\int dx dy d k_x^1 e^{-iE_{l^1}T_1-\frac{\sigma_1}{2}(k_x^1-P_x^1)^2-ik_x^1(x-X^1)}\phi_{l^1}^{*}(k_x^1,y)  \int d k_x^2 e^{iE_{l^2}T_2-%\frac{\sigma_2}{2}(k_x^2-P_x^2)^2+ik_x^2(x-X^2)}\phi_{l^2}(k_x^2,y) \nonumber \\
%& &=\int  dy d k_x^1 e^{-iE_{l^1}T_1-\frac{\sigma_1}{2}(k_x^1-P_x^1)^2-ik_x^1(-X^1)}\phi_{l^1}^{*}(k_x^1,y)  \int d k_x^2 e^{iE_{l^2}T_2-%\frac{\sigma_2}{2}(k_x^2-P_x^2)^2+ik_x^2(-X^2)}\phi_{l^2}(k_x^2,y) (2 \pi) \delta(k_x^1-k_x^2) \nonumber \\
%& &=\int  dy d k_x^1 e^{-iE_{l^1}T_1-\frac{\sigma_1}{2}(k_x^1-P_x^1)^2-ik_x^1(-X^1)}\phi_{l^1}^{*}(k_x^1,y)   e^{iE_{l^2}T_2-\frac{\sigma_2}{2}(k_x^1-P_x^2)^2+ik_x^1(-X^2)}\phi_{l^2}(k_x^1,y) (2 \pi) \nonumber \\
%& &=\int   d k_x^1 e^{-iE_{l^1}T_1-\frac{\sigma_1}{2}(k_x^1-P_x^1)^2-ik_x^1(-X^1)}   e^{iE_{l^2}T_2-\frac{\sigma_2}{2}(k_x^1-P_x^2)^2+ik_x^1(-X^2)}  (2 \pi) \delta_{l_1,l_2}\nonumber \\
%& &= (2 \pi) \delta_{l_1,l_2}e^{-iE_{l^1}(T_1-T_2)}\int   d k_x^1 e^{-\frac{\sigma_1}{2}(k_x^1-P_x^1)^2+ik_x^1(X^1-X_2)}   e^{-\frac{\sigma_2}{2}(k_x^1-P_x^2)^2}  \nonumber \\
= (2 \pi) \delta_{l_1,l_2}e^{-iE_{l^1}(T_1-T_2)} \nonumber \\
& & \times e^{i \phi_0}  e^{ -\frac{1}{2(\sigma_1+\sigma_2)}(X_1-X_2)^2-\frac{\sigma_1 \sigma_2}{2(\sigma_1+\sigma_2)}(P_x^1-P_x^2)^2},\nonumber \\
\end{eqnarray} 
 where $\phi_0$ is a constant. The overlap  is uniform in $T_1- T_2$ and decreases with differences $| X_1-X_2|$ and $|P_x^1-P_x^2|$.  
 \\
 \\
(II) Superposition of the continious $E$ is given by 
\begin{eqnarray}
& &|E_l,k_x,X,T_1 \rangle=\int d E e^{iE T_1-\frac{\sigma}{2}(E-E_0)^2+ik_x(x-X)}\phi_l (k_x,y)  \nonumber \\
\end{eqnarray}
and their overlap is
\begin{eqnarray}
& &\langle E_{l^1} ,k_x^1,X^1,T_1 | E_{l^2},k_x^2,X^2,T_2 \rangle  \nonumber \\
%& &=\int dx dy d  E_1 e^{-iE_1T_1-\frac{\sigma_1}{2}(E_1-E_0^1)^2-ik_x^1(x-X^1)}\phi_{l^1}^{*}(k_x^1,y)  \int d E_2 e^{iE_{l^2}T_2-%\frac{\sigma_2}{2}(E_2-E_0^2)^2+ik_x^2(x-X^2)}\phi_{l^2}(k_x^2,y) \nonumber \\
%& &=\int  dy d E_1 e^{-iE_{l^1}T_1-\frac{\sigma_1}{2}(E_1-E_0^1)^2-ik_x^1(-X^1)}\phi_{l^1}^{*}(k_x^1,y)  \int d E_2 e^{iE_{l^2}T_2-%\frac{\sigma_2}{2}(k_x^2-P_x^2)^2+ik_x^2(-X^2)}\phi_{l^2}(k_x^2,y) (2 \pi) \delta(k_x^1-k_x^2) \nonumber \\
%& &=\int  dy d E_1 dE_2 e^{-iE_{l^1}T_1-\frac{\sigma_1}{2}(E_1-E_0^1)^2-ik_x^1(-X^1)}\phi_{l^1}^{*}(k_x^1,y)   e^{iE_{l^2}T_2-\frac{\sigma_2}{2}%(E_2-E_2^0)^2+ik_x^1(-X^2)}\phi_{l^2}(k_x^1,y) (2 \pi) \delta(k_x^1-k_x^2) \nonumber \\
%& &=e^{ik_x^1(X_1-X_2)}\int   d E_1 d E_2 e^{-iE_{l^1}T_1-\frac{\sigma_1}{2}(E_1-E_0^1)^2   +iE_{l^2}T_2-\frac{\sigma_2}{2}(E_2-E_0^2)^2}  (2 \pi) \delta_{l_1,l_2}\delta(k_x^1-k_x^2) \nonumber \\
%& &= N (2 \pi) e^{ik_x^1(X_1-X_2)}\delta(k_x^1-k_x^2) \int   d E^1  e^{-iE_{l^1}(T_1-T_2)-\frac{\sigma_1}{2}(E_1-E_0^1)^2 -\frac{\sigma_2}{2}(E_1-E_0^2)^2}    \nonumber \\
& &= N (2 \pi) e^{ik_x^1(X_1-X_2)} e^{i \phi_0}\delta(k_x^1-k_x^2) \sqrt{ \frac{2 \pi}{\sigma_1+\sigma_2}} \nonumber \\
& &e^{- \frac{1}{\sigma_1+\sigma_2}(T_1-T_2)^2-\frac{\sigma_1\sigma_2}{\sigma_1+\sigma_2}(E_1-E_2)^2 },    
\end{eqnarray}
 where $f_{\phi_0}$ is a constant. The overlap  is uniform in $X_1-X_2$ and decreases with  differences $|T_1-T| $ and $|E_1-E_2|$.
%(III)Wave packet 3 (Symmetric gauge)
%\begin{eqnarray}
%& &|E_l,k_x,X,T_1 \rangle=\int d E e^{iE T_1-\frac{\sigma}{2}(E-E_0)^2+ik_x(x-X)}\phi_l (k_x,y) \nonumber \\
%& &\langle E_{l^1} ,k_x^1,X^1,T_1 | E_{l^2},k_x^2,X^2,T_2 \rangle   =\int dx dy d E_1 e^{-iE_1 T_1-\frac{\sigma_1}{2}(E_1-E_0^1)^2-ik_x^1(x-%X^1)}\phi_{l^1}^{*}(k_x^1,y)\nonumber  \\
%& &\times  \int d E_2 e^{iE_{2}T_2-\frac{\sigma_2}{2}(E_2-E_0^2)^2+ik_x^2(x-X^2)}\phi_{l^2}(k_x^2,y) \nonumber \\
%& &= N (2 \pi) \delta_{l_1,l_2}e^{ik_x^1(X_1-X_2)} \sqrt{ \frac{2 \pi}{\sigma_1+\sigma_2}} e^{- \frac{1}{\sigma_1+\sigma_2}(T_1-T_2)^2-%\frac{\sigma_1\sigma_2}{\sigma_1+\sigma_2}(E_1-E_2)^2 +i \phi }    \nonumber \\
%\end{eqnarray}
%The overlap decreases with  time  and energy differences. 
%%%%%%%%%%%%%%%%%%%%%%%%%%%%%%%%%%%%%%%%%%%%%%%%%%%%%%%%%%%%%%%
 \subsection*{ Wave packets in standard of physical quantities  } 
 %%%%%%%%%%%%%%%%%%%%%%%%%%%%%%%%%%%%%%%%%%%%%%%%%%%%%%%%%%%%%
(1) Wave packet of atoms.

(1-1) Light emitted from the transition of atoms of the energy difference $\delta E=E_A-E_{A'}$
\begin{eqnarray}
A \rightarrow A'+ \gamma
\end{eqnarray}
is used as a frequency standard. Relation of energies  
\begin{eqnarray}
{\delta E}= h \nu
\end{eqnarray}
is satisfied with  plane waves. Realistic experiments use wave packets  and the above relation may be modified. The magnitudes may depend on the wave packet sizes of $A$,  $A'$, and $\gamma$ .

(1-2)   The wave packet size of atom in solid may depend on many factors.  This  issue is  important for an absolute value of transition probability  in   precision experiments.

(2) Electron wave packet

In  precision tests of QED, such as lepton  $g-2$ in Penning trap, and other situations,   an electron is described by a wave packet. 
In Penning trap, one electron is confined in a space confined by  a magnetic and electric fields.  The size is determined by electric and magnetic fields. Owing to an electric charge of the electron, the finite electric field is present in a long distance, which satisfies boundary conditions. It is highly nontrivial to solve the electron  in a long range force.

(3) Laser wave packet

Laser plays special  roles in modern technology. What is  the wave packet size ?  The x-ray laser, and mirror vs  wave packets may be important.

%\newpage


\section*{References}
\begin{thebibliography}{99}
\bibitem{schroedinger} Von E.~Schr\"{o}dinger, Die Naturwissensehaften 14, 664(1926).
\bibitem{heisenberg} Von W. Heisenberg, Zeit. Physik 43, 172 (1927).
\bibitem{dirac}  P.~M.~Dirac, 
{\it Principle of quantum mechanics}~(Oxford University Press,  4-th ed. New York, 1988).
\bibitem{Reed-Simon} M.Reed and B.Simon,  {\it Methods of Modern Mathematical Physics, III:Scattering Theory}~(Academic Press,New York, 1979).
\bibitem{Kato} T. Kato,  {\it Perturbation Theory for Linear Operators}~(Springer-Verlag,Tokyo, 1980).
\bibitem{Feynman} R.~Feynman,Phy.Rev .76,749,(1949).

\bibitem{LSZ} H.~Lehman , K.~Symanzik, and W. ~Zimmerman, Nuovo Cimento,1,1425 (1960).

\bibitem{Goldberger}  M.~L.~Goldberger and Kenneth ~M.~Watson, 
{\it Collision Theory}~(John Wiley \& Sons, Inc. New York, 1965).

\bibitem{newton}  R.~G.~Newton, 
\textit{Scattering Theory of Waves and Particles}~(Springer-Verlag, New York, 1982).
\bibitem{taylor}  J.~R.~Taylor, 
\textit{Scattering Theory: The quantum theory of non-relativistic collisions}~(Dover-Publication, New York, 2006).
\bibitem{Sasakawa} T.~Sasakawa, Prog. Theor. Physics.  \textbf{ Suppl.11}, 69(1959).

%\bibitem{schwartz} 
%L.Schwartz,\textit{Theorie des distributions} (Hermann, paris, 1966); L. Schwartz, “Sur l'impossibilit\'e de la multiplications des distributions”, C.R.Acad. Sci. Paris 239: 847-848.
\bibitem{schwartz} 
L.Schwartz,\textit{Theorie des distributions} (Hermann, paris, 1966); L. Schwartz,   C.R.Acad. Sci. Paris 239: 847-848.
%\bibitem{dirac}  P.~M.~Dirac, 
%{\it Principle of quantum mechanics}~(Oxford University Press,  4-th ed. New York, 1988).
%\bibitem{EPR} A. Einstein, B. Podolsky, and N. Rosen,  Phy. Rev. 47, 777, (1935)
%\bibitem{Aspet}  A. Aspet, P.  Grangier, G. Roger,  Phy. Rev. Lett. 49, 91-4, (1982)  
\bibitem{Tonomura} A.Tonomura et al, Amer. J. Physics, 57(2) (1989) 117.
\bibitem{Tsuchiya} Y.Tsuchiya et al., Television J. 36(11),(1982) 1010.
%\bibitem{Ishikawa-Shimomura} K.~Ishikawa and T.Shimomura,
%	Prog. Theor. Physics.  \textbf{ 114}, (2005), 1201-1234.
%\bibitem{landau}  L. D. Landau, and E. M. Lifshitz , 
%{\it Nonrelativistic quantum mechanics}~(Butterwise and Heinemann, Oxford, 2003).

%\bibitem{Ishikawa-Tobita-ptp} K.~Ishikawa and Y.~Tobita ,`` Coherence 
%lengths of wave packets '' 	Prog. Theor. Physics.  \textbf{ 122}, November, (2009),1111-1136. 

%\bibitem{Ishikawa-Tobita} K.~Ishikawa and Y.~Tobita ,`` Coherence 
%length of cosmic background radiation enlarges the attenuation length of the
%ultra-high energy proton '' Hokkaido University preprint (2008); 
%`` Neutrino mass and mixing '' in the 10th Inter. Symp. on `` Origin of Matter
%and Evolution of Galaxies '' AIP Conf. proc. 1016, P.329(2008)

%\bibitem{Ishikawa-Shimomura} K.~Ishikawa and T.Shimomura,
%	Prog. Theor. Physics.  \textbf{ 114}, (2005), 1201-1234.
\bibitem{Ishikawa-Tobita-ptep} K.~Ishikawa and Y.~Tobita,
	Prog. Theor. Exp. Phys. 073B02, doi:10.1093/ptep/ptt049 (2013).
\bibitem{Ishikawa-Tobita-anp}  K.~Ishikawa and Y.~Tobita, Ann. of Phys. 344, 118(2014).doi:10.1016/j.aop.2014.02.007. 
\bibitem{Ishikawa-Tajima-Tobita-ptep} K.~Ishikawa,T.~Tajima,and Y.~Tobita,
	Prog. Theor. Exp. Phys. 013B02, doi:10.1093/ptep/ptu168 (2015).
\bibitem{Ishikawa-Jinnouchi} K.~Ishikawa, O.~Jinnouchi, A.~Kubota, T.~Sloan, T. H.~Tatsuishi, and R. ~Ushioda, `` On experimental confirmation of the correction to  Fermi's golden rule''
	Prog. Theor. Exp. Physics.  \textbf{2019}, 033B02,(2019).
\bibitem{Ushioda} R.~Ushioda, O.~Jinnouchi, K.~Ishikawa and T.~Sloan ,`` Search for the correction to Fermi's golden rule in positron annihilation '' 	Prog. Theor. Exp. Physics.  \textbf{2020}, April, (2020). 
\bibitem{Ishikawa-Nishiwaki-Jinnouchi-Oda} K.~Ishikawa, K.~Nishiwaki,  O.~Jinouchi, and K.~Oda, Eur. Jour. Phys. C83,978(2923)  doi.org/10.1140/epjc/s10052-023-12077-7
http://arxiv:2006.14159[hep-ph].
\bibitem{Maeda} N.~Maeda,T.~Yabuki,Y.~Tobita., and K.~.Ishikawa, Prog. Theor. Exp. Phys.2017,053J01, doi:10.1093/ptep/pty066 (2017).  
%\bibitem{Ishikawa-Oda}  K.~Ishikawa and K.~Oda 
%	Prog. Theor. Exp. Phys.123B01, doi:10.1093/ptep/pty127(2018).  arXiv:1809.04285[hep-ph],
%\bibitem{Ishikawa-Oda-Nishiwaki} K.~Ishikawa, K.~Nishiwaki,  and K.~Oda, Prog. Theor. Exp. Phys.2020,103 B04 doi:10.1093/ptep/ptta127 (2020)
%http://arxiv:2006.14159[hep-ph]
%\bibitem{Ishikawa-Nishiwaki-Jinnouchi-Oda} K.~Ishikawa, K.~Nishiwaki,  O.~Jinouchi, and K.~Oda, Eur. Jour. Phys. C83,978(2923)  doi.org/10.1140/epjc/s10052-023-12077-7
%http://arxiv:2006.14159[hep-ph]
%\bibitem{threshold} K.~Ishikawa, O.~Jinnochi, K.~Nishiwaki,  and K.~Oda, Euo.Phys. J.C 83(2023) 978  http://dx.do.org/10.1140/epj/s10052-0123-12077-7.
\bibitem{corona} K.~Ishikawa and Y.Tobita, ``Topological interaction of the neutrino with photon: Electro-weak Hall effect, Physics Open 17 (2023) 100174.
\bibitem{magnetization} K.Ishikawa,  ``Magnetization without  spin: Effective Lagrangin of itinerant electrons'',Nucl.Phys.B 1007(2024) 116663. 
\bibitem{sky} K.~Ishiakwa and M.Takesada, Hokkaido University preprint. 
\bibitem{Ishikawa-Shimomura} K.~Ishikawa and T.Shimomura,
	Prog. Theor. Physics.  \textbf{ 114}, (2005), 1201-1234.
\bibitem{Ishikawa-Tobita-ptp} K.~Ishikawa and Y.~Tobita ,`` Coherence 
lengths of wave packets '' 	Prog. Theor. Physics.  \textbf{ 122}, November, (2009),1111-1136. 
\bibitem{Ishikawa-Tobita} K.~Ishikawa and Y.~Tobita ,`` Coherence 
length of cosmic background radiation enlarges the attenuation length of the
ultra-high energy proton '' Hokkaido University preprint (2008); 
`` Neutrino mass and mixing '' in the 10th Inter. Symp. on `` Origin of Matter
and Evolution of Galaxies '' AIP Conf. proc. 1016, P.329(2008).
\bibitem{landau}  L. D. Landau, and E. M. Lifshitz , 
{\it Nonrelativistic quantum mechanics}~(Butterwise and Heinemann, Oxford, 2003).
%\bibitem{breene}
%	R.~G.~Breene,~Jr, Rev. Mod. Phys. \textbf{29},~94~(1957).
\bibitem{Ishikawa-Oda}  K.~Ishikawa and K.~Oda 
	Prog. Theor. Exp. Phys.123B01, doi:10.1093/ptep/pty127(2018).  arXiv:1809.04285[hep-ph].
\bibitem{Ishikawa-Oda-Nishiwaki} K.~Ishikawa, K.~Nishiwaki,  and K.~Oda, Prog. Theor. Exp. Phys.2020,103 B04 doi:10.1093/ptep/ptta127 (2020)
http://arxiv:2006.14159[hep-ph].
	\bibitem{ishikawa_1} K.~Ishikawa and Y.~Nishio, "Overlap integral of stationary continuum states " arXiv: [2305.16939 ]  Annals of Physics. \textbf{469}. October 2024, 169750.doi.org/10.1016/j.aap.2023.169750.
\bibitem{ishikawa_2}K.~Ishikawa, " Rigorous scattering probability of wave packets (2)"  arXiv: [2305. 16970]. Annals of Phys. \textbf{460}. January 2024, 169571.doi.org/10.1016/j.aap.2023.169571.
\bibitem{ishikawa_3}K.~Ishikawa, " Lecture on the quantum mechanics (I) (II)" Shokabou. Tokyo (2020). 
\bibitem{sdextension}D. M.Gitman,I.V.Tyutin, B.L.Voronov,``Self-adjoint Extensions in Quantum Mechanics, General Theory and Applications to Schrödinger and Dirac Equations with Singular Potentials'',(PMP,62) Birkhauser (2012)
\bibitem{sdextension_2} A.~Ibort, F.~Lledó, and J.~M~Pérez-P ,  On self-adjoint extensions and symmetries in quantum mechanics''  arXiv:1402.5537v3 [math-ph] 23 Sep 2014
\bibitem{particle_data_1} S.Navas et al.(Particle Data Group).Phys. Rev. D 110,030001 (2024) and 2025 update, Sec. 34.Passage of Particles through Matter, Fig.34.11.

\bibitem{breene}
	R.~G.~Breene,~Jr, Rev. Mod. Phys. \textbf{29},~94~(1957).	
%\bibitem{SK-Atom}
%J.~Hosaka, et~al,~Phys. Rev. Vol. D\textbf{74},~032002,~(2006).

%\bibitem{SK-Solar}
%The Super-Kamiokande Collaboration.~Phys.~Lett.~B\textbf{539},~179,~(2002).

%\bibitem{SNO-NC}
%S.~N. Ahmed, et~al.
%Phys. Rev. Lett.~\textbf{92}, 181301~(2004).


%\bibitem{KamLAND-Reactor}
%T.~Araki, et~al.
%Phys. Rev. Lett.~\textbf{94},~081801~(2005).

%\bibitem{Borexino}
%E.~A. Litvinovich.
%Phys. Atom. Nucl.~\textbf{72},~522--528~(2009).

%\bibitem{K2K}
%E.~Aliu, et~al.~Phys. Rev. Lett.~\textbf{94},~081802~(2005).

\bibitem{particle-data}
	C.~Amsler~et al.~[Particle Data Group],~Phys.~Lett.~B\textbf{667},~1~(2005).

\bibitem{Krauss} L.~Krauss,  F.Wilczek~Phys.~Rev.~Lett\textbf{55},~122(1985).
\bibitem{Loeb}A.Loeb,~Phys.~Rev.~D\textbf{39},~1009(1989). 
\bibitem{Kim_Versner} C.W.Kim, and A. Pevsner, Neutrino in Physics and Astrophysics, Harwood Academic Publishers, Chur.,1993(Ch.7 and 17). 
%\bibitem{Tritium}
%C.~Weinheimer, et~al.
%Phys. Lett.~B\textbf{460},~219--226~(1999).

%\bibitem{WMAP-neutrino}
%E.~Komatsu, et~al.
%Astrophys. J. Suppl.~\textbf{180},~330--376~(2009).

%\bibitem{KamLAND-Geo}
%T.~Araki, et~al.
%Nature,~\textbf{436},~499--503~(2005).


%\bibitem{Ishikawa-shimomura-lunar} K.~Ishikawa and T.~Shimomura ,`` Coherent lunar effect on solar neutrino '' Hokkaido University preprint (2005)

\bibitem{Kayser} B.~Kayser,~Phys.~Rev.~D\textbf{24},~110(1981);~Nucl.Phys.~B\textbf{19} ~(Proc.Suppl),~177(1991).
\bibitem{Stodolsky} L.~Stodolsky, Phys. Rev. \textbf{D58}, 036006(1998).

\bibitem{Giunti} C.~Giunti, C.~W.~Kim, and U.~W.~Lee, Phys. Rev. \textbf{D44}, 3635(1991).
\bibitem{Nussinov} S.~Nussinov, Phys. Lett. \textbf{B63}, 201(1976).

\bibitem{Kiers} K.~Kiers, N.~Nussinov and N.~Weisis,
	Phys. Rev. \textbf{D53}, 537(1996).

\bibitem{Anada_1} H.Anada and H.Nishimura,
Phys. Rev. \textbf{D37}, 552(1988).

\bibitem{Anada_2} H.Anada and H.Nishimura,
Phys. Rev. \textbf{D41}, 2379(1990).
\bibitem{Lipkin}~H.~J.~Lipkin,~Phys.~Lett.~B\textbf{642},~366(2006).
\bibitem{Akhmedov} E.K.Akhmedov and A.Y.Smirnov,Phys.Atom.Nucl. \textbf{72} (2009).
\bibitem{Giunti_2} C.~Giunti and  C.~W.~Kim, Phys. Rev. \textbf{D58}, 017301(1998).
\bibitem{Giunti_3} C.~Giunti, JHEP. \textbf{11}, 017(2002).
\bibitem{Beuthe} M.Beuthe, Phys.Rept. \textbf{375} (2003).
\bibitem{Jones} B.~J.~P.~Jones, Phys.Rev. \textbf{ D91},~053002 (2015).
%\bibitem{Ishikawa-Tobita-nu} K.~Ishikawa and Y.~Tobita, Ann. Phys. 344,118(2014)
\bibitem{Asahara} A.~Asahara, K.~Ishikawa, T.~Shimomura, and T.~Yabuki,
Prog. Theor. Phys. \textbf{113}, 385(2005); T.~Yabuki and K.~Ishikawa,
Prog. Theor. Phys. \textbf{108}, 347(2002).



%\bibitem{Wilson-OPE} K.~Wilson, in Proceedings of the Fifth International Symposium on Electron and Photon Interactions at High Energies, Ithaca, New York, 1971,~p.115~(1971).
% edited by N. B. Mistry (Cornell Univ. Press, Ithaca, New York, 1971 

%\bibitem{Ishikawa-Jinnouchi} K.~Ishikawa, O.~Jinnouchi, A.~Kubota, T.~Sloan, T. H.~Tatsuishi, and R. ~Ushioda, `` On experimental confirmation of the correction to  Fermi's %golden rule''
%	Prog. Theor. Exp. Physics.  \textbf{2019}, 033B02,(2019).
%\bibitem{Ushioda} R.~Ushioda, O.~Jinnouchi, K.~Ishikawa and T.~Sloan ,`` Search for the correction to Fermi's golden rule in positron annihilation '' 	Prog. Theor. Exp. Physics.  %\textbf{2020}, April, (2020). 
\bibitem{particle_data_2} S.Navas et al.(Particle Data Group).Phys. Rev. D 110,030001 (2024) and 2025 update, Sec. 34.Passage of Particles , Fig. 34.15.
\bibitem{threshold} K.~Ishikawa, O.~Jinnochi, K.~Nishiwaki,  and K.~Oda, Euo.Phys. J.C 83(2023) 978  http://dx.do.org/10.1140/epj/s10052-0123-12077-7.
%\bibitem{corona} K.~Ishikawa and Y.Tobita, ``Topological interaction of the neutrino with photon: Electro-weak Hall effect, Physics Open 17 (2023) 100174.
%\bibitem{magnetization} K.Ishikawa,  ``Magnetization without  spin: Effective Lagrangin of itinerant electrons'',Nucl.Phys.B 1007(2024) 116663. 
%\bibitem{sky} K.~Ishiakwa and M.Takesada, Hokkaido University preprint. 
\end{thebibliography}
\end{document}